\documentclass[11pt]{article}
\pdfoutput=1
\usepackage{graphicx}
\usepackage{subfigure,amssymb,amsmath,slashed}
\usepackage{cite,color,xcolor,url,cancel,ulem}
\usepackage{bm}
\usepackage{float}
\usepackage{jheppub}

\def\bea{\begin{eqnarray}}   \def\eea{\end{eqnarray}}

\def\gsim{\ ^>\llap{$_\sim$}\ }

\title{Probing Lepton Flavor Violating decays in MSSM with Non-Holomorphic Soft Terms}

\author[a]{Utpal Chattopadhyay,}
\affiliation[a]{School of Physical Sciences, Indian Association for the Cultivation of Science,\\  
	2A \& B Raja S.C. Mullick Road, Jadavpur, 
	Kolkata 700 032, India}

\author[b]{Debottam Das,}
\affiliation[b]{Institute of Physics, Bhubaneswar, Odisha 751005, and HBNI, Mumbai, India}
\author[a]{Samadrita Mukherjee}

\emailAdd{tpuc@iacs.res.in}
\emailAdd{debottam@iopb.res.in}
\emailAdd{tpsm9@iacs.res.in}

\abstract{The Minimal Supersymmetric Standard Model (MSSM)
  can be extended to include non-holomorphic trilinear
  soft supersymmetry (SUSY) breaking interactions that may have distinct signatures. We consider non-vanishing off-diagonal
  entries of the coupling matrices associated with holomorphic (of MSSM)
  and non-holomorphic trilinear terms corresponding to sleptons with elements $A^l_{ij}$ and $A^{\prime l}_{ij}$. We first improve the MSSM charge breaking minima
  condition of the vacuum to include the off-diagonal entries $A^l_{ij}$ (with $i \neq j$).   
  We further extend this analysis for non-holomorphic trilinear interactions.
  No other sources of lepton flavor violation
  like that from charged slepton matrices are considered. We constrain the interaction terms 
  via the experimental limits of processes like charged 
  leptons decaying with lepton flavor violation (LFV) and Higgs boson decaying to charged leptons with
  LFV. Apart from the leptonic decays we compute all the
  three neutral LFV Higgs boson decays of MSSM. We find that an analysis with non-vanishing $A^\prime_{e\mu}$ involving 
the first two generations of sleptons receives the dominant constraint
from $\mu \to e \gamma$. On the other hand, $A^\prime_{e\tau}$ or 
$A^\prime_{\mu\tau}$ can be constrained from the CMS 13 TeV analysis giving limits to the 
respective Yukawa couplings via considering 
SM Higgs boson decaying into $e\tau$ or $\mu\tau$ final states. Contributions from
$A^l_{ij}$ is too little to have any significance compared to the large effect from $A^{\prime l}_{ij}$.
}

\keywords{Supersymmetry, Non-holomorphic soft terms, Lepton flavor violation}
\begin{document}
\begin{flushright}
IP-BBSR-2019-8
\end{flushright}
\maketitle

\vspace{1cm}

\section{Introduction}

The Higgs data is gradually
drifting towards the Standard Model (SM) expectations \cite{Khachatryan:2016vau} ever since the 
first observation of a new resonance at the Large Hadron Collider (LHC)
\cite{Aad:2012tfa, Chatrchyan:2012xdj} in 2012. Still, SM is far away to be a complete description 
of particle physics in view of many theoretical aspects and a few experimental data.
Indeed, the unknown new physics (NP) has always been the driving force
in studying particle physics for many decades to explain e.g., 
the existence of dark matter, neutrino masses, or matter-anti matter asymmetries. Most of these NP models
offer new particles with new interactions that could possibly be tested at the LHC. Apart from this direct test, one can hope 
to find the NP signatures via indirect search experiments involving flavor physics, e.g., through the
dedicated experiments that search for quark
or charged lepton flavor violations (cLFV) like
$b\rightarrow s \gamma$ or $\mu \rightarrow e \gamma$. Among the flavor violating observables, 
the cLFV processes are of particular interest. The reason is that in the context of Standard Model (SM) or in the
minimal extension of the SM that includes the Yukawa interactions in the neutral lepton sector, the decay rates involving
cLFV processes are strongly suppressed (e.g., $BR (\mu\rightarrow e \gamma)\sim 10^{-55}$)\cite{petkov}.
This can be attributed to the tinyness of the neutrino masses which
is the only source of cLFV processes. However, any extension of SM, mainly
in the leptonic sector may offer new particles or new interactions with the
SM leptons. This can potentially change the cLFV decay
rates drastically \cite{Cheng:1976uq,Bilenky:1977du} (for a review see for instance \cite{Raidal:2008jk}).
The Minimal Supersymmetric Standard Model (MSSM), the supersymmetric extension 
of SM with the two Higgs doublets in general may have a large number of 
flavor violating couplings through soft SUSY breaking interactions \cite{SUSYreviews1,SUSYbook1,SUSYbook2, SUSYreviews2, Gabbiani:1996hi}.
The lepton sector is particularly important in this context. 
Interestingly, the flavor violating soft SUSY breaking parameters may also be generated radiatively.
In fact, guided by the origin of cLFV processes, one may broadly classify a few MSSM and beyond the MSSM scenarios as follows:

\begin{itemize}
\item
In the extensions of MSSM, one may connect the origin of the cLFV to the masses for 
neutrinos, the existence of which have been strongly established
through neutrino oscillation experiments
\cite {deSalas:2017kay,Esteban:2016qun}.
One of the most attractive possibilities would be
to consider a seesaw 
mechanism~\cite{seesaw:I, seesaw:II, seesaw:III} that can also be generalized
in the framework of SUSY models (i.e.~the SUSY seesaw) \cite{susy-seesaw}.
In the simplest example i.e, type-I SUSY seesaw, the right handed neutrino Yukawa couplings
that generate neutrino masses, may also radiatively induce the SUSY soft-breaking left handed slepton
mass matrices ($M_{{\tilde L}_{ij}}^2$), 
leading to flavor violations at
low energies \cite{Borzumati:1986qx,hisano}. This can substantially influence
lepton flavor violating decays of the types
($l_j \rightarrow l_i \gamma)$) or three-body lepton decays
$l_j \rightarrow 3 l_i$ through photon, $Z$ or Higgs penguins and the flavor
violating decays of the Higgs scalars
(see e.g., \cite{hisano,Hamzaoui:1998nu,DiazCruz:1999xe,Isidori:2001fv,Babu:2002et,masieroall,Paradisi:2005tk,
Brignole:2003iv,Brignole:2004ah,Han:2000jz,Arganda:2004bz,
Arganda:2005ji,Gomez:2015ila,Gomez:2017dhl,Aloni:2015wvn,Arhrib:2012ax,Arana-Catania:2013xma,Abada:2014kba,Hammad:2016bng, Evans:2018ewb}).
Other important probes are semileptonic $\tau$-decays, $\mu-e$ conversion of nucleus etc. The flavor changing processes involving
Higgs scalars potentially become large for large $\tan\beta$
\cite{Babu:2002et,Brignole:2003iv}.
However here 
the typical mass scales
of the extra particles (such as right handed neutrinos) are in
general very high, often close to the gauge coupling unification scale. An attractive alternative is the inverse seesaw 
scenario, where the presence
of comparatively light right-handed neutrinos and sneutrinos can enhance
the flavor violating decays
\cite{Deppisch,mypaper1,mypaper2,Arganda:2014dta,Arganda:2015naa,Arganda:2015uca}. In addition to generating $M_{{\tilde L}_{ij}}^2$ radiatively,
the right handed neutrino extended models
may as well be embedded in grand unified theory (GUT) framework
\cite{Barbieri:1994pv,Barbieri:1995tw,Hall:1985dx,Dev:2009aw}.

\item
 Though neutrino mass models, particularly in presence
of new states imply cLFV, the later does not necessarily imply neutrino mass
generation.
The simplest example is the R-parity
conserving MSSM. Here
direct sources of flavor violation are in the off-diagonal soft terms of the
slepton mass matrices and 
trilinear coupling matrices (see e.g., \cite{SUSYreviews2})(specifically through
$M_{\tilde e}^2, A_{f}$). One may probe the non-zero off-diagonal elements of all the soft SUSY breaking terms 
($A_{f}, M_{\tilde L}^2, M_{\tilde e}^2$) which may
induce cLFV processes through loops mediated by sleptons-neutralinos and/or sneutrinos-charginos.
This can also be realized in a High scale SUSY breaking
model, e.g., in a supergravity or superstring inspired scenarios, where
non-universal soft terms can be realized
in the high-scale effective Lagrangian (see, for example Ref:\cite{Brignole:1997dp} and references therein) 
apart from running via Renormalization Group (RG) evolution that itself may generate 
flavor violation \cite{SUSYreviews2}. \footnote{We also note in passing that
intricacies related to the large number of soft breaking parameters in the 
cLFV computations and also the inter-generation mixings in the general MSSM can be evaded completely in a High scale SUSY
model
where SUSY breaking is communicated in a flavor blind manner \cite{SUSYreviews2}. Popular examples are mSUGRA, anomaly-mediated supersymmetry
breaking (AMSB) or gauge-mediated supersymmetry breaking (GMSB). }
\end{itemize}

Although, in general, cLFV processes through radiatively induced $M_{\tilde L}^2$ may look more appealing,
in some cases it may be somewhat restrictive in obtaining any significant amount of flavor violating branching ratios.
On the contrary, being free from the
constraints of the masses of neutrinos, soft SUSY breaking parameters can lead to reasonably large decay rates 
which may be interesting and could be testable in near future. 
In this analysis, we would go one step further. We would limit ourselves within the MSSM field content 
augmented by most general soft SUSY breaking terms
without considering their high scale origin. 
In its generic form MSSM includes only holomorphic trilinear soft SUSY breaking
terms \cite{SUSYreviews1,SUSYbook1,SUSYbook2,SUSYreviews2}. However,
in a most general framework, it has been shown 
that certain non-holomorphic supersymmetry breaking terms may qualify as soft terms when there is no gauge singlet field present
\cite{Jack:nh, Martin:nh, Jack:nh1, Haber:wh, Hetherington:2001bk}. Such a
consideration can be phenomenologically interesting.
For example, if MSSM soft SUSY breaking 
sector is  extended to include $A_f^{\prime}\phi^2\phi^*$ type of interactions,
one may find that an SM like CP even Higgs boson with mass $\sim$125 GeV can be achieved with relatively 
lighter squarks with the help of the specific $A^{\prime}_{t}$
\cite{Chattopadhyay:2016ivr}, the relevant coupling from non-holomorphic trilinear interaction. Similarly, the non-holomorphic (NH)
terms may also be helpful 
to fulfill constraints 
from rare B-decays (viz. $Br(B \rightarrow X_s \gamma)$, $Br(B_s \rightarrow \mu^+ \mu^-)$ etc) both in a phenomenological MSSM (pMSSM) like scenario \cite{Chattopadhyay:2016ivr}
or in some high scale model like constrained MSSM (CMSSM) \cite{Un:2014afa, Ross:2016pml, Ross:2017kjc} or minimal Gauge Mediated Supersymmetry Breaking 
(mGMSB)\cite{Chattopadhyay:2017qvh}.
Another interesting feature is that a small NH trilinear coupling (namely $A^{\prime}_{\mu}$) may be capable to attune the inflexible constraints of $(g-2)_{\mu}$ 
\cite{Chattopadhyay:2016ivr}. Focusing only on the leptonic sector,
the playground associated with the NH soft
terms may not be completely free, rather there can be strong constraints appearing
from
different lepton flavor violating decays via their off-diagonal entries \cite{Crivellin:2018mqz} 
\footnote{For chirally-enhanced flavor violating loop processes with most general set of soft terms see \cite{Crivellin:2011jt}.}.
This is of-course similar to holomorphic trilinear
interactions of MSSM, apart from an enhancement by $\tan\beta$ with $A'_i$ associated with down type of quark and lepton.
We will consider the slepton mass squared matrix to be diagonal, and consider that the only source of cLFV
to be the holomormorphic and non-holomorphic trilinear coupling matrices, namely $A_{f} ~ \& ~ A_f^{\prime}$. For the sake of 
explicit understanding we will scan either $A_{f}$ or $A_f^{\prime}$ at a time
to find the allowance of associated off-diagonal elements
under the present and future experimental sensitivities of
different cLFV observables. In order to perform this analysis, a more important checkpoint is avoidance of any
dangerous charge and color breaking global minima (CCB). It is known that a large trilinear coupling (diagonal) in general,
lead to unphysical or metastable CCB minima.
For lepton flavor it is only the charge breaking (CB) of vacuum that is of concern, but it requires the 
involvement of the off-diagonal entries of the trilinear couplings. 
In this analyses we will necessarily improve the charge breaking constraint for a general 
trilinear coupling matrix having non-vanishing entries in its diagonal and 
off-diagonal elements.
So, we should survey if there is any violation in 
charge breaking minima condition and always ensure that the electroweak symmetry breaking minimum i.e., the global minimum is a charge conserving one.

The rest of the work is ordered as follows. In Section 2 we discuss the theoretical framework that includes the slepton mass matrices in presence of the 
NH terms along with the cLFV observables which we would consider in the analysis.
Section 3 covers the examination of analytical structure of charge breaking minima in the context of our model.
Numerical results are presented in Section 4.
Here we take the inter-generational mixings in the trilinear couplings of the slepton fields for both holomorphic
and non-holomorphic couplings. The free rise of these couplings are limited via charge breaking minima and 
also from the non-observation of any signals at the LHC apart from the same experiments that search for cLFV processes. Here
we also consider the future sensitivity of the experiments.
In closing we conclude in Section 5.

\section{Theoretical Framework}

We focus on the general lepton flavor mixing through left and right 
slepton mixing in the MSSM with R-parity conserved. We will further include contributions  
from non-standard soft supersymmetry breaking interactions.
The superpotential is given by,
\begin{equation}
\begin{split}
\label{superpotential}
W_{MSSM} &= \bar{U} {\bf y_u} Q \cdot H_u - \bar{D} {\bf y_d} Q \cdot H_d  - \bar{E} {\bf y_e} L \cdot H_d  + \mu H_u \cdot H_d .
\end{split}
\end{equation}
\\
where $y$'s are the Yukawa matrices in flavor space. 
The MSSM soft terms read \cite{SUSYbook1, SUSYbook2, SUSYreviews2},

\begin{equation}
\label{MSSMsoft}
\begin{split}
-\mathcal{L}_{soft}^{MSSM} &= \frac{1}{2} (M_3 \tilde g\tilde g + M_2 \tilde W \tilde W + M_1 \tilde B\tilde B + c.c )\\
& \quad +({\tilde{u}^*}_{iR} {\bf A_u}_{ij} \tilde{q}_{jL}\cdot H_u + {\tilde{d}^*}_{iR} {\bf A_d}_{ij}\tilde{q}_{jL}\cdot H_d 
+ {\tilde{e}^*}_{iR} {\bf A_{l}}_{ij} \tilde{\ell}_{jL}\cdot H_d +h.c.) \\
& \quad +\tilde q^\dagger_{iL} {\bf M^2_{\tilde q}}_{ij} \tilde q_{jL}
+\tilde \ell^\dagger_{iL} {\bf M^2_{\tilde L}}_{ij} \tilde \ell_{jL}
+{\tilde {\tilde u}^*}_{iR} {\bf M^2_{\tilde u}}_{ij} {\tilde {u}^{*\dagger}}_{jR}
+{\tilde {d}^*}_{iR} {\bf M^2_{\tilde d}}_{ij} {\tilde {d}^{*\dagger}}_{jR}\\
& \quad +{\tilde {e}^*}_{iR} {\bf M^2_{\tilde e}}_{ij} {\tilde {e}^{*\dagger}}_{jR}
 + m_{H_u}^2 H_u^* H_u + m_{H_d}^2 H_d^* H_d + ( B_{\mu} H_u.H_d + c.c ).
\end{split}
\end{equation}
MSSM can be extended to include a set of possible additional 
non-holomorphic trilinear soft SUSY breaking terms as given below \cite{Jack:nh, Jack:nh1, Martin:nh},
\begin{equation}
\begin{split}
\label{nh_lagrangian}
 -\mathcal{L'}_{soft}^{NH} &\supset \tilde{u}^*_{iR} {\bf A_{u}'}_{ij} \tilde{q}_{jL}\cdot H_{d}^* + 
 \tilde{d}^*_{iR}{\bf A_{d}'}_{ij}\tilde{q}_{jL}\cdot H_{u}^*  + \tilde{e}^*_{iR}
       {\bf A_{e}'}_{ij} \tilde{\ell}_{jL}\cdot H_{u}^*  + h.c . ~ .\\
\end{split}
\end{equation}
Here $i,j = 1,2,3$ denote indices in the fermion (f) family space.
In contrast to eq.\ref{MSSMsoft} here the sfermions ${{\tilde u}^*}_{iR}$
and ${\tilde q}_{iL}$ 
couple with $H_d^*$ instead of $H_u$.
The trilinear coupling matrices are typically scaled and
characterized by the quark/charged lepton Yukawa couplings.
The effect of above non-standard or non-holomorphic terms (Eq:\ref{nh_lagrangian}) are reflected in the mass matrices
and mixing angles of physical sparticles. Since we are concerned with general flavor mixing in the slepton sector, 
the general form of $6 \times 6$ 
slepton mass squared matrix is written in the electroweak basis ($\tilde{e}_L, \tilde{\mu}_L, \tilde{\tau}_L, 
\tilde{e}_R, \tilde{\mu}_R, \tilde{\tau}_R$)
in terms of left and right handed blocks as given below.
\begin{eqnarray}
\label{slepton_mass}
  M^2_{\tilde{l}}=&\left(\begin{matrix}
                                 M^2_{\tilde{l}LL} & M^2_{\tilde{l}LR} \\
                                 M^2_{\tilde{l}RL} & M^2_{\tilde{l}RR}\\
                                \end{matrix}\right).
\end{eqnarray}
In the above, each block is a $3 \times 3$ matrix where one has,
\begin{align}
  M^2_{{\tilde{l}LL}_{ij}} &= M^2_{\tilde{L}_{ij}}+
  (M_Z^2(-\frac{1}{2}+\sin^2\theta_{W})\cos2\beta +m_{\ell_i}^2)\delta_{ij},\\ 
 M^2_{{\tilde{l}RR}_{ij}} &= M^2_{\tilde{e}_{ij}}+(-M_Z^2\sin^2\theta_{W}\cos2\beta +m_{\ell_i}^2)\delta_{ij}.
\end{align}
Here $\beta$ is defined via $\tan\beta = \frac{v_u}{v_d}$, the ratio of Higgs vacuum expectation values. 
$\theta_W$ and $M_Z$ refer to the Weinberg angle and the Z-boson mass respectively 
whereas $m_{\ell_i}$ refers to lepton masses respectively.
The non-holomorphic trilinear couplings modify slepton left-right mixings.
For MSSM with non-holomorphic soft 
terms one has the following \footnote{Flavor mixing through trilinear couplings may be generated radiatively in presence
  of right handed neutrinos (see e.g.,\cite{hisano,Gomez:2015ila}).}.

\footnotesize{
\begin{eqnarray}
 \label{M_LR}
  M^2_{\tilde{l}LR} = &\left(\begin{matrix}
                               (A_e - (\mu+A_e^{\prime})\tan\beta) & A_{e\mu} - A_{e\mu}^{\prime}\tan\beta & A_{e\tau} - A_{e\tau}^{\prime}\tan\beta \\
                              A_{\mu e} - A_{\mu e}^{\prime}\tan\beta &  (A_{\mu} - (\mu+A_{\mu}^{\prime})\tan\beta) & A_{\mu\tau} - A_{\mu\tau}^{\prime}\tan\beta \\
                            A_{\tau e} - A_{\tau e}^{\prime}\tan\beta & A_{\tau\mu} - A_{\tau\mu}^{\prime}\tan\beta & (A_{\tau} - (\mu+A_{\tau}^{\prime})\tan\beta)\\
                             \end{matrix}\right)
\end{eqnarray}
\normalsize{
\begin{equation}
 M^2_{\tilde{l}RL} = (M^2_{\tilde{l}LR})^{\dagger}.
\end{equation}
}
%
With only three sneutrino eigenstates, ${\tilde \nu}_{L}$ with $\nu=\nu_e, \nu_\mu, \nu_ \tau$ in MSSM, the sneutrino mass matrix corresponds
to a $3 \times 3$ matrix. 
We note that the non diagonality in flavor comes exclusively from the soft SUSY-breaking parameters.
The main non-vanishing sources for $i \neq j$ are: the masses $M_{\tilde L \, ij}$ for the slepton $SU(2)$ doublets
$(\tilde \nu_{Li}\,\,\, \tilde l_{Li})$, the masses $M_{\tilde e \, ij}$ for the slepton $SU(2)$ singlets $(\tilde l_{Ri})$, and the trilinear couplings ${A}_{ij}$.
Our analysis however would only explore the effects of non-diagonal holomorphic or 
non-holomorphic trilinear couplings that induce mixing in the slepton mass square
matrices ($M^2_{\tilde{\ell}LR}$)\footnote{We are taking the off-diagonal soft mass-squared matrix elements in $M^2_{\tilde{\ell}LL}$ \& $M^2_{\tilde{\ell}RR}$
to be zero at the input scale to probe the effect of trilinear parameters with more clarity.
Still, RGE running would generate non-vanishing off-diagonal 
mass-square matrix elements at a lower scale.}. Regarding sneutrinos, we may write down 
 a corresponding $3\times 3$ mass matrix, with respect to the 
 $(\tilde \nu_{eL}, \tilde \nu_{\mu L}, \tilde \nu_{\tau L})$ electroweak interaction basis 
 in the sneutrino sector, and we have
\begin{equation}
{\mathcal M}_{\tilde \nu}^2 =\left( \begin{array}{c}
M^2_{\tilde \nu \, LL}  
\end{array} \right),
\label{eq:sneu-3x3}
\end{equation} 
where
\begin{equation} 
 M_{\tilde \nu \, LL \, ij}^2 
 =  M_{\tilde L \, ij}^2 + \left( 
   \frac{1}{2} M_Z^2 \cos 2\beta \right) \delta_{ij} \,.
\label{eq:sneu-matrix}
\end{equation} 
 In the above, due to $SU(2)_L$ gauge invariance, the same soft mass $M_{\tilde L \, ij}$ 
 occurs in both the slepton and sneutrino $LL$ mass matrices.  
 
\normalsize{
As is known, within the MSSM, LFV decays
get no tree-level contribution, just like other FCNC decays. They obtain leading 
order contributions at loop level via mediation of sleptons (sneutrino)-neutralinos (charginos).
Here the source of lepton flavor violation can be from any one (or all) entries-
$ M^2_{\tilde{l}LL},  M^2_{\tilde{l}LR},  M^2_{\tilde{l}RR}$ of Eq.\ref{slepton_mass}. But, we will focus on studying 
the impacts of the non-holomorphic trilinear couplings on the cLFV 
observables in the NH-MSSM (which would henceforth be called NHSSM), specially
in comparison to their holomorphic counterparts. Thus
the only source of lepton flavor violation which we would consider here
is associated with the
left-right slepton mixing. This means that sneutrino-chargino loops will hardly carry any importance in our analysis.

\subsection*{\boldmath{$l_j \rightarrow l_i  \gamma$ :}}
Supersymmetric contributions to lepton flavor violating decays $l_j \rightarrow l_i \gamma$
can be sizable and potentially quite large compared to the same for various other BSM physics models. 

The slepton-neutralino and sneutrino-chargino loops mostly 
contribute to the amplitude of $l_j \rightarrow l_i \gamma$,
through charged particles appearing in the
loops.  The general amplitude can be written as \cite{hisano},
\begin{align}
  i\mathcal{M} = ie\epsilon^{\mu*} \overline{u}_i(p-q) \left[
    q^2\gamma_{\mu}(A_1^LP_L + A_1^R P_R) + m_{l_j} i\sigma_{\mu\nu}
    q^{\nu}(A_2^LP_L + A_2^R P_R)\right] u_j(p) \ ,
\end{align}
where $\epsilon^*$ is the photon polarization vector and $q$ being its momentum.
If the photon is
on-shell, the first part of the off-shell amplitude vanishes. Thus, we only need to focus on $A_2^L$ \& $A_2^R$. 
The coefficients $A_2^{L,R}$ that consist of chargino and neutralino contributions are as given below,
\begin{align}
  A_2^{L,R} = A_2^{(\tilde{\chi}^0)L,R} + A_2^{(\tilde{\chi}^\pm)L,R}. 
\end{align}
For the case of our current interest, only 
the NH trilinear couplings may change the slepton mass matrices. So, the flavor violating effects would directly 
enter from the slepton mass matrix elements into the elements of the diagonalizing matrices. 
The $\tilde{\chi}^{\pm} - \tilde{\nu}$ loops are hardly of any importance here because NH couplings do not affect the 
sneutrino mass matrix.
So $A^L_2(\tilde{\chi}^0)$ which corresponds to the contribution from real photon emission, is given by \cite{hisano},
\begin{align}
  A_2^{(\tilde{\chi}^0)L} & = \frac{1}{32 \pi^{2}} \sum_{A=1}^{4}
  \sum_{X=1}^{6} \frac{1}{M_{\tilde{l}_{X}}^{2}} \left[ N_{iAX}^{L}
    N_{jAX}^{L*} \frac{1}{12}F_1^N(x_{AX}) + N_{iAX}^{L} N_{jAX}^{R*}
    \frac{m_{\tilde{\chi}^{0}_{A}}}{3m_{l_{j}}} F_2^N(x_{AX}) \right],
\end{align}
where, $x_{AX} =\frac {m_{\tilde{\chi}^{0}_{A}}^2}{M^2_{\tilde{l_X}}}$. One obtains $A^R$
by simply interchanging $L\leftrightarrow R$.
The loop functions denoted by $F_1$, $F_2$ and
the couplings $N_{AX}^{L,R}$ can be read from \cite{hisano,Rosiek:1995kg}.

\begin{align}
 F_1^N(x) & = \frac{2}{(1-x)^4}\left[1-6x+3x^2-6x^2\log x\right] ,
      \nonumber
      \\
      \& ~~F_2^N(x) & = \frac{3}{(1-x)^3}\left[1-x^2+2x\log x\right],
      \label{loop-functions}
\end{align}
and
  \begin{align}
    N_{iAX}^{L} & = -\sqrt{2} g_{1} \left(Z_{L}^{i+3,X}\right)^{*}
    Z_{N}^{1A} + Y_{l_{i}} \left(Z_{L}^{i,X}\right)^{*} Z_{N}^{3A} ,
    \nonumber \\
    N_{iAX}^{R} & = \frac{(Z_{L}^{i,X})^{*}}{\sqrt{2}} \left(g_{1}
      \left(Z_{N}^{1A}\right)^{*} + g_{2} \left(Z_{N}^{2A}\right)^{*}
    \right) + Y_{l_{i}} \left(Z_{N}^{3A}\right)^{*}
    \left(Z_{L}^{i+3,X}\right)^{*}  .
    \label{liljicoupling}
  \end{align}

Clearly, the couplings $N_{AX}^L$ involve
$Z_N$ \& $Z_L$ the neutralino and slepton mixing matrices respectively  
that transform them from the electroweak basis to the 
mass basis. $Z_L$ is the 
$6\times6$ slepton mixing matrix that allows for flavor changes in the loop, leading to 
the flavor violation. One may also evaluate the process using the mass
insertions in the slepton mixing matrix (through Eq: \ref{M_LR}) which
depend on the diagonal and non-diagonal entries of the NH trilinear coupling matrix
$A_{\ell}^{(')}$. The associated Feynmann diagram are shown in fig. \ref{liljgammanew}. For example,
in case of $\mu \to e \gamma$, in the slepton mixing  
would be induced by $A_{e\mu}-A'_{e\mu}\tan\beta$. This indicates a typical domination
of $A'_{e\mu}$ unless $A_{e\mu}$ is too large or there is much cancellation.
Finally, the decay rate is given by,
\begin{align}
  \Gamma(l_j \rightarrow l_i\gamma) = \frac{e^2}{16 \pi} m^5_{l_j}
  \left( \left|A^L_2\right|^2 + \left|A^R_2\right|^2 \right) .
\end{align}

\smallskip

\begin{figure}[!htb]
 \begin{center}
  \includegraphics[width=0.50\textwidth]{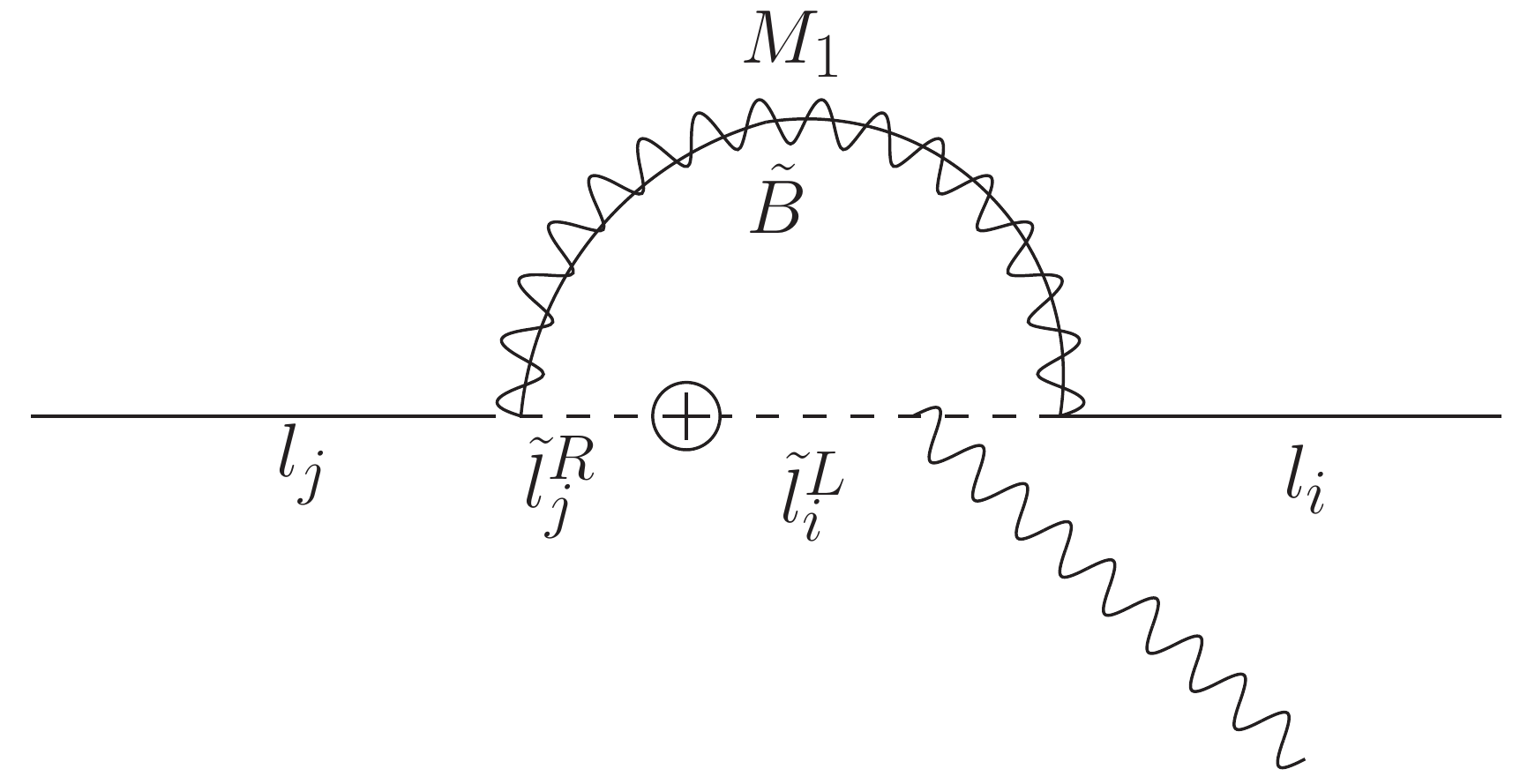}
  \caption{Slepton and neutralino induced Feynman diagram for the process $l_j \rightarrow l_i \gamma$.}
  \label{liljgammanew}
 \end{center}
\end{figure}

\subsection*{\boldmath{$l_j \rightarrow 3l_i$ :}}

In the Standard Model (SM), $l_j \rightarrow 3l_i$ has a vanishingly small branching fraction, e.g. $Br(\tau \rightarrow 3\mu)< 10^{-14}$ \cite{Pham:1998fq},
while various models of beyond the SM may predict this particular process
to be of the order of $10^{-10} - 10^{-8}$.
The current experimental limit of the same BR
is of the order of few times
$10^{-8}$ \cite{Lees:2010ez, Hayasaka:2010np, Aaij:2014azz, Amhis:2014hma} which has much better sensitivity compared to the 3-body decays of $\mu$.
The main experimental obstacle to improve
the sensitivity with $\tau$ leptons is the fact that $\tau$ is not produced in large numbers.
The amplitude for $l_j \rightarrow 3l_i$ comprises of contributions from $\gamma, Z, \phi(=h,H,A)$ penguin diagrams 
and the box diagram as
shown in fig. \ref{lito3lj}
with slepton (sneutrino)-neutralino (chargino) appearing inside the loop. Detailed expressions for 
the diagrams may be found in 
\cite{hisano,Arganda:2005ji,Babu:2002et}. The effects of NH off-diagonal elements toward LFV appear via slepton mass matrices.
This induces an effective vertex $\phi, \gamma, Z -l_i-\bar{l}_j$ which in turn leads to
processes like $l_j \rightarrow 3l_i$. All the penguin processes would further be
boosted in case of non-holomorphic couplings via additional $\tan\beta$ factor 
(see Eq: \ref{M_LR})).
  Similar to the MSSM case the
  dominance of $\gamma$-penguins in the cLFV
  processes in NHSSM also holds good. This is irrespective
  of the fact that Higgs penguin contributions that scale as
  $\tan^6\beta$ are expected 
  to be large for large $\tan\beta$ 
  \cite{Babu:2002et, Dedes:2002rh}.
  The dominance of $\gamma$-penguins is also valid in relation to the  
  Z and box contributions\footnote{For relative contributions see Fig.5 of Ref.\cite{Arganda:2005ji}.}.
 
\begin{figure}[!htb]
 \begin{center}
 \subfigure[]{
  \includegraphics[width=0.22\textwidth]{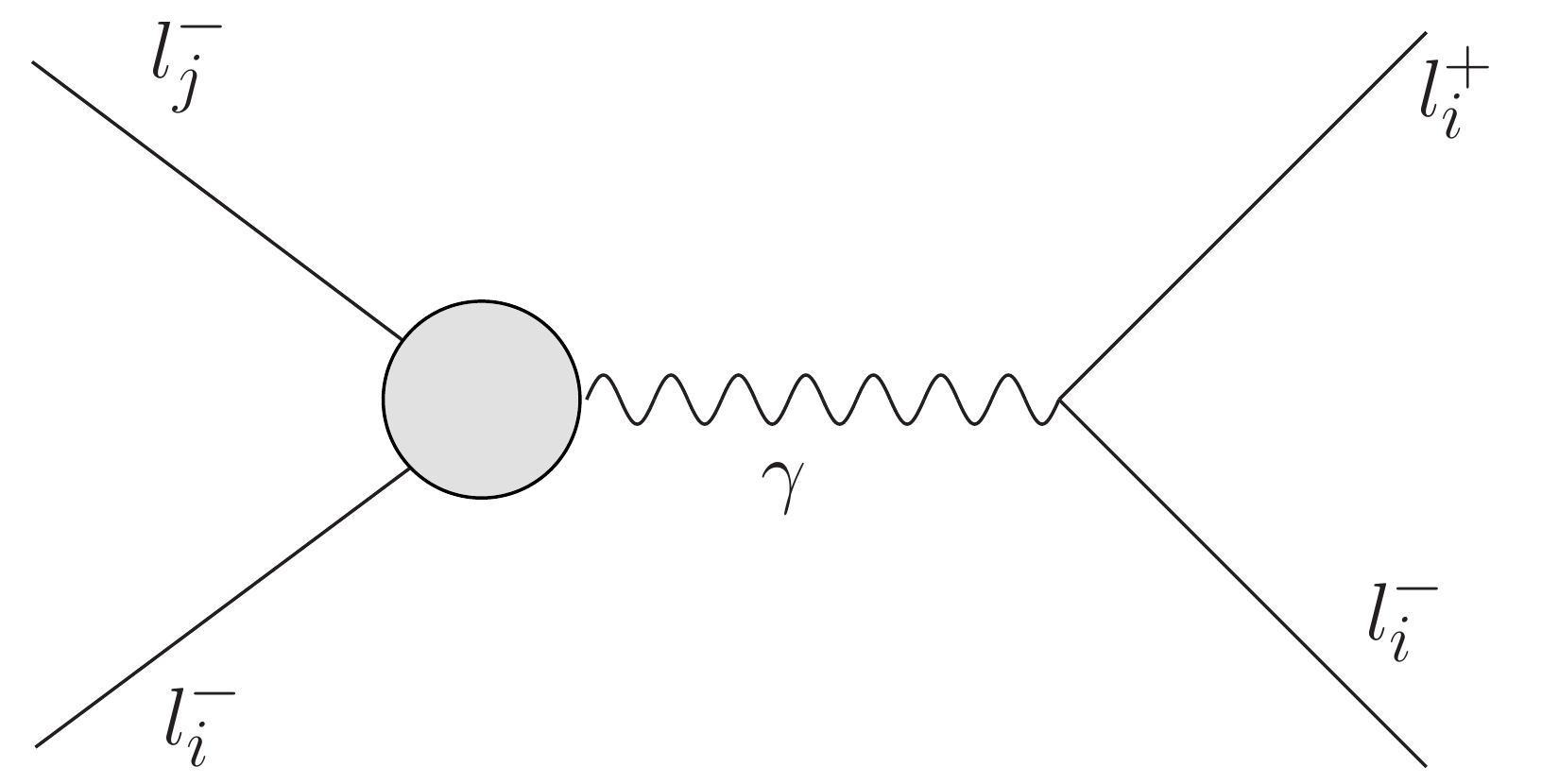}
  }
  \subfigure[]{
  \includegraphics[width=0.22\textwidth]{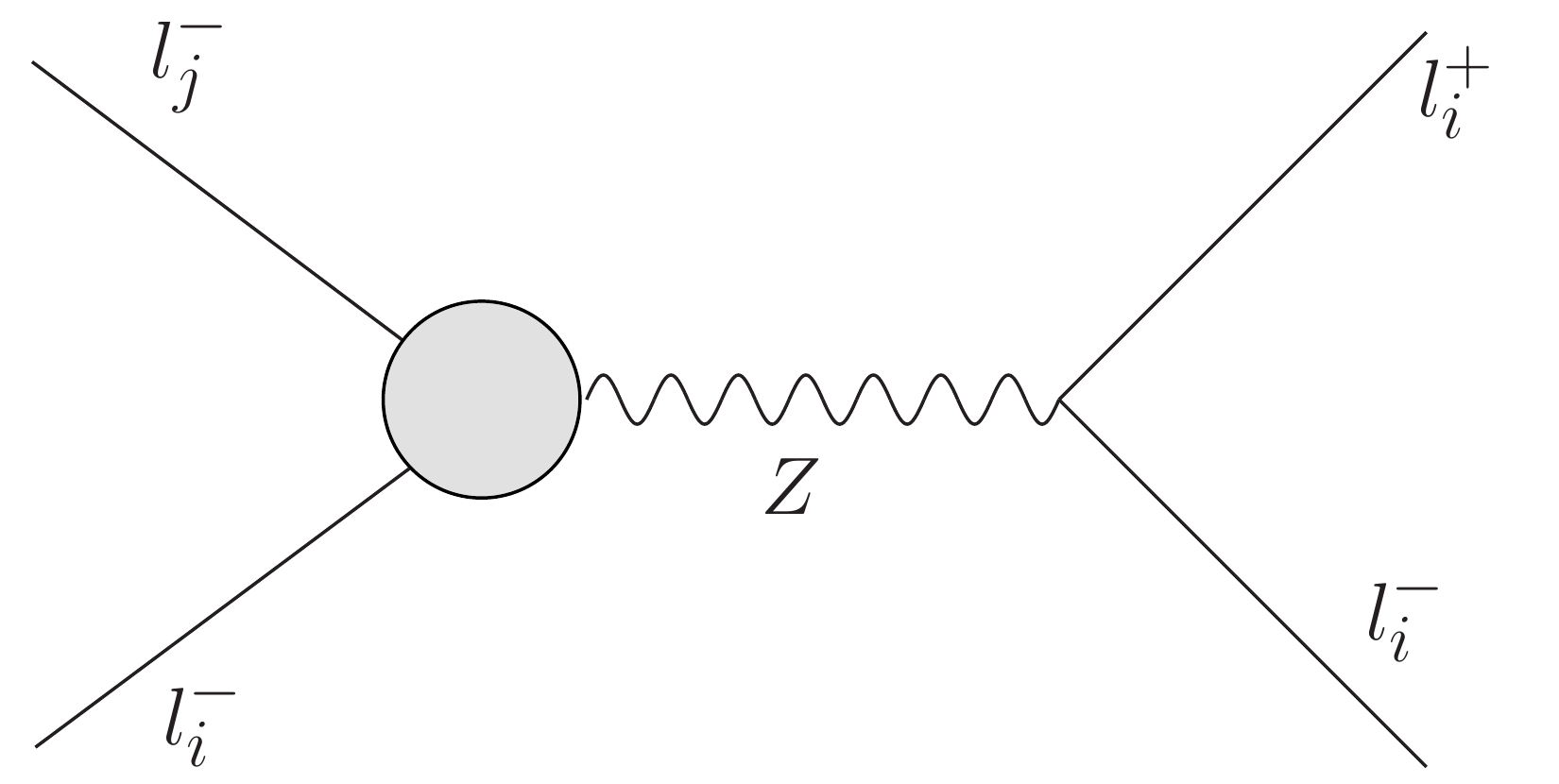}
  }
  \subfigure[]{
  \includegraphics[width=0.22\textwidth]{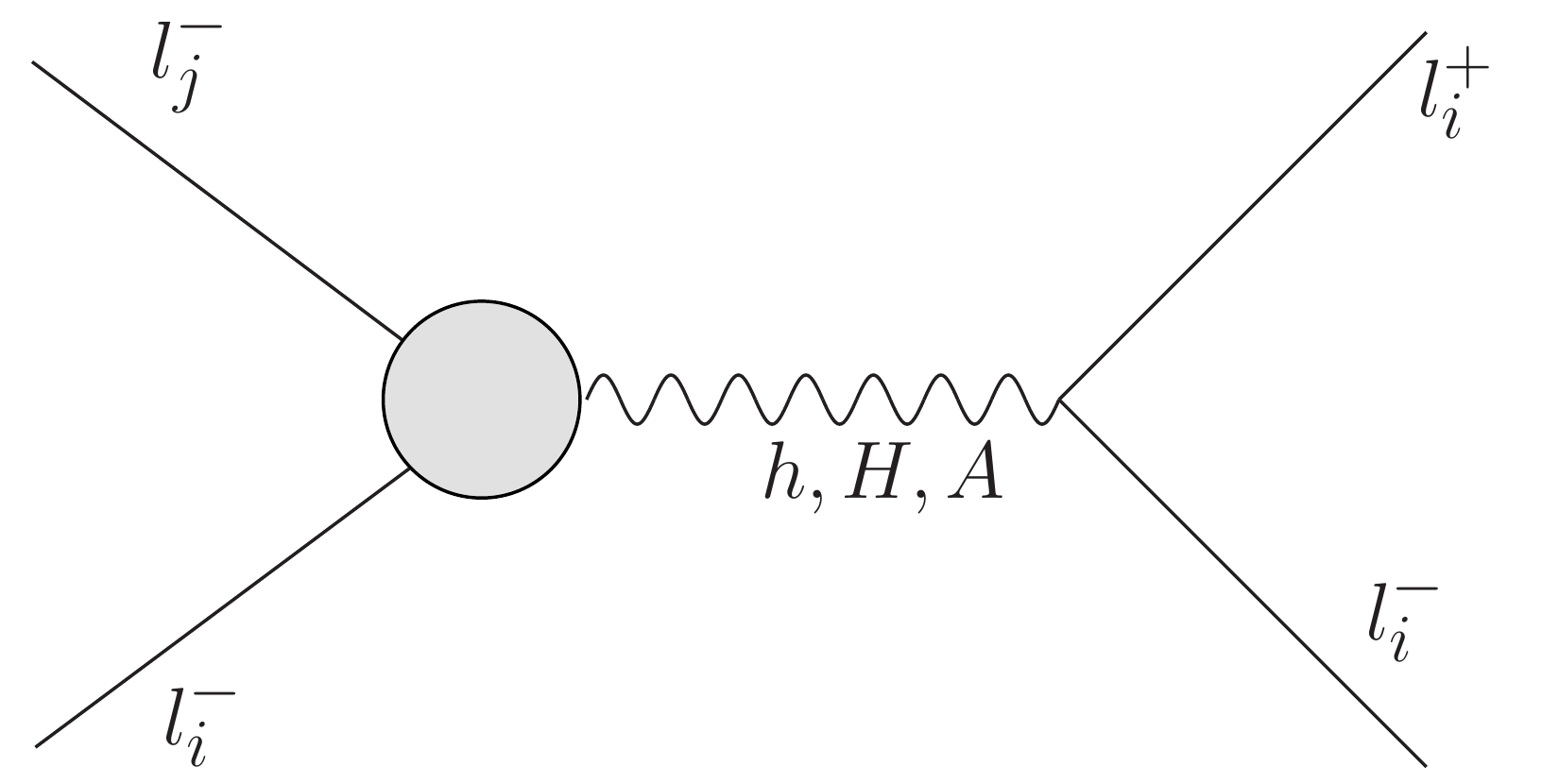}
  }
  \subfigure[]{
  \includegraphics[width=0.22\textwidth]{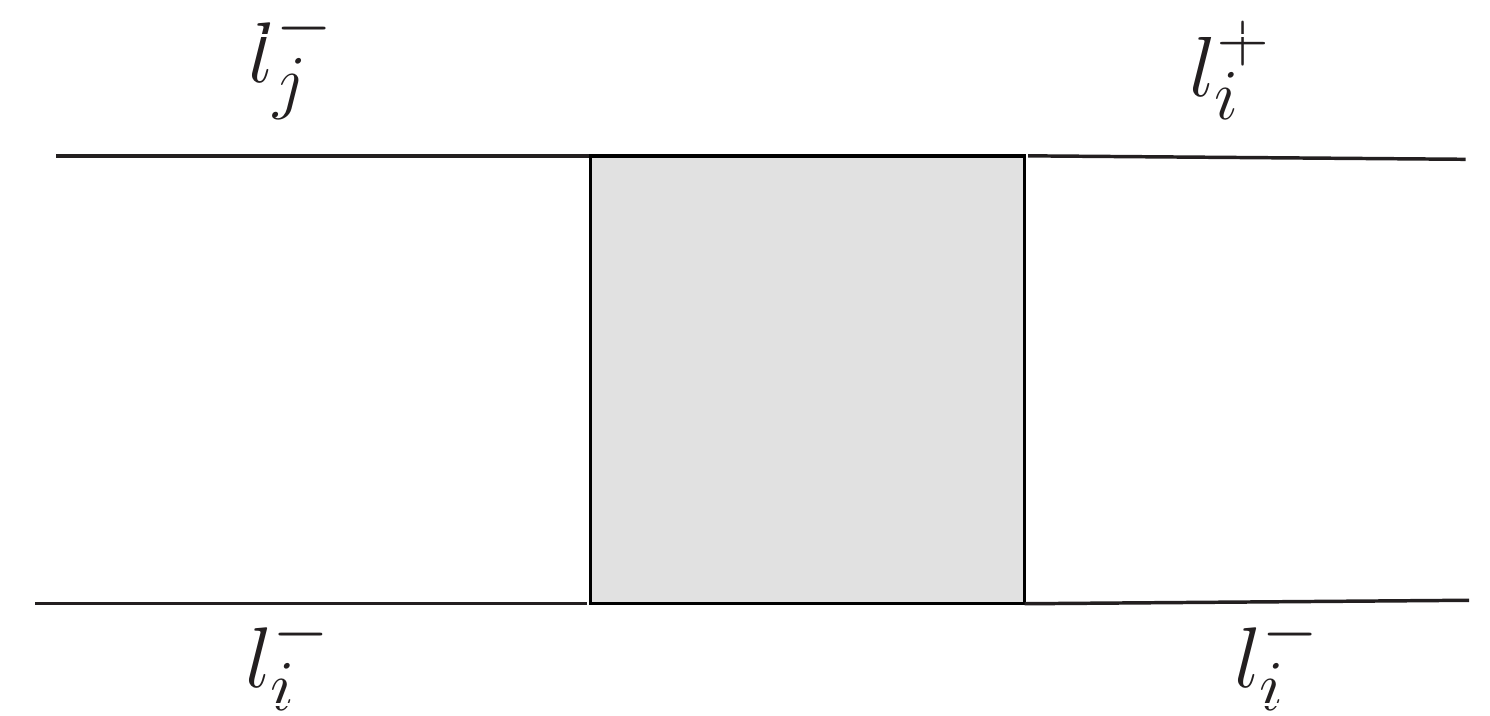}
  }
  \caption{Photon, Z-boson and neutral Higgs boson mediated penguin diagrams and box-type of diagram contributing to $l_j \rightarrow 3l_i$ decay.}
  \label{lito3lj}
 \end{center}
\end{figure}

\subsection*{\boldmath{$\phi(h,H,A) \rightarrow l_i \bar{l}_j$ :}}
Flavor changing Higgs decays can play significant roles for investigating lepton flavor violation.
The same Higgs mediated penguin diagrams, induced by $\phi-\tilde{l}_i-\bar{\tilde{l}}_j$ vertex, may effectively contribute in 
$\phi-l_i-\bar{l}_j$ vertex through loops leading to Higgs flavor violating decays.
The effective Lagrangian representing the interaction between neutral Higgs boson and charged leptons is given by \cite{Dedes:2002rh, Babu:2002et},
\begin{align}
 -\mathcal{L}_{eff} = \bar{e}^i_R y_{e_{ii}}\big[ \delta_{ij} H^0_d + (\epsilon_1\delta_{ij}(A_{ij}H_d^0 - (\mu+A'_{ij})H_u^{0*}))
 + \epsilon_{2ij}(A_{ij}H_d^0 - A'_{ij}H_u^{0*})\big]l_L^j + h.c.
 \label{hlleff}
\end{align}

\begin{figure}[!htb]
 \begin{center}
  \includegraphics[width=0.89\textwidth,height=0.20\textheight]{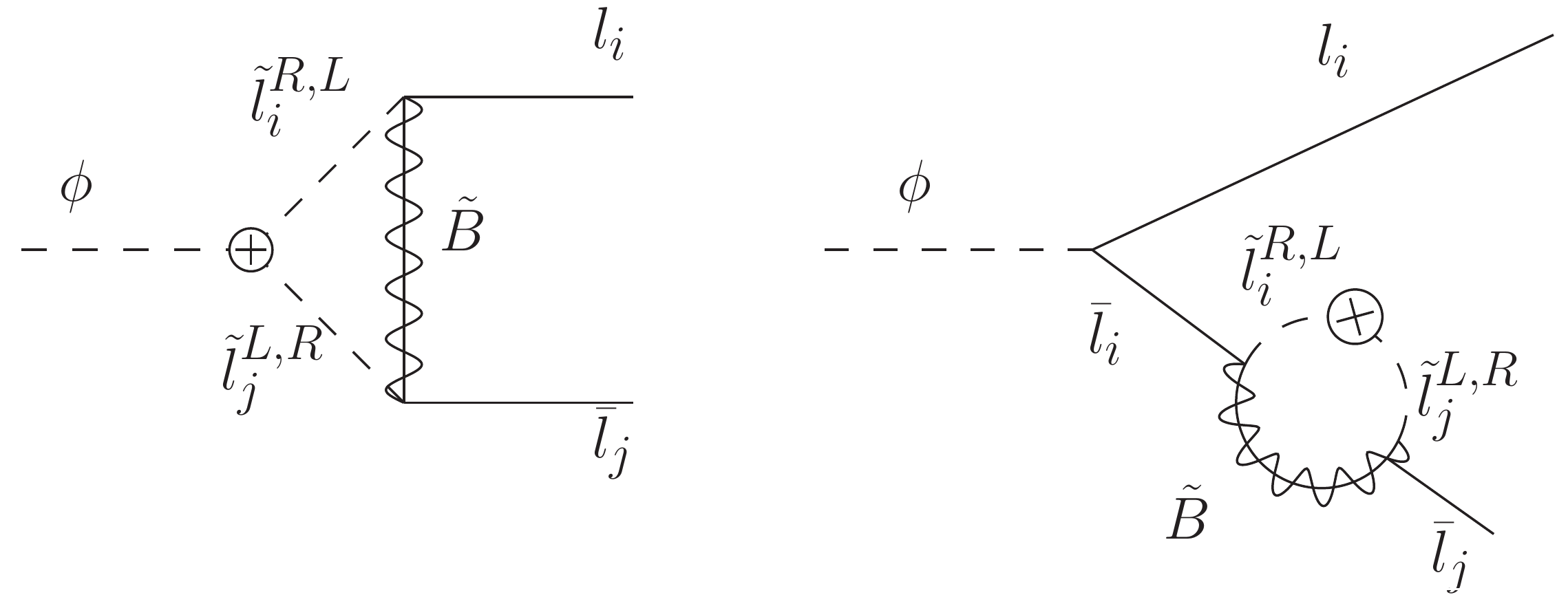}
  \caption{One loop diagrams contributing to computation of $Br(\phi \rightarrow l_i \bar{l_j})$ with LR \& RL mixing.}
 \end{center}
 \label{fig:htomutau_1}
\end{figure}


The first term of the above equation denotes the Yukawa interaction whereas $\epsilon_1$ encodes the corrections to the charged lepton Yukawa couplings from flavor 
conserving loops \cite{Babu:2002et}.
The last term in 
Eq.\ref{hlleff} corresponds to the source of flavor violation through the insertion of $(A_{ij}-A'_{ij}\tan\beta)$ in the slepton arms inside the loops. 
$\epsilon_2$ arises out of loop functions involving neutralino and slepton masses owing to various cLFV processes. 
The effective Lagrangian in Eq:\ref{hlleff} esentially generates all the off-diagonal Yukawa couplings radiatively if the respective 
holomorphic (non-holomorphic) trilinear couplings $A_{ij} ~ (A'_{ij})$ are non zero. This in turn produces flavor violating decays of 
Higgs scalars or lepton 3-body decays induced by the Higgs penguins. Among the Higgs mediated diagrams, typically
dominant contributions come from the CP-odd Higgs exchange $A$ for large $\tan\beta$. This may be understood from the effective Lagrangian describing
couplings of the physical Higgs bosons to the leptons, which can be derived from  Eq.~(\ref{hlleff})\cite{Dedes:2002rh,Babu:2002et}.

\begin{align}
-{\cal L}^\text{eff}_{i\neq j} =
(2G_F^2)^{1/4} \,
\frac{m_{E_i} \kappa^E_{ij}}{\cos\beta}
\left(\bar e^i_{R}\,l^j_{L}\right)
\left[\cos(\alpha-\beta) h + \sin(\alpha-\beta) H - i A\right]+\text{h.c.}.
\label{Leffl}
\end{align}
Here,
$\alpha $ is the CP-even Higgs mixing angle and $\tan\beta=v_u/v_d$, and 
 \begin{align}
  \kappa^E_{ij} &= \frac{\epsilon_2 (A_{ij}-A'_{ij}\tan\beta)} { [1+(\epsilon_1(A_{ii}-(\mu + A'_{ii})\tan\beta))
 +\epsilon_2 (A_{ii}-A'_{ii}\tan\beta))] }
 \label{kappa}
 \end{align}

Since the cLFV branching ratios are proportional to $({\kappa^E_{ij}})^2$, from
the above equation it is clear that the non-holomorphic trilinear couplings via $\tan\beta$ 
enhancement may
have greater importance towards Higgs mediated processes. In fact,
all the Higgs mediated flavor violating observables may receive large boost, 
while assuming no cancellation in eq. \ref{kappa} or if holomorphic $A_{ij}$
is negligible compared to $A'_{ij}$. In case of flavor violating 
Higgs decays the branching fraction $\phi_k\to \mu\tau$ where Higgs bosons
$h,H,A$ are denoted as $\phi_k$ for $k=1,2,3$ can be related to the flavor
conserving decay $\phi_k\to \tau \tau$ as follows\cite{Brignole:2003iv}.  
\bea
\label{br_heavy}
{\text{Br}(\phi_k\to \mu\tau)}
=  \tan^2\beta~ (|\kappa_{\tau\mu}^{E}|^2 ) 
 ~ C_\Phi~ { \text{Br}(\phi_k\to \tau\tau)}  \, ,
\eea
where we used $1/\cos^2\beta \simeq \tan^2\beta$. 
The coefficients $C_\Phi$  are given by,
\bea
\label{C_phi}
C_h = \left[\frac{\cos(\beta - \alpha)}{\sin\alpha}\right]^2, ~~~~
C_H = \left[\frac{\sin(\beta - \alpha)}{\cos\alpha}\right]^2, ~~~~
C_A = 1.
\eea
\section{Analysis of Charge Breaking Minima }

Absence of any flavor changing neutral current (FCNC) significantly constrains the off-diagonal elements in the mass and trilinear 
coupling matrices.
However, the Charge and Color Breaking(CCB) constraints are more robust than the 
corresponding FCNC data \cite{Casas:1996de}. In a multi-scalar theory, the existence of several vacua and choice of the desired electroweak symmetry breaking 
put strong constraints on the allowed parameter space. In this context, we first put the effort to analyze the charge
breaking bounds for two generations of sleptons associated with the 
$(\tilde{\mu} - \tilde{\tau})$ sector. Here trilinear couplings can accommodate 
the off-diagonal entries of both the holomorphic and
non-holomorphic soft SUSY breaking terms. Then we will generalize it for all the three generations of sleptons.

Three basic components of tree level scalar potential $V_0$ are the F-term, D-term and the soft breaking terms, $V_0 = V_F + V_D + V_{soft}$.
The constituents of $V_0$ are given as follows,
\begin{align}
 V_F &= \sum_a \Big|\frac{\partial W}{\partial \phi_a}\Big|^2, \\ \nonumber
\end{align}
here the superpotential W is given by Eq. \ref{superpotential}. The D-term part gives additional quartic terms for scalar potential
associated with gauge couplings $g_a$.
\begin{align}
 V_D &= \frac{1}{2} \sum_a g_a^2 ~ (\sum_a \phi_a^{\dagger} T^a \phi_a)^2 \\ \nonumber
\end{align}
The Holomorphic and non-holomorphic soft terms in V (and hence in $-\mathcal{L}$) can be written as,
\begin{align}
 V_{soft} &= \sum M^2_{\phi_a} |\phi_a|^2 + ({\tilde{u}^*}_{iR} {\bf A_u}_{ij} \tilde{q}_{jL}\cdot H_u + {\tilde{d}^*}_{iR} {\bf A_d}_{ij}\tilde{q}_{jL}\cdot H_d + {\tilde{e}^*}_{iR} {\bf A_{e}}_{ij} \tilde{\ell}_{jL}\cdot H_d + h.c.), \\
 V_{soft}^{NH} &= \tilde{u}^*_{iR} {\bf A_{u}'}_{ij} \tilde{q}_{jL}\cdot H_{d}^* + 
 \tilde{d}^*_{iR}{\bf A_{d}'}_{ij}\tilde{q}_{jL}\cdot H_{u}^*  + \tilde{e}^*_{iR} {\bf A_{e}'}_{ij} \tilde{\ell}_{jL}\cdot H_{u}^*  +  h.c .\\  \nonumber           
\end{align}

In the above, $\phi_a$ runs over all the scalar components of chiral superfields.
The full MSSM scalar potential may indeed have several minima where squarks
or sleptons may additionally acquire non-zero vevs which may in turn lead to charge and/or color breaking vacua. Since the violation of charge and/or
color quantum number is yet to be observed, it is understood that the universe at present
is at a ground state which is Standard Model like (SML) \cite{Chattopadhyay:2014gfa}, with only neutral
components of the Higgs scalars acquiring vevs.
A priori, it indicates that those parts of the multi-dimensional parameter space
corresponding to MSSM scalar potential that allow a deeper charge and color breaking
(CCB) minima \cite{Frere1983,Casas:1995pd, Gunion1988, Dress1985, Komatsu1988, Langacker:1994bc, Strumia:1996pr, Chattopadhyay:2014gfa} should be excluded.
The dangerous directions could be
associated with unacceptably large trilinear couplings, in particular $A_t$, $A_b$, the ones associated with top and bottom Yukawa couplings.
Thus, one may have a CCB vacuum that is deeper than the desired EWSB vacuum.
Analyses of CCB constraints in MSSM may be seen in Refs \cite{Hollik:2016dcm, Casas:1995pd, Casas:1996de, Chattopadhyay:2014gfa, Chowdhury:2013dka} 
a related study on non-holomorphic soft terms was made in Ref.\cite{Beuria:2017gtf}.
%
Here one should note that the rate of tunneling from SML false vacuum to such CCB true vacuum
is roughly proportional to $e^{-a/y^2}$, where
$a$ is a constant of suitable dimension that can be
determined  via field theoretic calculations and $y$ 
is the Yukawa coupling. The tunneling rate is enhanced for large Yukawa couplings 
\cite{Kusenko:1996jn,Kusenko:1996xt,Kusenko:1996vp,Kusenko:1995jv,
Brandenberger:1984cz,LeMouel:2001ym}, thus leading to large effects from the third generation of sfermions.

But it is not always true that, trilinear terms of third generations of squarks/sleptons
are the most important for charge and color breaking minima. With the 
variation of non-holomorphic 
soft terms, there can be significant changes in all of the corresponding Yukawa couplings through loops \cite{Chattopadhyay:2018tqv}
\footnote{In this context we note that the
  radiative effects will be larger in the strongly interacting sector as in the analysis of \cite{Chattopadhyay:2018tqv} involving $Y_b$. In the context of the
  present analysis, both the relevant diagonal and
  off-diagonal
  Yukawa couplings involving $\tau$
  may get affected. However, a change in the 
  off-diagonal Yukawas will be
  manifestly more relevant in Eq.\ref{kappa} via 
  the appearance of $\epsilon_2$ in the numerator. In contrast to the above, the
  denominator can hardly be different from unity unless
  the diagonal NH trilinear soft parameters are too large.} 
and other two generations of squarks/sleptons can have notable effect in charge and color breaking condition \cite{Beuria:2017gtf}.
Here we will study the analytic expressions for only charge breaking minima for slepton soft masses and
slepton trilinear couplings (both holomorphic and non-holomorphic) considering all the three generations.
We will particularly generalize our studies to include first the effect of non-vanishing, non-diagonal trilinear soft terms. 
Furthermore, we will consider only the case of absolute stability of the vacuum, i.e. without trying to analyze any tunneling effect.

\subsection{Charge breaking with flavor violation in MSSM}

In this subsection we first analyze the effect of non-vanishing off-diagonal entries $A^{(')}_{ij}$ on charge breaking in MSSM.
The relevant terms in $V_F$, $V_D$ and $V_{soft}$ of MSSM related to the slepton sector are as given below,
\begin{align*}
  V_F &= |\mu^{*} H_u^{-} - \tilde{\nu}^{*}_{iL} y_{ij} \tilde{e}_{jR}|^2 + |\mu^{*} H_u^0 - \tilde{e}^{*}_{iL} y_{ij} \tilde{e}_{jR}|^2 
        + \sum_l |y_{ij} H_d^0 \tilde{l}_{jL}|^2 + y_{ij} y^\star_{ij'} \tilde{e}_{jR}^{*} \tilde{e}_{j'R} (|H_d^0|^2 + H_d^{+}H_d^{-}), \\
  V_D &= V_{D_{Y}} + V_{\vec{D}}\\
      &= \frac{g^2_1}{8}(|H_u^0|^2 - |H_d^0|^2 - |\tilde{l}_{iL}|^2 + 2|\tilde{e}_{iR}|^2)^2
        + \frac{g^2_2}{8}(|H_u^0|^2 - |H_d^0|^2 + |\tilde{l}_{iL}|^2 )^2, \\
  V_{soft}^{\tilde{l}} &= \tilde{l}_{iL}^{\dagger} (M^2_{\tilde{L}})_{ij} \tilde{l}_{jL} + \tilde{e}_{iR}^{*} (M^2_{\tilde{e}})_{ij} \tilde{e}_{jR}^{\dagger}
  + [\tilde{e}_{iR}^{*} {\bf A_{lij}} \tilde{l}_{jL} \cdot H_d + h.c ]~,\\
V_{soft}^{H} &=   m^2_{H_d}|H_d^0|^2 + m^2_{H_u}|H_u^0|^2 - 2Re(B\mu H_d^0 H_u^0).
\end{align*}

Below, we collect the terms, originating from $V_F$, $V_D$, $V^{\tilde{l}}_{soft}$ and $V_{soft}^{H}$ appearing in the diagonal and off-diagonal elements
for the 2nd and 3rd generations of sleptons (viz. smuon and stau). 

\begin{align*}
 V_{\mu}^{diag} &= \tilde{\mu}_{L}^{*} (M^2_{{\tilde{L}}_{22}} + |y_{\mu}H_d^0|^2)\tilde{\mu}_{L} + \tilde{\mu}_{R}^{*} (M^2_{{\tilde{e}}_{22}} + |y_{\mu}H_d^0|^2)\tilde{\mu}_{R}
                 + [\tilde{\mu}_{L}^{*} (A_{\mu}h_d - \mu^{*} y_{\mu}h_u)\tilde{\mu}_{R} + h.c] + |y_{\mu}|^2 |\tilde{\mu}_{L}|^2|\tilde{\mu}_{R}|^{2} ~,\\
\\                 
 V_{\tau}^{diag} &= \tilde{\tau}_{L}^{*} (M^2_{{\tilde{L}}_{33}} + |y_{\tau}H_d^0|^2)\tilde{\tau}_{L} + \tilde{\tau}_{R}^{*} (M^2_{{\tilde{e}}_{33}} + 
                  |y_{\tau}H_d^0|^2)\tilde{\tau}_{R}
               + [\tilde{\tau}_{L}^{*} (A_{\tau}H_d^0 - \mu^{*} y_{\tau}H_u^0)\tilde{\tau}_{R} + h.c] + |y_{\tau}|^2 |\tilde{\tau}_{L}|^2|\tilde{\tau}_{R}|^{2} ~ ,\\
\\               
 V_{\mu \tau} &= \tilde{\mu}_{L}^{*} M^2_{{\tilde{L}}_{23}} \tilde{\tau}_{L} + \tilde{\mu}_{R}^{*} [M^2_{{\tilde{e}}_{23}}+|{\bf y_{\mu\tau}} H_d^0|^2 ]\tilde{\tau}_{R}
                 + [\tilde{\mu}_{L}^{*} (A_{\mu\tau}H_d^0 - \mu^{*} y_{\mu\tau}H_u^0)\tilde{\tau}_{R} + h.c] + |y_{\mu\tau}|^2 |\tilde{\mu}_{L}|^2|\tilde{\tau}_{R}|^{2}
               + \tilde{\tau}_{L}^{*} M^2_{{\tilde{e}}_{32}}\tilde{\mu}_{L} \\
&+ \tilde{\tau}_{R}^{*} [M^2_{\tilde{e}_{32}} + |{\bf y_{\tau\mu}} H_d^0|^2 ] \tilde{\mu}_{R}
                  + [\tilde{\tau}_{L}^{*} (A_{\tau\mu}H_d^0 - \mu^{*} y_{\tau\mu}H_u^0)\tilde{\mu}_{R} + h.c] + |y_{\tau\mu}|^2 |\tilde{\tau}_{L}|^2|\tilde{\mu}_{R}|^{2} ~ ,\\
                  \\
                  V_H &=(m^2_{H_u}+|\mu|^2)|H_u^0|^2+(m^2_{H_d}+|\mu|^2)|H_d^0|^2 ~,\\
 V_{D} &= \frac{g_1^2}{8}(|H_u^0|^2 - |H_d^0|^2 - |\tilde{\mu}_{L}|^2 - |\tilde{\tau}_{L}|^2 + 2|\tilde{\mu}_{R}|^2 + 2|\tilde{\tau}_{R}|^2)^2 
                + \frac{g_2^2}{8}(|H_u^0|^2 - |H_d^0|^2 + |\tilde{\mu}_{L}|^2 + |\tilde{\tau}_{L}|^2)^2.               
\end{align*}
In the first place we consider non-vanishing \textit{vev}s for the neutral
components of the two Higgs 
scalars and the stau and smuon fields. The latter are responsible for the 
generation of charge breaking minima. Allowing both $H_u^0$ and $H_d^0$ to fluctuate in the positive and negative directions, we choose to constrain the slepton 
fields with a particular scalar field value $\phi$.
In this specific direction one has,
\begin{align}
\label{flat-direction}
|\tilde{\tau_{L}}| = |\tilde{\tau_{R}}| = \alpha\phi~ ,\nonumber \\
|\tilde{\mu_{L}}| = |\tilde{\mu_{R}}| = \beta\phi~ , \nonumber \\
H_d^0 = \phi~ ,  \\
H_u^0 = \eta\phi. \nonumber  
\end{align}
with $\eta$ being any real number and $\alpha$, $\beta$ to be real and positive. The total tree-level scalar potential involving Higgs, smuon and stau fields,
assuming $\mu$ to be real and $y_{ij}$ or $A_{ij}$ referring to real symmetric matrices, reduces to,
\begin{equation}
V_{\tilde{l},H} = A\phi^2 + B\phi^3 + C\phi^4,
\end{equation}
where, 
\begin{align*}
 A &= \alpha^2 (M_{\tilde{L}_{33}}^2 + M_{\tilde{e}_{33}}^2) + \beta^2 (M_{\tilde{L}_{22}}^2 + M_{\tilde{e}_{22}}^2) 
 + 2\alpha\beta (M_{\tilde{L}_{23}}^2 + M_{\tilde{e}_{23}}^2) + m^2_{H_d} + \eta^2m^2_{H_u} 
     + (1 + \eta^2)|\mu|^2 - 2B\mu\eta~, \\
 B &=  2\alpha^2 (A_{\tau} - \mu y_{\tau}\eta) + 2\beta^2 (A_{\mu} - \mu y_{\mu}\eta) + 4\alpha\beta (A_{\mu\tau} - \mu y_{\mu\tau}\eta)~ , \\
 C &=  \frac{g_1^2 + g_2^2}{8}(\eta^2 -1 + \beta^2 + \alpha^2)^2 + (2 + \alpha^2)\alpha^2 y_{\tau}^2 + (2 + \beta^2)\beta^2 y_{\mu}^2
       + 2\alpha^2\beta^2 y_{\mu\tau}^2.
\end{align*}

We require that the minima at $\langle\phi\rangle = 0$ should be deeper than a minima with $\langle\phi\rangle \neq 0$ and this is possible when 
$B^2(\alpha, \beta, \eta) < 4A(\alpha, \beta, \eta) ~ C(\alpha, \beta, \eta)$.
Here we consider
a scenario with 6 \textit{vev}s corresponding to L and R components of smuon and stau fields apart from the neutral Higgs fields corresponding to eq. \ref{flat-direction}.

We want to have the most stringent condition that would avoid the charge breaking minima.
Thus, in the D-flat direction, which explicitly demands that all the $g^2_i$ terms in the tree level scalar potential to be absent, we choose,

\begin{align*}
 \alpha &= \frac{1}{\sqrt{2}}, ~
 \beta = \frac{1}{\sqrt{2}}, ~
 \eta = 0,
\end{align*}
so that, $(\eta^2 -1 + \alpha^2 +\beta^2) = 0$.

Thus we obtain,
\begin{align*}
 A &= \frac{1}{2} (M_{\tilde{L}_{33}}^2 + M_{\tilde{e}_{33}}^2) + \frac{1}{2} (M_{\tilde{L}_{22}}^2 + M_{\tilde{e}_{22}}^2) 
 +  (M_{\tilde{L}_{23}}^2 + M_{\tilde{e}_{23}}^2) + m^2_{H_d} + |\mu|^2 ~,\\
 B &=  A_{\tau} + A_{\mu} + 2A_{\mu\tau}~, \\
 C &=  \frac{5}{4}(y_{\tau}^2 + y_{\mu}^2 + \frac{2}{5}y_{\mu\tau}^2).
\end{align*}
With $B^2(\alpha,\beta,\eta) < 4A(\alpha, \beta, \eta)C(\alpha,\beta,\eta)$ one obtains the following that would avoid a charge breaking minima 
\begin{equation}
\label{Holo_CCB}
 \Big(A_{\tau} + A_{\mu} + 2A_{\mu\tau}\Big)^2 < 5(y_{\tau}^2 + y_{\mu}^2 + \frac{2}{5}y_{\mu\tau}^2) \times 
      \Big[\frac{1}{2} (M_{\tilde{L}_{33}}^2 + M_{\tilde{e}_{33}}^2) + \frac{1}{2}
        (M_{\tilde{L}_{22}}^2 + M_{\tilde{e}_{22}}^2) +  (M_{\tilde{L}_{23}}^2 + M_{\tilde{e}_{23}}^2) + m^2_{H_d} + |\mu|^2\Big].
\end{equation}

Including all the three generations of leptons, Eq. \ref{Holo_CCB} generalizes into the following.

\begin{equation}
 \Big( \sum_{e, \mu, \tau} A_i + 2 \sum_{i \neq j} A_{ij} \Big)^2 < 5 \big( \sum_{e, \mu, \tau} y_i^2 + \frac{2}{5} \sum_{i \neq j} y_{ij}^2 \big) \times 
             \Big[ \frac{1}{2} \sum_{e, \mu, \tau} (M^2_{\tilde{L}_{ii}} + M^2_{\tilde{e}_{ii}}) + \sum_{i \neq j} (M^2_{\tilde{L}_{ij}} + M^2_{\tilde{e}_{ij}})
             + m^2_{H_d} + |\mu|^2 \Big].	
             \label{Holo_CCB_total}
\end{equation}

\subsection{Charge breaking condition in NHSSM} 

With non-holomorphic terms in $V_{soft}$ involving only the appropriate trilinear NH couplings we will have following extra terms,
\begin{align*}
 V^{\tilde{l}}_{NH} = - [\tilde{\mu}_{L}^{*} (A^{\prime}_{\mu}H_u^0)\tilde{\mu}_{R}  + \tilde{\tau}_{L}^{*} (A^{\prime}_{\tau}H_u^0)\tilde{\tau}_{R} 
    + \tilde{\mu}_{L}^{*} (A^{\prime}_{\mu\tau}H_u^0)\tilde{\tau}_{R}  + \tilde{\tau}_{L}^{*} (A^{\prime}_{\tau\mu}H_u^0)\tilde{\mu}_{R} + h.c].
\end{align*}

Considering the direction mentioned in \ref{flat-direction}, we find the following for NHSSM. 

\begin{align}
\label{NHpotential}
V_{\tilde{l},H} &= \{ \alpha^2 (M_{L_{33}}^2 + M_{e_{33}}^2) + \beta^2 (M_{L_{22}}^2 + M_{e_{22}}^2) + 2\alpha\beta (M_{L_{23}}^2 + M_{e_{23}}^2) 
     +  m^2_{H_d} + \eta^2m^2_{H_u} + (1 + \eta^2)|\mu|^2 - 2B\mu\eta \}\phi^2 \\ 
     & + \{ 2\alpha^2 (A_{\tau} - A^{\prime}_{\tau}\eta - \mu y_{\tau}\eta) + 2\beta^2 (A_{\mu} - A^{\prime}_{\mu}\eta- \mu y_{\mu}\eta) \nonumber
     + 4\alpha\beta (A_{\mu\tau} - A^{\prime}_{\mu\tau}\eta- \mu y_{\mu\tau}\eta)\}\phi^3 \\ \nonumber
     & + \{ \frac{g_1^2 + g_2^2}{8}(\eta^2 -1 + \beta^2 + \alpha^2)^2 + (2 + \alpha^2)\alpha^2 y_{\tau}^2 + (2 + \beta^2)\beta^2 y_{\mu}^2
       + 2\alpha^2\beta^2 y_{\mu\tau}^2 \} \phi^4.
\end{align}
Earlier, we obtained the most stringent bound along D-flat direction which requires $\eta=0$. 
But, the same choice is insufficient to provide any bound on the NH trilinear couplings. 
This is simply because $\eta$ gets multiplied with the NH trilinear parameters in Eq. \ref{NHpotential}.
Instead we assume $ \alpha = \frac{1}{\sqrt{2}}, ~ \beta = \frac{1}{\sqrt{2}}, ~ \eta = 1$, which leads to a stringent bound on the
$A_{\mu}^{\prime}$ and $A^{\prime}_{\tau}$ for avoiding a deeper charge breaking minima in NHSSM. The bound can be read as shown below:

\begin{equation}
 \begin{split}
 \label{NH_CCB}
  \Big[A_{\tau} - (\mu y_{\tau} + A^{\prime}_{\tau}) + A_{\mu} - (\mu y_{\mu} + A^{\prime}_{\mu}) + 2\{A_{\mu\tau} - (\mu y_{\mu\tau} +A^{\prime}_{\mu\tau})\}\Big]^2 
  < 4.\frac{1}{4}(\frac{g^2_1 + g^2_2}{2} + 5y_{\tau}^2 + 5y_{\mu}^2 + 2y_{\mu\tau}^2) \\
  \times \Big[\frac{1}{2} (M_{\tilde{L}_{33}}^2 + M_{\tilde{e}_{33}}^2)
    + \frac{1}{2} (M_{\tilde{L}_{22}}^2 + M_{\tilde{e}_{22}}^2) 
 +  (M_{\tilde{L}_{23}}^2 + M_{\tilde{e}_{23}}^2) + m^2_{H_u} + m^2_{H_d} + 2|\mu|^2 - 2B\mu\Big].
 \end{split}
\end{equation}


Again dealing with all three generations of sleptons Eq \ref{NH_CCB} becomes, 

\begin{equation}
\begin{split}
\label{NH_CCB_total}
 \Big(\sum_{e, \mu, \tau}\{ A_i -(A_i^{\prime} + \mu y_i)\} + 2 \sum_{i \neq j} \{A_{ij} - (A^{\prime}_{ij} + \mu y_{ij})\}\Big)^2 
                     <  \Big(\frac{g^2_1 + g^2_2}{2} + 5\sum_{e, \mu, \tau} y_i^2 + 2 \sum_{i \neq j} y_{ij}^2 \Big)
                       \times  
             \Big[ \frac{1}{2} \sum_{e, \mu, \tau} (M^2_{\tilde{L}_{ii}}\\
             + M^2_{\tilde{e}_{ii}}) 
             + \sum_{i \neq j} (M^2_{\tilde{L}_{ij}} + M^2_{\tilde{e}_{ij}})
             + m^2_{H_u} + m^2_{H_d} + 2|\mu|^2 - 2B\mu \Big].
 \end{split}
\end{equation}

Eq.\ref{NH_CCB_total} represents the most general condition to avoid a charge breaking minima considering all kinds
of soft breaking terms for three generations of fermions. Stringent constraints
on individual non-holomorphic diagonal and non-diagonal
trilinear couplings may be derived from \ref{NH_CCB_total}. However, rather
than following the bounds on the individual couplings,
hereafter we will use eq. \ref{Holo_CCB_total} and \ref{NH_CCB_total} in all our results to constrain the trilinear parameters.

\section{Status of different LFV decays}

Here we would summarize the experimental efforts and the degree of current and future sensitivities of several cLFV processes.

In the radiative decay of $l_j \rightarrow l_i \gamma$, the  experiment
leading  to  the  most
stringent constraint is MEG \cite{TheMEG:2016wtm}, which is currently operational at
the Paul Scherrer Institute in
Switzerland. This searches for the radiative process $\mu \rightarrow e \gamma$. 
The MEG collaboration proclaimed  a  new  limit  on  the  rate  for  this  process  based  on  the  analysis  of  a  data set
with $3.6 \times 10^{14}$ stopped muons.  The non-observation of the cLFV process leads to
$Br(\mu \rightarrow e \gamma) < 4.2 \times 10^{-13}$ \cite{TheMEG:2016wtm},
which is four times more stringent 
than the earlier one,  obtained by the same collaboration.
Moreover, the MEG collaboration has announced plans for
future upgrades leading to a sensitivity of about $6 \times 10^{-14}$ after 3 years of data
acquisition \cite{Baldini:2013ke}.

The most interesting results in the near future are
expected in $\mu \rightarrow 3e$ and $\mu - e$ conversion in nuclei.
The Mu3e experiment \cite{Blondel:2013ia, Perrevoort:2016nuv} is designed to search for charged lepton flavor violation 
in the process $\mu \rightarrow 3e$ with
a branching ratio sensitivity of $10^{-16}$.
The present limit on the $\mu \rightarrow 3e$ has been set by the SINDRUM experiment \cite{Bellgardt:1987du}. 
As no signal was observed, branching fractions larger than $1.0 \times 10^{-12}$ were excluded
at 90\% confidence limit (CL). For the upcoming Mu3e experiment,  in phase I, a branching
fraction of $5.2 \times 10^{-15}$ can be measured or excluded at 90\% CL \cite{Perrevoort:2018ttp}.

In the recent times, the most actively studied cLFV processes are
the rare $\tau$ decays. $\tau$-pairs are abundantly produced at
the $B$ factories 
e.g., in the BELLE \cite{Hayasaka:2007vc} \& BABAR \cite{Aubert:2009ag} collaborations. 
There are significant improvement on most of the cLFV modes of the
$\tau$ decays, though any of them has not been discovered yet.
The LHCb collaboration also announced the first ever
bounds on $\tau \rightarrow 3\mu$ in a hadron collider \cite{Aaij:2013fia}. The current experimental upper limits on the LFV
radiative decays \cite{TheMEG:2016wtm, Aubert:2009ag, Hayasaka:2007vc} are collected in the table \ref{tableofconstraints} with references.

\begin{table}[!htb]
	\centering
	\begin{tabular}{|c|c|c|}
		\hline\hline 
		LFV Process  &  Present Bound &  Future Sensitivity   \\ [0.5ex]
		\hline
		$Br(\mu \rightarrow e  \gamma)$ & $ 4.2 \times 10^{-13}$ \cite{TheMEG:2016wtm} & $6 \times 10^{-14}$ \cite{Baldini:2013ke} \\
		$Br(\tau \rightarrow e \gamma)$ & $3.3 \times 10^{-8}$ \cite{Aubert:2009ag} & $ \sim 3 \times 10^{-9}$ \cite{Aushev:2010bq} \\
		$Br(\tau \rightarrow \mu \gamma)$ & $4.4 \times 10^{-8}$ \cite{Aubert:2009ag} & $ \sim 3 \times 10^{-9}$ \cite{Aubert:2009ag}\\
		$Br(\mu \rightarrow 3e)$ & $1.0 \times 10^{-12}$ \cite{Bellgardt:1987du} & $10^{-16}$ \cite{Blondel:2013ia, Perrevoort:2018ttp} \\
		$Br(\tau \rightarrow 3e)$ & $2.7 \times 10^{-8}$ \cite{Hayasaka:2010np} & $\sim 10^{-9}$ \cite{Aushev:2010bq} \\
		$Br(\tau \rightarrow 3\mu)$ & $3.3 \times 10^{-8}$ \cite{Hayasaka:2010np} & $\sim 10^{-9}$ \cite{Aushev:2010bq}\\
		$Br(\tau^- \rightarrow e^- \mu^+ \mu^-)$ & $2.7 \times 10^{-8}$ \cite{Hayasaka:2010np} & $ \sim 10^{-9}$ \cite{Aushev:2010bq}\\
		$Br(\tau^- \rightarrow \mu^- e^+ e^-)$ & $1.8 \times 10^{-8}$ \cite{Hayasaka:2010np} & $ \sim 10^{-9}$ \cite{Aushev:2010bq}\\
		$Br(\tau^- \rightarrow e^+ \mu^- \mu^-)$ & $1.7 \times 10^{-8}$ \cite{Hayasaka:2010np} & $ \sim 10^{-9}$ \cite{Aushev:2010bq}\\
                $Br(\tau^- \rightarrow \mu^+ e^- e^-)$ & $1.5 \times 10^{-8}$ \cite{Hayasaka:2010np} & $\sim 10^{-9}$ \cite{Aushev:2010bq}\\
                $Br(\tau \rightarrow \mu\eta)$ & $2.3 \times 10^{-8}$ \cite{collaboration:2010ipa} & $\sim 10^{-10}$ \cite{OLeary:2010hau}\\
                $Br(\tau \rightarrow \mu\eta')$ & $3.8 \times 10^{-8}$ \cite{collaboration:2010ipa} & $\sim 10^{-10}$ \cite{OLeary:2010hau}\\
                $Br(\tau \rightarrow \mu\pi^0)$ & $2.2 \times 10^{-8}$ \cite{collaboration:2010ipa} & $\sim 10^{-10}$ \cite{OLeary:2010hau}\\  
		\hline
	\end{tabular}
	\caption{Current Experimental situation and future sensitivities for principal LFV processes.}
	\label{tableofconstraints}
\end{table}

Apart from the leptonic decays with LFV there are bounds from LFV Higgs decays.
The first direct search of LFV Higgs decays were performed by CMS and ATLAS Collaborations \cite{Khachatryan:2015kon, Aad:2015gha}.
A slight excess of signal events with a significance of $2.4\sigma$ was observed by CMS at 8 TeV data
but, that early peak by CMS is not supported at 13 TeV anymore, finding
$Br(h \rightarrow \mu \tau) < 1.20\%$ with 2.3 $fb^{-1}$ data \cite{LFVCMS13TeV}. 
Subsequently CMS confirmed the disappearance of that excess \cite{Sirunyan:2017xzt,Aad:2019ugc}.
Additionally, at 13 TeV with integrated luminosity of $35.9 fb^{-1}$, no significant excess over
the Standard Model expectation is observed. The observed (expected) upper limits
on the lepton flavor violating branching fractions of the Higgs boson are
$Br(h \rightarrow \mu \tau) < 0.28$\% (0.37\%) and 
$Br(h \rightarrow e \tau) < 0.47$\% (0.34\%) at 95\% confidence level. These results are used to derive upper limits on the
off-diagonal $\mu \tau$ and $e \tau$ Yukawa couplings \cite{Aad:2019ugc}. These limits on the lepton flavor violating branching
fractions of the Higgs boson and on the associated Yukawa couplings are the most stringent to date (see table \ref{tableofhiggsdecays}). Similarly, the null search results of 
$Br(\tau \rightarrow e/ \mu + \gamma)$ and $Br(\tau \rightarrow 3e/ \mu)$ \cite{pdg} translate into bounds on corresponding 
objects like $\sqrt{Y_{ij}^2 + Y_{ji}^2}$ \cite{Harnik:2012pb}.

\begin{table}[!htb]
	\centering
	\begin{tabular}{|c|c|}
		\hline\hline 
		LFV Higgs decays  &  Present upper limit    \\ [0.5ex]
		\hline
		$Br(h \rightarrow  \mu \tau)$ & $ 2.5 \times 10^{-3}$ \cite{Sirunyan:2017xzt, Aad:2019ugc} \\
		$Br(h \rightarrow e \tau)$ & $ 6.1 \times 10^{-3}$ \cite{Sirunyan:2017xzt, Aad:2019ugc}  \\
		$Br(h \rightarrow e \mu)$ & $6.2 \times 10^{-5}$ \cite{Aad:2019ojw} \\
		\hline
	\end{tabular}
	\caption{Current experimental situation of LFV Higgs decays.}
	\label{tableofhiggsdecays}
\end{table}

Limits on LFV Higgs decay processes closely follow the search results of $h \rightarrow \mu\mu / \tau\tau$ channels. Evidence for the 125 GeV Higgs boson 
decaying to a pair of $\tau$ (or $\mu$) leptons are presented in Refs. \cite{Aad:2012an, Aad:2012cfr, Chatrchyan:2014nva, Aad:2015vsa}.
Furthermore, dedicated searches are conducted for additional neutral Higgs bosons decaying to the $\tau^+ \tau^-$ final state in proton-proton
collisions at the 13 TeV LHC \cite{Sirunyan:2017khh, Aaboud:2017sjh, Sirunyan:2018zut} which lead exclusion plots in the $m_A - \tan\beta$ plane and also give limits on 
$\sigma(gg \rightarrow \phi) \times Br(\phi \rightarrow \tau^+ \tau^-)$ with ($\phi = h, H, A$). Since the cLFV processes of heavier Higgs bosons are 
proportional to $Br(\phi \rightarrow \tau^+ \tau^-)$ \cite{Brignole:2003iv} these limits are extremely important for any analysis of $Br(\phi \rightarrow \mu \tau)$.
Similar results for $\tau^+ \tau^-$ finals states at $\sqrt{s} = 8$ TeV 
are available in \cite{Khachatryan:2014wca, Aad:2014vgg}.
Recently, some model independent analyses of heavier Higgs boson decaying into $\mu\tau$ channel have been performed \cite{Arganda:2019gnv} and it is shown that,
at $\sqrt{s}=14$ TeV with $\mathcal{L}=300$ $fb^{-1}$ the sensitivities to the experimental probes increase with heavier Higgs boson 
masses. Lepton flavor violating decays of Higgs have also been searched for in the
first-second and first-third generations of leptons, i.e  $e\mu$ and $e\tau$ \cite{Khachatryan:2016rke, Aad:2016blu} channels in the LHC at $\sqrt{s}=8$ TeV.
These classes of LFV processes are also being studied in LHC through the decays of
neutral heavy Higgs like bosons for different supersymmetric and non-supersymmetric models
\cite{Aad:2015pfa, Aaboud:2018jff, Aaij:2018mea}.

\section{Results}
 We divide our LFV decay analyses into three parts namely,
 $l_i \rightarrow l_j  \gamma$, $l_i \rightarrow 3l_j$ and
 $\phi(=h, H, A) \rightarrow l_i \bar{l}_j$.  As mentioned before, we try to explore the LFV effects 
 of the relevant trilinear coupling matrices related to the holomorphic and
 non-holomorphic trilinear interactions.
Simply, out of their 
 association with the type of interactions we will label the
 couplings themselves as ``holomorphic'' (like $A_{ij}$ or $A_{ii} \equiv A_i$)
 or ``non-holomorphic'' (like $A^\prime_{ij}$ or $A^\prime_{ii} \equiv A^\prime_i$).
 Furthermore, as pointed out earlier, for simplicity,
 our analysis probes either (i) a non-vanishing set of 
 [$A_{ij}, A_i$] while having vanishing values for the set [$A^\prime_{ij}, A^\prime_i$] or (ii) vice-versa.
 We will particularly identify the region of parameter space in relation to
 a given constraint from an LFV decay where non-holomorphic couplings may
 have prominent roles.    
 Considering the fact that only the trilinear couplings associated with sleptons are
 of importance in this analysis, we will either consider the matrices $A_f=0$ and $A^\prime_f \neq 0$
 and vice-versa, where $f \equiv e, \mu, \tau$.
 Combining the holomorphic and non-holomorphic couplings, we will often use   
 $A_f^{(\prime)}$ to mean either $A_f$ or $A^\prime_f$ depending on the context.
 We vary both the diagonal and off-diagonal components of
 ${\bf {A^{(\prime)}}_{e}}_{ij}$ over broad range of values
 (see table \ref{table-data})
 subject to the condition $|\delta^{(\prime)}_{ij}|<1$ with $i \ne j$, where 
 $\delta^{(\prime)}_{ij}=\frac{{{A^{(\prime)}}_{e}}_{ij}}{{{A^{(\prime)}}_{e}}_{ii}}$. Here, we remind that all the off-diagonal entries of the slepton mass matrices
 are considered to be zero leading to flavor violation possible only from the trilinear coupling sources. 
  We will impose the charge breaking
  constraint and select the
  possible diagonal and off-diagonal
  trilinear coupling values that would obey Eq.\ref{NH_CCB_total}.
  We must emphasize here that in an analysis that scans
  the diagonal and off-diagonal elements of the trilinear coupling matrices,
  such as the present one, we do not expect severe restrictions on the off-diagonal
  entries $A^{(\prime)}_{ij}$ simply because of the possible cancellation of
  terms in the left hand side of Eq.\ref{NH_CCB_total}. On the other hand, 
  a choice of fixed signs of diagonal and off-diagonal entries of
  tri-linear coupling matrix elements would show prominent effects of
  charge breaking.    
 Apart from the above, we impose the experimental
 bounds of different cLFV observables and LHC direct search results for 
$\phi \rightarrow l_i \bar{l}_j$.   
 The values/ranges of relevant soft parameters used in this analysis
 are listed in table \ref{table-data}.
\begin{table}[!htb]
  	\centering
	\begin{tabular}{|c|c|c|c|}
		\hline\hline 
		Parameters  &  Value &  Parameters & Value   \\ [0.5ex]
		\hline
		$M_1$ & [100, 1000] & $M_{2}$  &  1500\\
		$M_3$ & 2800 & $\mu$ & 800 \\
		$m_A$ & 1500 & $\tan\beta$ & 40 \\
		$M_{\tilde{q}_{33}} / M_{\tilde{u}_{33}}$ & 2000 &  $M_{\tilde{d}_{33}}$  & 2000 \\
		$M_{\tilde{q}_{11,22}}/ M_{\tilde{u}_{11,22}}$ & 2000 & $M_{\tilde{d}_{11,22}}$  & 2000 \\
		$M_{\tilde{L}_{11,22,33}}$ & [1000, 10000] & $M_{\tilde{e}_{11,22,33}}$ & [1000, 10000]\\
		\hline
		$A_{t}, A_{b}$  & -2200, 0 & $A_{t}^{\prime}, A_b^{\prime}$ & 0, 0  \\
		\hline
                ${{A}_{(e,\mu, \tau)}}_{ij}$, $|\delta_{ij}|<1$ &
                    [-8000, 8000] & ${{A^{\prime}}_{(e,\mu, \tau)}}_{ij}$, $|\delta^{\prime}_{ij}|<1$
                 & [-8000, 8000] \\
		\hline
	\end{tabular}
	\caption{Soft masses and trilinear parameters are listed here. All the masses and trilinear couplings are in GeV and 
	there is no off-diagonal entries in the soft bilinear mass matrices.}
	\label{table-data}
\end{table}
 We compute the SUSY mass spectra and related branching fractions from SARAH (v4.10.0)\cite{Staub:2013tta, Staub:2015kfa} generated 
FORTRAN codes that are subsequently used in SPheno (v4.0.3) \cite{Porod:2011nf}. All the flavor observables are calculated using 
FlavorKit which is inbuilt within the SARAH-SPheno 
framework.
In regard to a few relevant SM parameters, we use  
$m_t^{pole}=173.5$~GeV, $m_b^{\overline{MS}}=4.18$~GeV and $m_\tau=1.77$~GeV \cite{Olive:2016xmw} and we use the SUSY mass scale as
$M_{\rm SUSY} = \sqrt{m_{\tilde{t}_{1}} m_{\tilde{t}_{2}}}$.
We further impose the following limits for Higgs mass $m_h$\footnote{We consider a $\pm$3 GeV theoretical uncertainty in computing lighter 
Higgs mass\cite{loopcorrection} that arises
    from the uncertainties from radiative corrections up to
three loops, top quark pole mass, renormalization scheme and scale dependencies
etc.}, branching ratios like $Br(b \rightarrow s \gamma)$ and $Br(B_s \rightarrow \mu^{+} \mu^{-})$ at $2\sigma$ level, lighter chargino mass bound
from LEP, along with an LHC limit for the direct lighter top squark searches\cite{Olive:2016xmw, Aaboud:2017ayj}.  
\begin{align}
  122.1 ~ {\rm GeV} \leqslant m_h \leqslant 128.1 ~ {\rm GeV},  \nonumber \\
2.99 \times 10^{-4} \leqslant Br(b \rightarrow s \gamma) \leqslant 3.87 \times 10^{-4}, \nonumber \\
1.5 \times 10^{-9} \leqslant Br(B_s \rightarrow \mu^{+} \mu^{-}) \leqslant 4.3 \times 10^{-9}, \nonumber \\
m_{\tilde{\chi}_1}^{\pm} \geq 104 ~ {\rm GeV}, m_{\tilde{t}_1} \geq 1000 ~ {\rm GeV}. 
\label{mh_bsg}
\end{align}

\subsection{$\mathbf{l_j \rightarrow l_i \gamma}$}
  
We show the results of computation of $Br(l_j \rightarrow l_i \gamma)$ in fig.~\ref{litoljgamma}. Here, the cyan and blue colored zones correspond to
the scanning selected for the relevant holomorphic and non-holomorphic
trilinear coupling matrix elements $A_{ij}$ and $A^{\prime}_{ij}$
respectively.  Only the parameter points that satisfy the
charge breaking vacuum constraint of 
Eq. \ref{NH_CCB_total} are shown where
the domain of variations of
$A_{ij}$ and $A^{\prime}_{ij}$  are mentioned in table~\ref{table-data}.
Apart from $A_{ij}^{(\prime)}$, the
mass parameters varied are $M_1$, the mass of bino and the diagonal soft masses 
of the left and right handed sleptons namely, $M_{\tilde{L}_{ii}}$  and $M_{\tilde{e}_{ii}}$ whose ranges are as given in table~\ref{table-data}.
As we mentioned earlier, in this analysis
we do not consider any non-vanishing off-diagonal entry for the scalar mass matrices.
The off-diagonal trilinear couplings $A'_{e\mu}$ has stronger influence on $Br(\mu \rightarrow e \gamma)$
compared to $A_{e\mu}$. The resulting influence superseed the present upper bound of $Br(\mu \rightarrow e \gamma)$ 
(see fig.\ref{litoljgammaA}).   The excluded regions that are unavailable via experimental limits are shown as gray bands in the top. 
Additionally, the black horizontal lines in fig.\ref{litoljgammaA} to fig.\ref{litoljgammaC} display the upcoming sensitivities
for the respective channels of $l_j \rightarrow l_i \gamma$ (see table~\ref{tableofconstraints}). 
We remind that the effect of a non-vanishing $A'_{e\mu}$ is associated
with an enhancement
by $\tan\beta$ that in general pushes up the branching ratio 
$Br(\mu \rightarrow e \gamma)$ over 2 to 3 orders of magnitude compared to the same
arising out of $A_{e\mu}$. This is also true for the other decay modes namely
$\tau \rightarrow e \gamma $ and $\tau \rightarrow \mu \gamma $ in relation to the corresponding trilinear couplings.
However,
only the present $Br(\mu \rightarrow e \gamma)$ bound
is strong enough to exclude a significant amount of NHSSM 
parameter space
depending on the values of $A'_{e\mu}$ and the mass parameters 
that would simultaneously satisfy the charge breaking constraint.
\begin{figure}[H]
 \begin{center}
 \subfigure[]{
   \includegraphics[width=5.1cm,height=4.5cm]{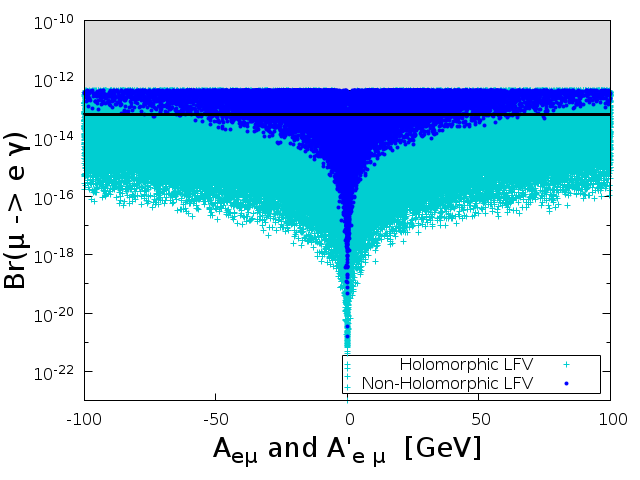}
   \label{litoljgammaA}
  }
  \subfigure[]{
    \includegraphics[width=5.1cm,height=4.5cm]{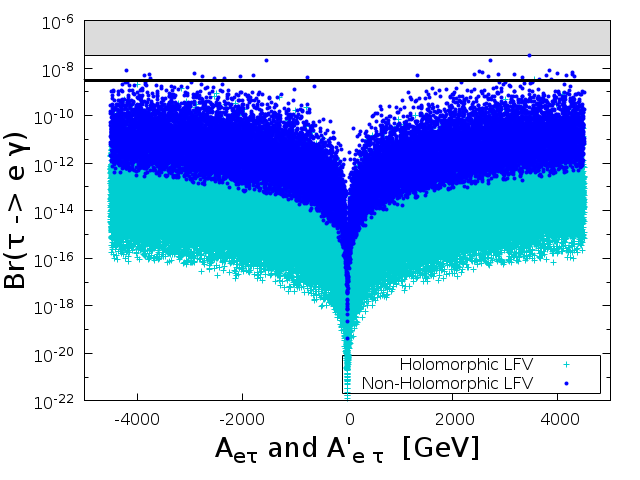}
    \label{litoljgammaB}
  }
  \subfigure[]{
    \includegraphics[width=5.1cm,height=4.5cm]{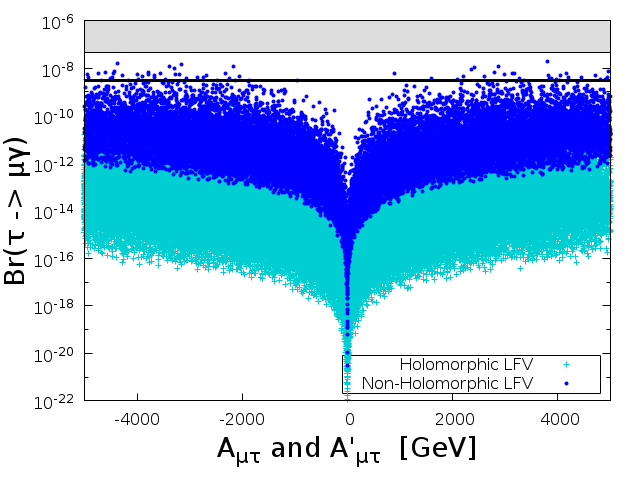}
    \label{litoljgammaC}
  }
  \caption{Branching ratios of $l_j \rightarrow l_i \gamma$ as a function of $A_{ij}$ in cyan and $A'_{ij}$
  in blue after satisfying the charge breaking bound and are checked through the other collider bounds \textit{viz.} eq. \ref{mh_bsg} and table 
  \ref{tableofconstraints}.
  The gray shaded regions and 
  black horizontal lines in each plot denote the present exclusion region and future announced sensitivity of the corresponding channels.}
  \label{litoljgamma}
 \end{center}
\end{figure}


\subsection{$\mathbf{l_j \rightarrow 3l_i}$}
We will now try to investigate the influence of non-vanishing trilinear couplings $A_{ij}$ and $A'_{ij}$ on $Br{(l_j \rightarrow 3l_i)}$ in fig.~\ref{fig:tau_3mu}.
This is in spite of the fact that any influence is yet too far to be tested in experiments such as that for $Br(\tau \rightarrow 3 e )$ and $Br(\tau \rightarrow 3 \mu)$.
Similar to fig.~\ref{litoljgamma}, the cyan and blue colored regions refer to parameter
points with given $A_{ij}$ and $A'_{ij}$ values respectively that would avoid the charge breaking minima.
The present bounds of $Br(\tau \rightarrow 3 e )$ and $Br(\tau \rightarrow 3 \mu)$ are of the order of $10^{-8}$ 
by BELLE collaboration\cite{Hayasaka:2010np}. These are expected to reach $10^{-9}$ in Super B \cite{Aushev:2010bq}. Clearly as seen in the figures both the limits are significantly
larger compared to the level of contributions under discussion. Regarding $Br(\mu \rightarrow 3e)$, a non-vanishing $A'_{e\mu}$ can push it up to 
$10^{-12}$ but $Br(\mu \rightarrow e\gamma)$ is more effective a constraint to limit $A'_{e\mu}$.

In figure \ref{fig:tau_3mu} we see the stretch of $Br(\tau \rightarrow 3 e/\mu)$ with holomorphic and non-holomorphic trilinear couplings respectively. The color 
coding is the same as in the figure \ref{litoljgamma}. The cyan and blue colored points are shown after avoiding the charge breaking minima.
The current sensitivity of these two channels is of the order of $10^{-8}$ 
by BELLE collaboration \cite{Hayasaka:2010np} which is expected to reach $10^{-9}$ in Super B \cite{Aushev:2010bq}. In the following, 
after satisfying all the respective bounds as mentioned in Eq:\ref{mh_bsg} and in table \ref{tableofconstraints} and the radiative decays $Br(\tau \rightarrow e \gamma)$,
$Br(\tau \rightarrow \mu \gamma)$ in particular, we find
both $Br(\tau \rightarrow 3e)$ and $Br(\tau \rightarrow 3\mu)$ can reach up to $\sim 10^{-11}$. This is again two orders of magnitude smaller than 
the future proposed sensitivity. The maximum reaches are $10^{-10}$ and $10^{-11}$ for $Br(\tau \rightarrow 3 e)$ and $Br(\tau \rightarrow 3\mu)$ respectively, 
while for $Br(\mu \rightarrow 3e)$ may become $10^{-12}$ with non-holomorphic $A'_{e\mu}$. 
$Br(\mu \rightarrow e\gamma)$ limits the free increase of $Br(\mu \rightarrow 3e)$ in particular.

\begin{figure}[H]
       \begin{center}
       \subfigure[]{%
         \includegraphics[width=5.5cm,height=4.5cm]{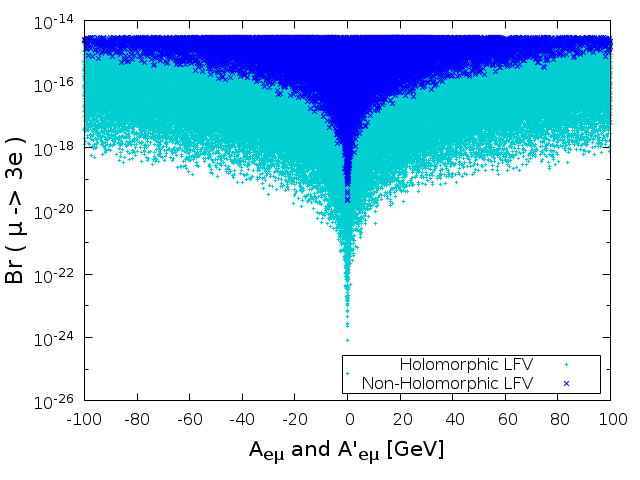}
            \label{fig:tau_3muA}
           }%
          \subfigure[]{%
            \includegraphics[width=5.5cm,height=4.5cm]{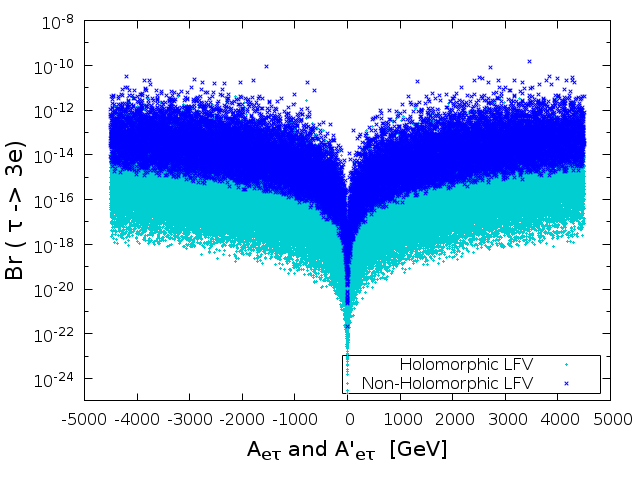}
            \label{fig:tau_3muB}
           }%
           \subfigure[]{%
          \includegraphics[width=5.5cm,height=4.5cm]{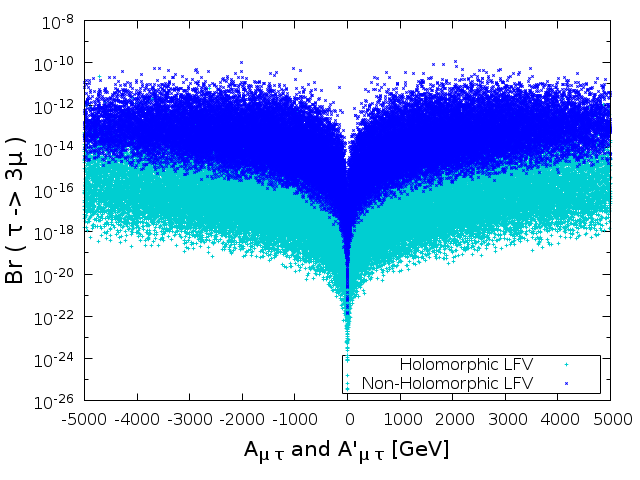}
          \label{fig:tau_3muC}
          }
          \caption{The variation of $Br(\mu \rightarrow 3e)$ with $A_{e\mu}$ and $A_{e\mu}^{\prime}$ has been shown in \ref{fig:tau_3mu}(a),
          \ref{fig:tau_3mu}(b) depicts $Br(\tau \rightarrow 3e)$ with $A_{e\tau}$ 
          and $A_{e\tau}^{\prime}$ and in \ref{fig:tau_3mu}(c) depicts $Br(\tau \rightarrow 3\mu)$ with $A_{\mu\tau}$ 
          and $A_{\mu\tau}^{\prime}$. They are passed through the checks of \ref{mh_bsg} and 
          table \ref{tableofconstraints}, and after avoiding the overall charge breaking minima given in \ref{Holo_CCB_total} 
          and \ref{NH_CCB_total}
          for holomorphic and non-holomorphic trilinear couplings respectively.}
        \label{fig:tau_3mu}
  \end{center}
\end{figure}

We like to comment that virtual Higgs exchange can also induce $\tau$ decaying to $\mu$ along with psedoscalar meson like $\tau \rightarrow \mu (\pi / \eta / \eta')$
though the later decay fractions lie
much below the future sensitivity presented in table \ref{tableofconstraints}.
With CP conservation in Higgs sector, only the exchange of CP-odd Higgs is expected to be present dominantly because of its enhanced couplings to the 
down-type quarks. These interactions and relations to the other LFV processes can be found in \cite{Brignole:2004ah, Sher:2002ew}.


\subsection{$\mathbf{\phi(h,H,A) \rightarrow l_i \bar{l}_j}$}
  We focus now on flavor violating Higgs decays that lead to
  two oppositely charged leptons where a Higgs boson can be
  the SM-like Higgs boson ($h$), CP-even 
  heavier Higgs ($H$) or CP-odd Higgs ($A$).
  We however note that the current experimental level to probe light higgs
  decay branching ratios as seen in table~\ref{tableofhiggsdecays} is way too
  large compared to the branching ratio values shown below. This is unlike
  the decays of the heavier Higgs where non-holomorphic parameters may give
  large contributions.\\
\noindent
$\bullet$  $\mathbf{h \rightarrow l_i \bar{l}_j}$:  
  We discuss the
  light higgs decay here for completeness before moving to the discussion
  of heavy Higgs leptonic decays with flavor violations. 
  fig.\ref{fig:h_mu_tau} shows the
  scatter plots of the branching ratios $Br(h \rightarrow e \mu)$, $Br(h \rightarrow e\tau)$ and $Br(h \rightarrow \mu\tau)$ when $A^{(')}_{e\mu}$, $A^{(')}_{e\tau}$
  and $A^{(')}_{\mu\tau}$ respectively are varied. All the points satisfy the
  mass bounds and flavor constraints of Eq.\ref{mh_bsg} and 
  table~\ref{tableofconstraints} apart from avoiding any charge breaking
  minima (Eqs.\ref{Holo_CCB_total},\ref{NH_CCB_total}).   
  The color convention is same as before, i.e.
  the points in blue and cyan are for varying non-holomorphic and holomorphic 
  trilinear couplings respectively. The spread in the colored points
  is the consequence of the random scanning of soft masses
  as mentioned in table \ref{table-data}. 
  One notes that, irrespective of the source of flavor violation
  the branching ratio
$Br(\phi \rightarrow l_i \bar{l}_j)$ itself is proportional to
  $\tan^2\beta$  \cite{Brignole:2003iv}. An appropriate
  NH coupling additionally multiplies the result with  $\tan^2\beta$
  potentially leading to an enhancement by a factor of $\sim 10^3$ when
  compared to corresponding the MSSM scenario for $\tan\beta \gsim 30$.  
 Our results for the holomorphic case closely  
 agree with the results of \cite{Arganda:2015uca} with LR type flavor
 violation in the slepton sector of MSSM. 
 $Br(h \rightarrow e\tau)$ is  
 in the ballpark of $\sim 10^{-10} ~ (\sim 10^{-13})$ for
 non(-holomorphic) trilinear parameters.
 We note that as seen in fig.\ref{litoljgamma}(a),
 stringency due to $Br(\mu \rightarrow e\gamma)$ 
 causes the extent of allowed
 variation $A_{e\mu}^{(')}$ to be much smaller than the other trilinear couplings
 corresponding to those with $e$-$\tau$ or $\mu$-$\tau$. Consequently, 
 $Br(h \rightarrow e \mu)$ of fig.\ref{hdecaygraphemu} is smaller by 3 to 4 orders of magnitude with respect to the branching ratios of
 fig.\ref{hdecaygraphetau} and fig.\ref{hdecaygraphmutau}.
 
\begin{figure}[H]
       \begin{center}
       \subfigure[]{%
         \includegraphics[width=5.5cm,height=4.5cm]{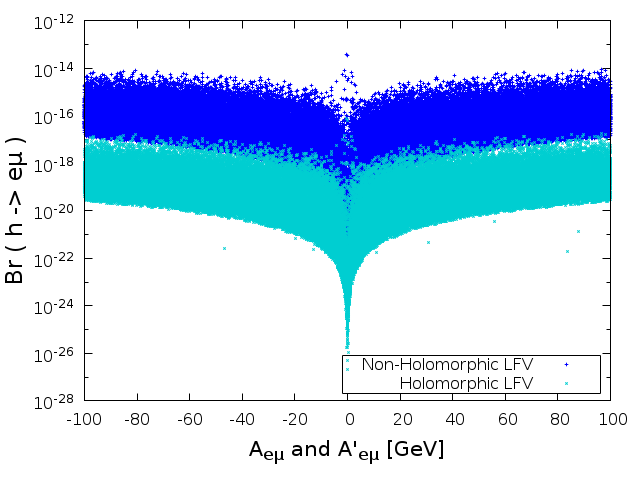}
          \label{hdecaygraphemu}
           }%
          \subfigure[]{%
            \includegraphics[width=5.5cm,height=4.5cm]{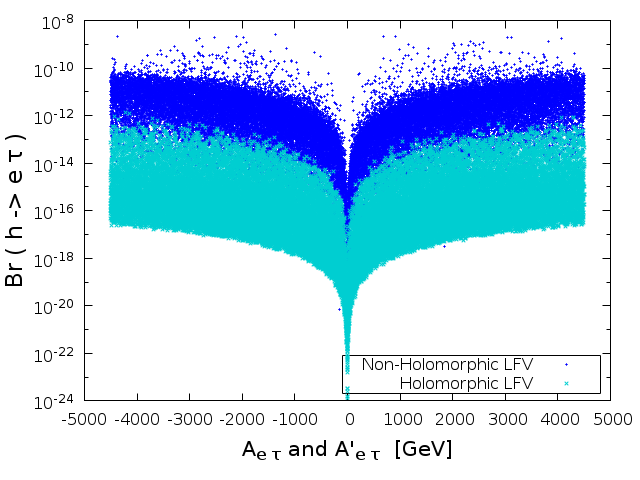}
            \label{hdecaygraphetau}
           }%
           \subfigure[]{%
          \includegraphics[width=5.5cm,height=4.5cm]{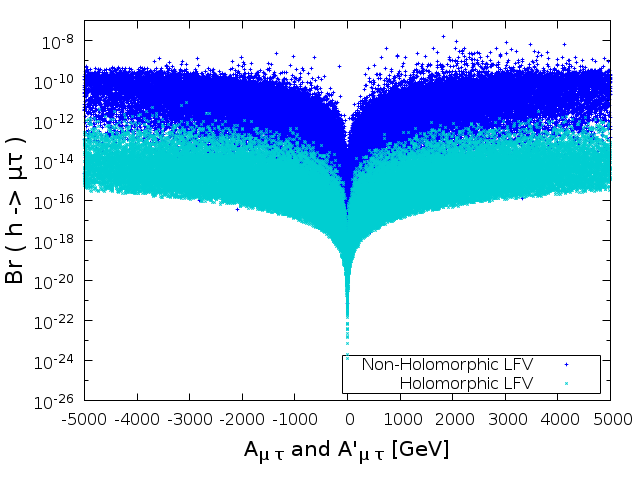}
          \label{hdecaygraphmutau}
          }
          \caption{figure \ref{fig:h_mu_tau}(a) shows the variation of $Br(h \rightarrow e\mu)$ with $A_{e\mu}$ and $A_{e\mu}^{\prime}$, in figure \ref{fig:h_mu_tau}(b) 
          dependence of $Br(h \rightarrow e  \tau)$  with $A_{e\tau}$ and $A_{e\tau}^{\prime}$ and \ref{fig:h_mu_tau}(c) shows the variation of $Br(h \rightarrow \mu  \tau)$ with $A_{\mu\tau}$ and $A_{\mu\tau}^{\prime}$. All points fulfill the analytical expressions of charge 
          breaking minima given in \ref{Holo_CCB_total} and \ref{NH_CCB_total} and the color coding is same as previous.}
        \label{fig:h_mu_tau}
  \end{center}
\end{figure}

 We will now briefly study the level of dependence of the associated
 sparticle masses  
 on the above LFV $h$-decays in fig.\ref{fig:h_mu_tau_M1_stau}. 
 The relevant loops contain neutralinos and sleptons.
 The left and the right panels show $Br(h \rightarrow \mu\tau)$ when
 soft masses of bino ($M_1$) and the lighter stau ($m_{\tilde{\tau}_1}$)
 respectively are varied.  All the input parameters and ranges are as given 
 in table~\ref{table-data}. The branching ratio profile of Fig.\ref{fig:h_mu_tau_M1_stau}b shows
 that the radiative corrections associated with the decay expectedly
 fall with $m_{\tilde{\tau}_1}$ while the same of  
 fig.\ref{fig:h_mu_tau_M1_stau}a hardly shows any correlation with $M_1$.
 We note that for NHSSM, the regions of larger $Br(h \rightarrow \mu\tau)$ 
 that are typically associated with large $A'_{\mu\tau}$ may
 lead to tachyonic staus due to large L-R slepton mixing. This is consistent
 with what we find to be a small discarded zone  
 with $m_{\tilde{\tau}_1}\sim$ near 1~TeV that could otherwise
 have a large branching ratio.
\begin{figure}[H]
       \begin{center}
          \subfigure[]{%
           \includegraphics[width=7.5cm,height=4.5cm]{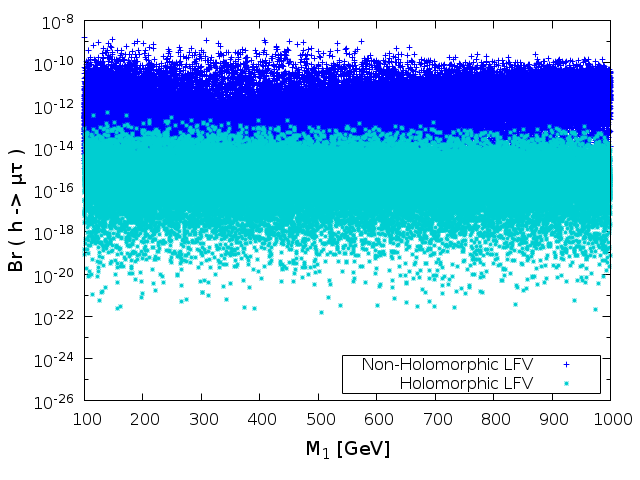}
           }%
          \subfigure[]{%
          \includegraphics[width=7.5cm,height=4.5cm]{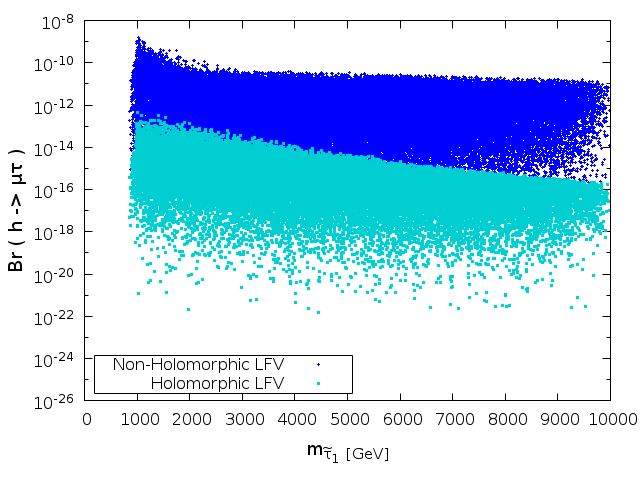}
          }
          \caption{In left panel of figure \ref{fig:h_mu_tau_M1_stau} we can see the dependence of $Br(h \rightarrow \mu\tau)$ on $M_1$. 
            In the right side the same branching ratio is plotted against the lightest slepton mass. All other soft mass parameters and trilinear couplings are according to table 
          \ref{table-data} and are checked through the charge breaking minima condition.} 
        \label{fig:h_mu_tau_M1_stau}
  \end{center}
\end{figure}
    
     $\bullet$  $\mathbf{H/A \rightarrow l_i \bar{l}_j}$:
 \noindent
 We now turn to explore the dependence on $A_{ij}^{\prime}$'s when
 heavier Higgs bosons decay into 
 leptons with LFV namely, $H/A \rightarrow l_i \bar{l}_j$.
 figure~\ref{fig:H_A_mu_tau}(a), figure~\ref{fig:H_A_mu_tau}(b) and figure~\ref{fig:H_A_mu_tau}(c) refer to the scatter plots of the associated
 branching ratios for the decay of H or A bosons
 into  $ e \mu$, $ e \tau$ and $\mu \tau$
 respectively where $A_{ij}^{\prime}$ are varied.
 The figures have the same color convention as
 before.  Here, all the points satisfy the mass bounds and
 flavor constraints of Eq.~\ref{mh_bsg} and table~\ref{tableofconstraints}.
 As before, the figures show only the points that satisfy the charge breaking
 minima constraints (Eqs.~\ref{Holo_CCB},~\ref{NH_CCB_total}) for varying
 holomorphic or non-holomorphic trilinear coupling parameters.
 The branching ratios are expected to be large because
 the couplings of H/A to down-type fermions grow with $\tan\beta$.
 As we discussed earlier, the LFV branching fraction
$Br(\phi_k \rightarrow e \tau)$ or $Br(\phi_k \rightarrow \mu \tau)$
 can be cast in
 terms of flavor conserving di-tau branching ratio
 $Br(\phi_k \rightarrow \tau \tau)$ 
 following Eq. \ref{br_heavy} where $\phi_k$ refers to $h,H,A$ for $k=1,2,3$
 respectively.  
$Br(\phi_k \rightarrow \tau \tau)$ 
 in general depends on slepton and
 neutralino masses\cite{Brignole:2003iv} with hardly any dependence on
 trilinear couplings at the lowest order.
 We must use the LHC data here, particularly for the
 heavier Higgs bosons decaying into the di-tau channels\cite{Aaboud:2017sjh, Aaboud:2016cre}. The constraint in the
 $[m_A, \tan\beta]$ plane is rather stringent in the large $\tan\beta$ and small $m_A$ region. With our choice of $m_A$, that is rather high, 
 the flavor conserving branching ratio satisfies the LHC
 limit\cite{Aaboud:2017sjh}.  Apart from this, our computed results of 
 $\sigma(gg\phi_k)Br(\phi_k \rightarrow \tau \tau)$ for all $k$  
 fall in the allowed 
 zones of the LHC limit\cite{Aaboud:2016cre}. The flavor violating decay rates  
 of $H/A$ that are computed by using Eq.~\ref{br_heavy}
 become large because of large $C_H$ or $C_A$ (Eq.\ref{C_phi}) when $h$ is chosen
 to be SM like in its couplings. The decay rates may potentially
 get amplified by a factor of $\sim 10^3$ (via $\propto \tan^2\beta$)
 in presence of non-vanishing non-holomorphic trilinear couplings. Thus,
 unlike the case of LFV $h$-decay of figure~\ref{fig:h_mu_tau},
 the LFV branching ratios of $H/A$ as shown in
 figures~\ref{fig:H_A_mu_tau}(b) and \ref{fig:H_A_mu_tau}(c)
 may scale as high as $10^{-4}$. This may be of significance in relation
 to a future high energy
 collider.

\begin{figure}[H]
       \begin{center}
       \subfigure[]{%
           \includegraphics[width=5.5cm,height=4.5cm]{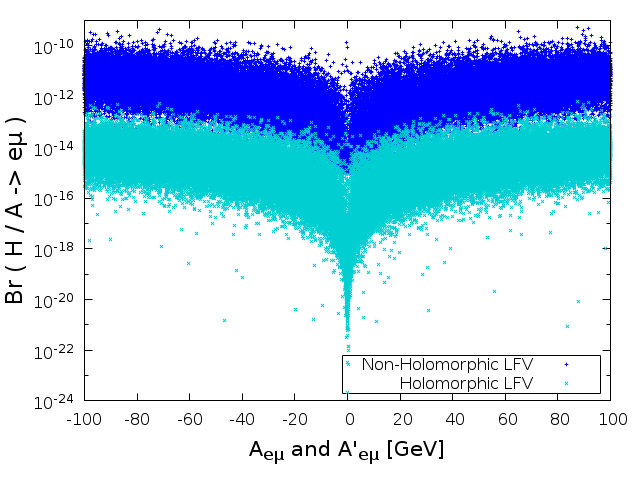}
           }%
          \subfigure[]{%
           \includegraphics[width=5.5cm,height=4.5cm]{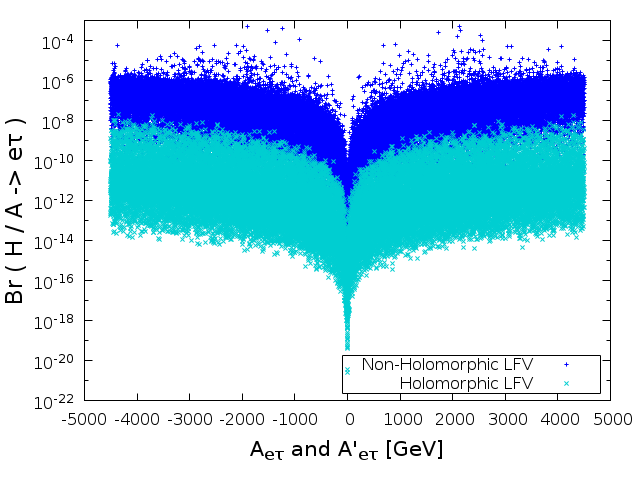}
           }%
            \subfigure[]{%
           \includegraphics[width=5.5cm,height=4.5cm]{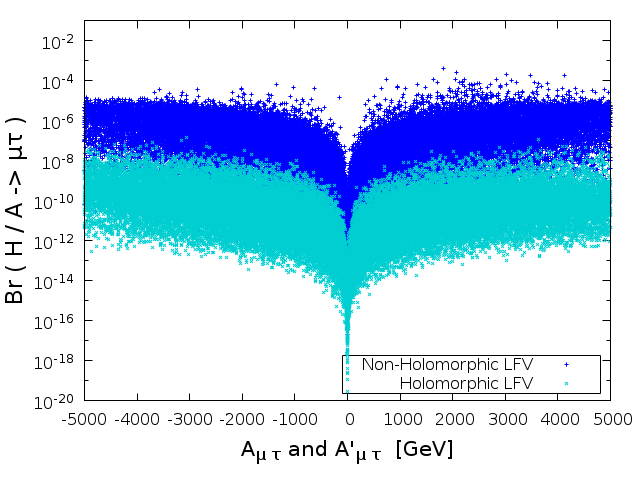}
           }%
          \caption{$Br(H/A \rightarrow e\mu)$ as a function of $A^{(')}_{e\mu}$ in the left most side, $Br(H/A \rightarrow e \tau)$ as a function of $A^{(')}_{e\tau}$ in the middle and $Br(H/A \rightarrow \mu \tau)$ as a function of $A^{(')}_{\mu\tau}$ in the right side are shown with $m_H, m_A = 1.5$ TeV.
          Color coding is same as explained in figure \ref{fig:h_mu_tau}.}
        \label{fig:H_A_mu_tau}
  \end{center}
\end{figure}

\subsection{Direct constraints on $Y_{ij}$ from LFV Higgs decay limits of LHC}
  
 With identical Yukawa couplings that MSSM inherits from SM, we will now use the constraints on relevant off-diagonal Yukawa
  couplings arising out of 
  the null limits of SM-like higgs decays like
  $h \rightarrow \mu\tau$, $h \rightarrow e\tau$, and $h \rightarrow e\mu$ as given by the LHC data. 
  The CMS collaboration performed direct exploration of
  $h \rightarrow \mu\tau$, followed by the hunt
  for $h \rightarrow e\tau / e\mu$ decays with 8~TeV
  corresponding to an integrated luminosity
  of 19.7 $fb^{-1}$\cite{Khachatryan:2015kon, Khachatryan:2016rke}.
  The hadronic, electronic and muonic decay channels for the 
  $\tau$-leptons were also explored for the above mentioned LFV processes
  with 13 TeV LHC data\cite{Sirunyan:2017xzt, Aad:2019ugc, Aad:2019ojw}. 
The null results can effectively put upper limits on the off-diagonal
namely $\mu\tau$ and $e\tau$ Yukawa couplings. We relate this to the
outcome of having non-vanishing Yukawa couplings arising out of radiative
corrections due to the trilinear soft terms of both holomorphic and
non-holomorphic origins. The current LHC
bounds\cite{Sirunyan:2017xzt, Aad:2019ugc} are $\sqrt{(Y_{\mu\tau}^2 + Y_{\tau\mu}^2)} < 1.50 \times 10^{-3}$ and
$\sqrt{(Y_{e\tau}^2 + Y_{\tau e}^2)} < 2.26 \times 10^{-3}$.
Regarding $Y_{e \mu}$, one finds that the null observation of 
$\mu \rightarrow e \gamma$ implies a very stringent limit:
$\sqrt{(Y_{e\mu}^2 + Y_{\mu e}^2)} < 3.6 \times 10^{-6}$ \cite{Harnik:2012pb}. 
We note that in
general for an LFV scenario there can be two independent
Yukawa couplings $Y_{ij}$ and $Y_{ji}$\cite{Calibbi:2017uvl}. For example, this arises from different possible Yukawa couplings with higgs
for $e_L$ with $\mu_R$ and $\mu_L$ with $e_R$ superfields. However, 
we consider them to be identical in this analysis for simplicity and
this is consistent with our assumption of a single   
trilinear soft parameter $A_{ij}$ which is same as $A_{ji}$.

\subsubsection{Results for varying bino and slepton masses}
We now impose the Yukawa coupling bounds to update
figures \ref{fig:tau_3mu}, \ref{fig:h_mu_tau} and
\ref{fig:H_A_mu_tau} keeping all other
constraints unchanged. This will help
us in understanding the extent of influence of the above
constraints on the trilinear soft paramaters. The color
conventions in these figures are same as before. Below,
we will
only emphasize the essential differences between the
figures focusing only on NHSSM. This is simply
because the above bounds are hardly
effective for MSSM involving only the trilinear
holomorphic couplings. 

Compared to figure \ref{fig:tau_3muA},
figure \ref{fig:tau_3muA_afterYukawa} discards a significant
amount of 
large $Br(\mu \rightarrow 3e)$ zone with large $|A_{e\mu}^{\prime}|$. Thus, approximately a region 
with $|A_{e\mu}^{\prime}|$ above 60 GeV is eliminated.
The above only demonstrate the stringency of the constraints on 
the first two generations of the off-diagonal Yukawa couplings
arising from the null observation of 
$\mu \rightarrow e \gamma$ \cite{Harnik:2012pb} along with the 
CMS results on the first two generations \cite{Aad:2019ojw, Sirunyan:2019shc}. 
Figure \ref{fig:tau_3muB_afterYukawa} is not so 
different from figure \ref{fig:tau_3muB} indicating
only an insignifant influence. 
Regarding figure \ref{fig:tau_3muC_afterYukawa}
vs figure \ref{fig:tau_3muC}
we see that the upper limit of $Br(\tau \rightarrow 3\mu)$ 
is truncated by approximately an order of
magnitude over the parameter space, still being a mild
effect. 

\begin{figure}[H]
       \begin{center}
       \subfigure[]{%
            \includegraphics[width=5.5cm,height=4.5cm]{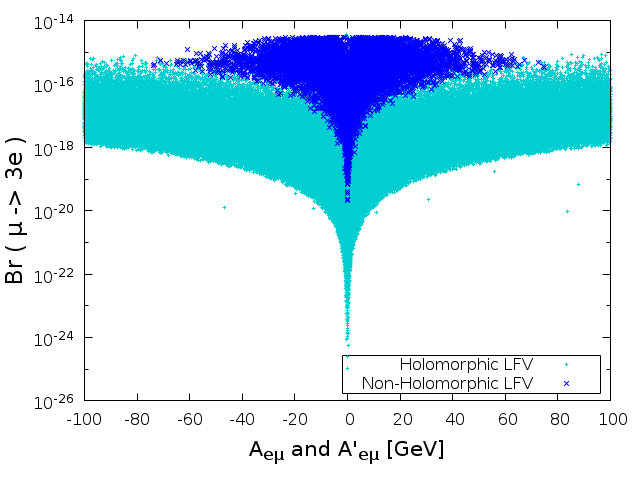}
            \label{fig:tau_3muA_afterYukawa}
           }%
          \subfigure[]{%
            \includegraphics[width=5.5cm,height=4.5cm]{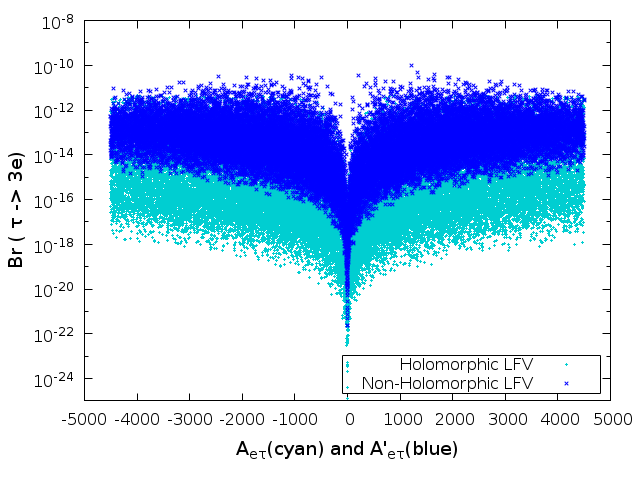}
            \label{fig:tau_3muB_afterYukawa}
           }%
           \subfigure[]{%
          \includegraphics[width=5.5cm,height=4.5cm]{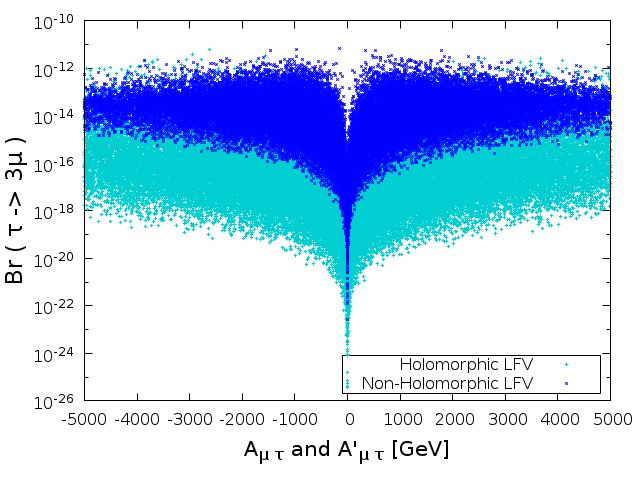}
          \label{fig:tau_3muC_afterYukawa}
          }
          \caption{Effect of imposing the LHC constraints on $Y_{ij}$ on the analysis of figure \ref{fig:tau_3mu}:  
            Scatter plots of $Br(\mu \rightarrow 3e)$ with $A_{e\mu}$ and $A_{e\mu}^{\prime}$ are shown in \ref{fig:tau_3mu_after_Yukawa}(a). Similar
            plots for $Br(\tau \rightarrow 3e)$ with $A_{e\tau}$ 
          and $A_{e\tau}^{\prime}$ are shown in figure 
          \ref{fig:tau_3mu_after_Yukawa}(b). Figure
          \ref{fig:tau_3mu_after_Yukawa}(c) is for $Br(\tau \rightarrow 3\mu)$ with $A_{\mu\tau}$ 
          and $A_{\mu\tau}^{\prime}$.
}
        \label{fig:tau_3mu_after_Yukawa}
  \end{center}
\end{figure}

The effect of the off-diagonal Yukawa coupling constraints 
from the CMS results is shown for the cLFV $h$-decays in  
figure \ref{fig:h_mu_tau_afterYukawa} in which the
scattered plots of the two colors for each
of the sub-figures are subsets of the same
of figure \ref{fig:h_mu_tau}. In the context of limiting 
$A_{ij}^{\prime}$ the conclusion is essentially similar to
the discussion made in relation to each of the sub-figures
of figure \ref{fig:tau_3mu_after_Yukawa}. 

\begin{figure}[H]
       \begin{center}
       \subfigure[]{%
         \includegraphics[width=5.5cm,height=4.5cm]{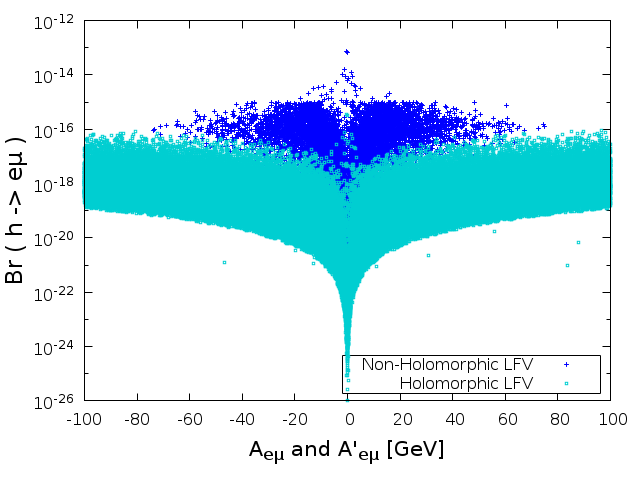}
          \label{hdecaygraphemu}
           }%
          \subfigure[]{%
            \includegraphics[width=5.5cm,height=4.5cm]{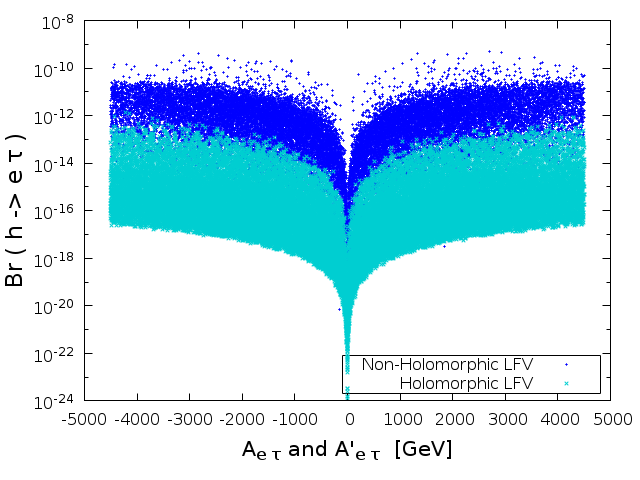}
            \label{hdecaygraphetau}
           }%
           \subfigure[]{%
          \includegraphics[width=5.5cm,height=4.5cm]{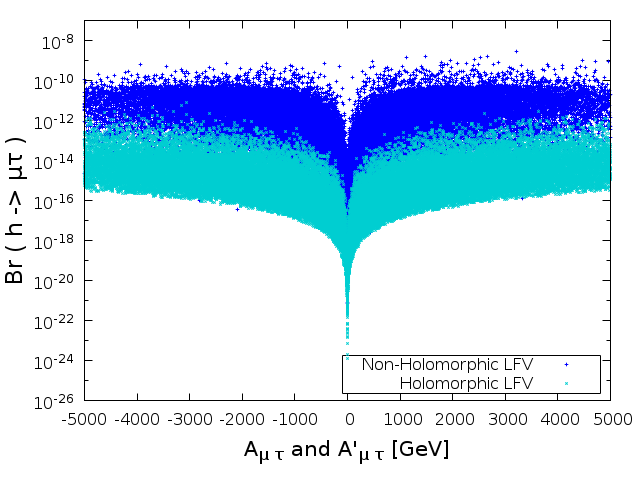}
          \label{hdecaygraphmutau}
          }
           \caption{
             Effect of imposing the LHC constraints on $Y_{ij}$ on the
             analysis of figure \ref{fig:h_mu_tau}:
            Scatter plots of $Br(h \rightarrow e\mu)$ with $A_{e\mu}$ and $A_{e\mu}^{\prime}$ are shown in \ref{fig:h_mu_tau_afterYukawa}(a). Similar
            plots for  $Br(h \rightarrow e  \tau)$ with $A_{e\tau}$ 
          and $A_{e\tau}^{\prime}$ are shown in figure 
          \ref{fig:h_mu_tau_afterYukawa}(b). Figure
          \ref{fig:h_mu_tau_afterYukawa}(c) is for  $Br(h \rightarrow \mu  \tau)$  with $A_{\mu\tau}$ 
          and $A_{\mu\tau}^{\prime}$.
           }
        \label{fig:h_mu_tau_afterYukawa}
  \end{center}
\end{figure}

Coming to cLFV $H$-decays, each of the scattered plots of $Br(H/A \rightarrow l_i \bar{l}_j)$ vs $A_{ij}$ and $A_{ij}^\prime$ as shown in figure
\ref{fig:H_A_mu_tau_afterYukawa} incorporates 
the off-diagonal Yukawa coupling constraints 
and these are subsets of the appropriate sub-figures of
figure \ref{fig:H_A_mu_tau}. Unlike before
$|A_{e\mu}^\prime|$ is further limited, down to 30 GeV
as may be seen 
in figure (\ref{fig:H_A_mu_tau_afterYukawa})a, signifying an appreciable
degree of constraint when compared with figure \ref{fig:tau_3muA_afterYukawa}
or \ref{hdecaygraphemu}. While
figure (\ref{fig:H_A_mu_tau_afterYukawa})b is not much
different from figure (\ref{fig:H_A_mu_tau})b, the
upper limit of $Br(H/A \rightarrow \mu \tau)$ in figure
(\ref{fig:H_A_mu_tau_afterYukawa})c is approximately
reduced by an order of magnitude over the parameter
space when compared with figure (\ref{fig:H_A_mu_tau})c.

\begin{figure}[H]
       \begin{center}
       \subfigure[]{%
           \includegraphics[width=5.5cm,height=4.5cm]{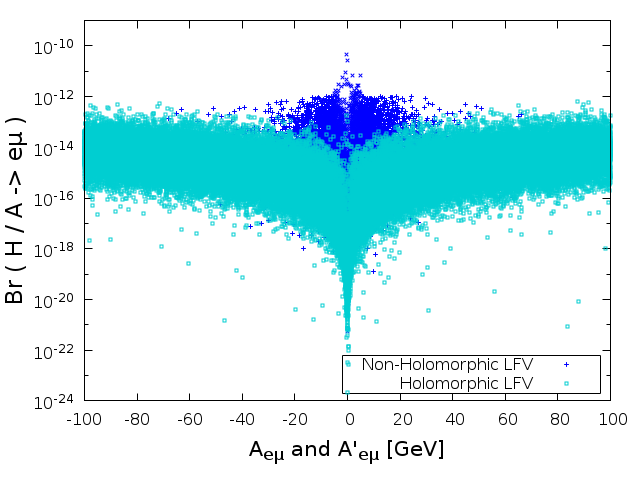}
            \label{Hdecaygraphemu_afterYukawa}
           }%
          \subfigure[]{%
           \includegraphics[width=5.5cm,height=4.5cm]{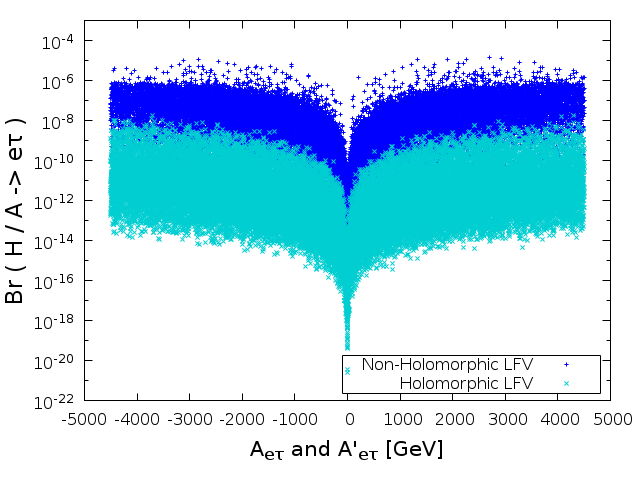}
           }%
            \subfigure[]{%
           \includegraphics[width=5.5cm,height=4.5cm]{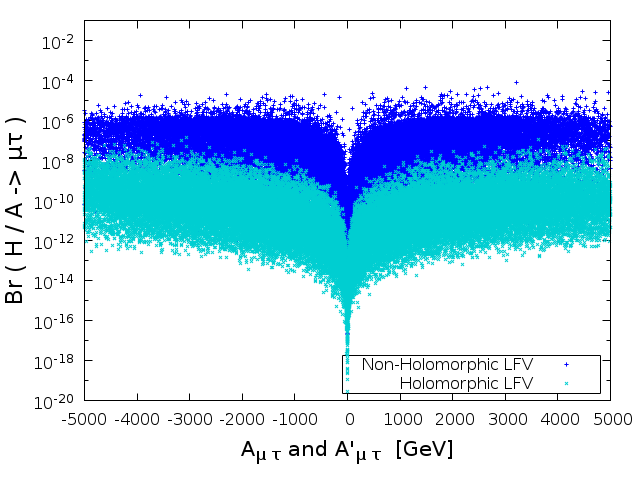}
           }%
            \caption{
              Effect of imposing the LHC constraints on $Y_{ij}$ on the
              analysis of figure \ref{fig:H_A_mu_tau}:
            Scatter plots of $Br(H/A \rightarrow e\mu)$ with $A_{e\mu}$ and $A_{e\mu}^{\prime}$ are shown in \ref{fig:H_A_mu_tau_afterYukawa}(a). Similar
            plots for $Br(H/A \rightarrow e \tau)$ with $A_{e\tau}$ 
          and $A_{e\tau}^{\prime}$ are shown in figure 
          \ref{fig:H_A_mu_tau_afterYukawa}(b). Figure
          \ref{fig:H_A_mu_tau_afterYukawa}(c) is for $Br(H/A \rightarrow \mu \tau)$ 
          and $A_{\mu\tau}^{\prime}$.
}
        \label{fig:H_A_mu_tau_afterYukawa}
  \end{center}
\end{figure}


%
\subsubsection{Results for fixed bino and slepton masses}
We will see now that the multi-parameter scattered plots like 
figure \ref{fig:tau_3mu_after_Yukawa} to \ref{fig:H_A_mu_tau_afterYukawa}
are quite limited in emphasizing the degree of importance of the
Yukawa coupling bounds. 
Hence, in this subsection we fix $M_1$ at 400~GeV and consider 
the diagonal soft slepton soft mass parameters  
($M_{\tilde{L}}$ \& $M_{\tilde{e}}$) to have the
specific values 1, 2, 3 and 5
TeV for MSSM  and 2, 3 and 5 TeV for NHSSM.
The resulting figure \ref{fig:Aij_Yij} depicts the behavior of 
$\sqrt{Y_{ij}^2+Y_{ji}^2}$ with $A_{ij}$ or $A^{(')}_{ij}$ corresponding
to the MSSM and NHSSM cases respectively. 
In the left panel we show 
the plots for the holomorphic off-diagonal trilinear terms and in the right panel we show the similar terms for the non-holomorphic ones. Clearly, larger radiative corrections are induced
in the case of non-vanishing $A^{'}_{ij}$, particularly, when sleptons are light.
For smaller soft masses of sleptons $M_{\tilde{L}}$ \& $M_{\tilde{e}} \simeq 1$ TeV,
with our choice of high $\tan\beta$, non holomorphic trilinear couplings may even generate unacceptable tachyonic states of sleptons. 
The black horizontal lines in each plot relates to the upper bound on respective 
$\sqrt{Y_{ij}^2+Y_{ji}^2}$. 
As we will see next that for the first two generations, the black line
corresponds to the upper limit of $Br(\mu \rightarrow e \gamma)$, and for the other two generations they refer to 
 the null results of $Br(h \rightarrow l_i \bar{l}_j)$ searched by the LHC experiment.

\begin{figure}[H]
       \begin{center}
          \includegraphics[width=7.5cm,height=5.5cm]{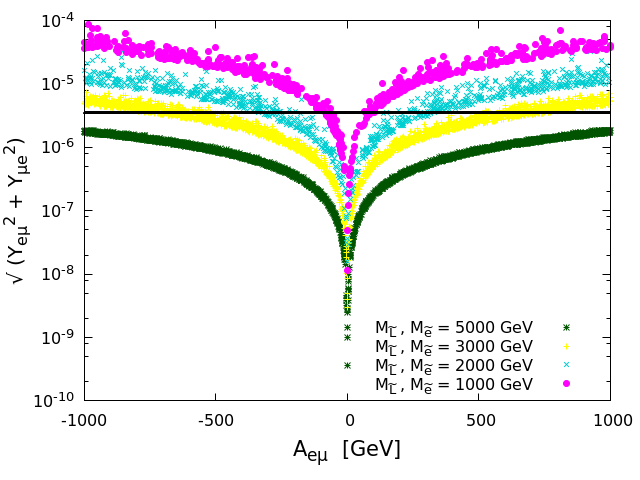}
           \label{fig:Aij_Yij_a}     
            \includegraphics[width=7.5cm,height=5.5cm]{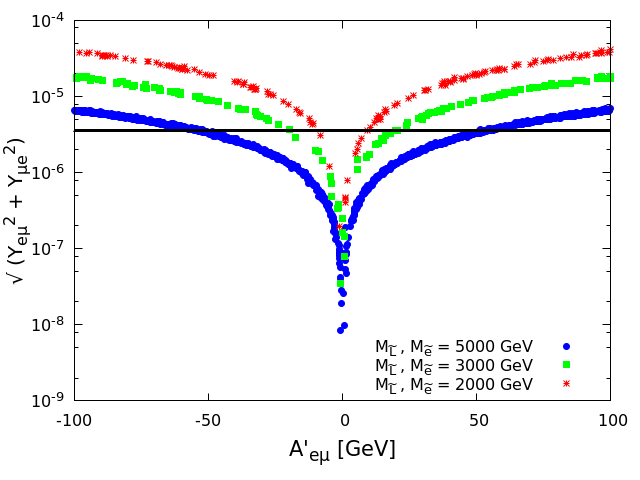}
            \label{fig:Aij_Yij_b}
             \includegraphics[width=7.5cm,height=5.5cm]{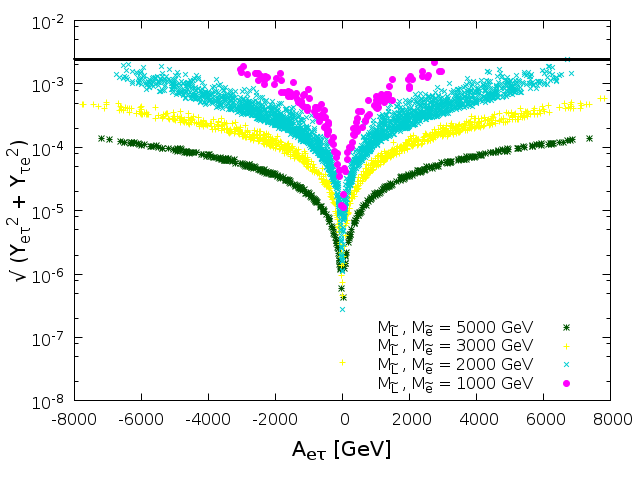}
             \label{fig:Aij_Yij_c}    
             \includegraphics[width=7.5cm,height=5.5cm]{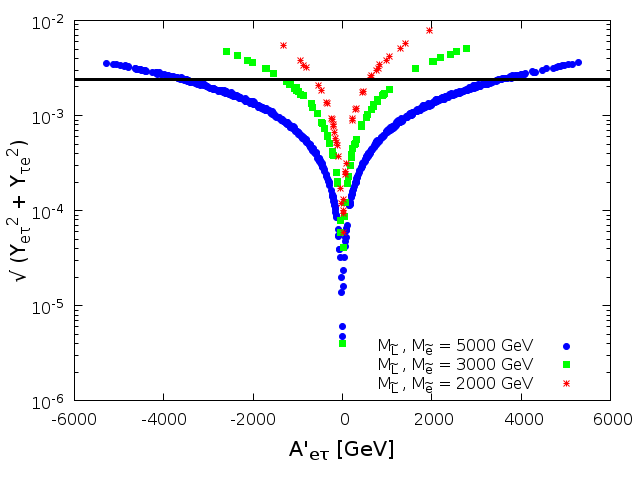}
             \label{fig:Aij_Yij_d}
              \includegraphics[width=7.5cm,height=5.5cm]{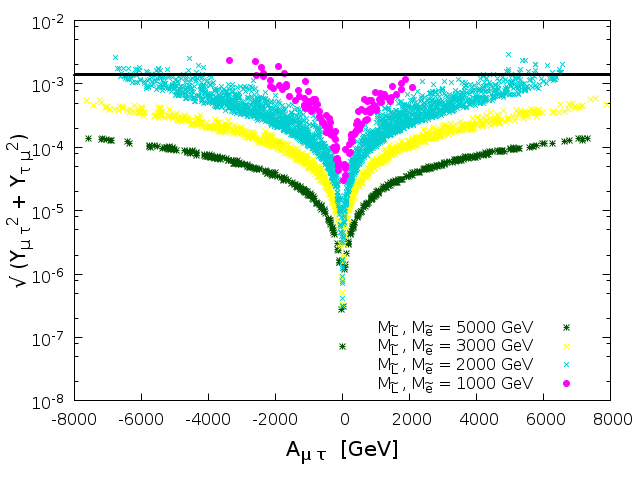}
              \label{fig:Aij_Yij_e}
	      \includegraphics[width=7.5cm,height=5.5cm]{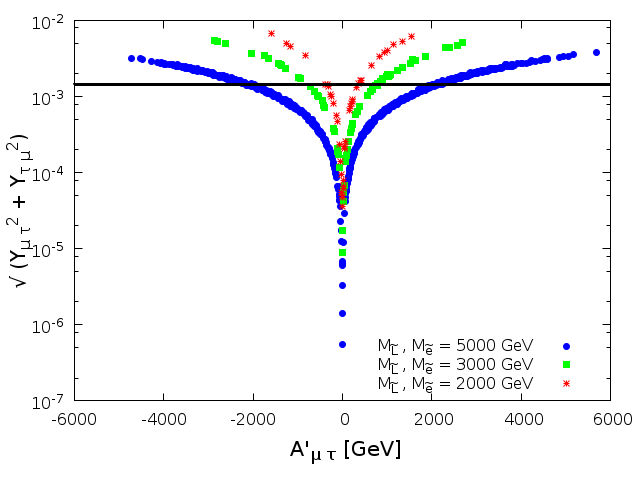}
              \label{fig:Aij_Yij_f}
          \caption{$\sqrt{Y_{ij}^2+Y_{ji}^2}$ vs. $A_{ij}$ (left panel) and $A'_{ij}$ (right panel) are shown
          for fixed slepton bilinear soft masses and $M_1 = 400$ GeV. Other fixed parameters are as stated 
          in table \ref{table-data}. 
          For holomorphic trilinear couplings, slepton soft masses are fixed at 1, 2, 3 and 5 TeV and the same for NH couplings are at 2, 3 and 5 TeV.
          Black horizontal lines in each plot denote the corresponding upper limits on the off-diagonal Yukawa couplings. }
          \label{fig:Aij_Yij}
  \end{center}
\end{figure}

\color{black}
With the understanding on how off-diagonal Yukawa couplings can be directly influenced by trilinear parameters, we now present figure \ref{Ymutau_plot} 
that shows the derived bounds on $l_i \rightarrow 3l_j$, $l_i \rightarrow l_j \gamma$ and $\sqrt{(Y_{ij}^2 + Y_{ji}^2)}$ in the 
$(|Y_{ij}| - |Y_{ji}|)$ plane. The observed LHC limits on $\sqrt{(Y_{ij}^2 + Y_{ji}^2)}$ for 
$\sqrt{s} = 13$ TeV which are derived from the direct searches of $Br(h \rightarrow \mu\tau)$ and 
$Br(h \rightarrow e\tau)$ \cite{Sirunyan:2017xzt} are shown as black solid curves in figure \ref{Ymutau_plot}(b) and (c). These indeed constitute
the most stringent limits concerning the 2nd and 3rd generations of sleptons.
As mentioned earlier, the Higgs boson going to $\mu\tau$ channel gives 
$\sqrt{(Y_{\mu\tau}^2 + Y_{\tau\mu}^2)} < 1.50 \times 10^{-3}$ and the same for $e\tau$ channel leads to 
$\sqrt{(Y_{e\tau}^2 + Y_{\tau e}^2)} < 2.26 \times 10^{-3}$ at 95\% confidence level.
These limits constitute a significant improvement in the $\mu\tau$ channel over the previously obtained limits by CMS
and ATLAS using 8 TeV proton-proton collision data corresponding to an integrated luminosity of about $20 ~ fb^{-1}$ 
\cite{Khachatryan:2015kon,Khachatryan:2016rke,Aad:2019ojw} shown by the green curves above the black curves. For the $e\tau$ mode, in the 
8 TeV analysis, the allowed value of $\sqrt{(Y_{e\tau}^2 + Y_{\tau e}^2)}$ was less than $2.4 \times 10^{-3}$, so the 13 TeV limit is seen to be 
almost overlapped with the 8 TeV one in the middle plot of figure \ref{Ymutau_plot}. We observe that some of the blue points originated from the 
non-holomorphic trilinear couplings exceed the current LHC limits but almost all the cyan points from the holomorphic couplings are safe here.
Needless to mention again, all the data points shown here 
respect charge breaking minima condition.
For the first two generations, most dominant constraint comes from the absence of $\mu \rightarrow e\gamma$,
which is shown by the solid black curve in $|Y_{e\mu}| - |Y_{\mu e}|$ parameter space.
One may find that, the LHC $\sqrt{s}=13$ TeV data puts
95\% confidence level constraints on Yukawa couplings derived from $Br(h \rightarrow e\mu) < 6.2 \times 10^{-5}$ which yields 
$\sqrt{(Y_{e\mu}^2 + Y_{\mu e}^2)}$ to be less than $2.24 \times 10^{-4}$ \cite{Aad:2019ojw}. Absence of $\mu \rightarrow e \gamma$ regulates it even more, 
implying a limit of $\sqrt{(Y_{e\mu}^2 + Y_{\mu e}^2)} < 3.6 \times 10^{-6}$ \cite{Harnik:2012pb}.
Such a tiny Yukawa coupling, with $\cos(\beta - \alpha) \sim 10^{-3} / 10^{-4}$ leads to $Br(h \rightarrow e\mu) \sim \mathcal(O)(10^{-13})$ or so.
\footnote{Here we should state that for non-decoupling Higgs, $m_A\gtrsim m_h$, $\cos(\beta - \alpha)$ can be large which may enhance
$Br(h \rightarrow e\mu / e\tau / \mu\tau)$. For maximum $Br(h \rightarrow l_i \bar{l}_j)$ one may see the reference \cite{Aloni:2015wvn}.}.


\begin{figure}[H]
       \begin{center}
         \subfigure[]{%
           \includegraphics[width=6.0cm,height=5.0cm]{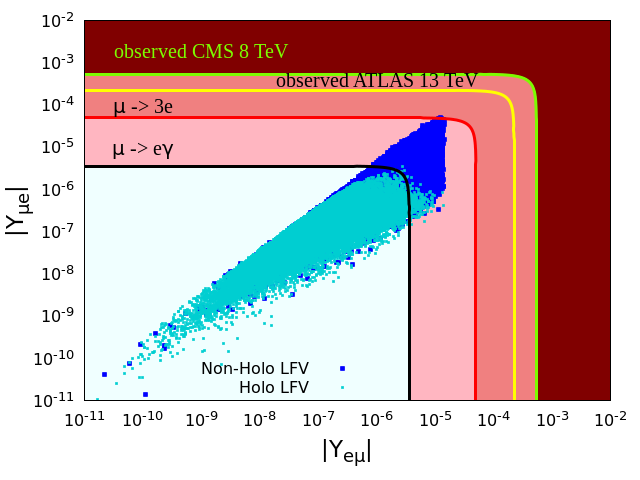}
           }%
            \subfigure[]{%
           \includegraphics[width=6.0cm,height=5.0cm]{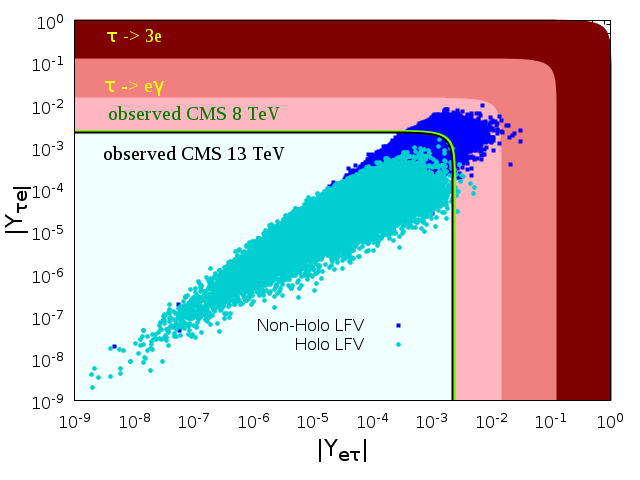}
           }%
          \subfigure[]{%
          \includegraphics[width=6.0cm,height=5.0cm]{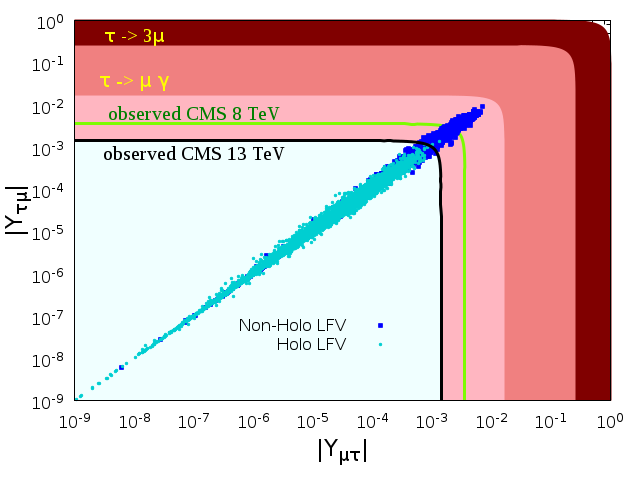}
          }
          \caption{Constraints on the flavour violating Yukawa couplings, $|Y_{ij}|$ and $|Y_{ji}|$ from CMS results of $\sqrt{s}=8$ TeV 
          and $13$ TeV. 
          The different light and deep red shaded regions are restricted by the 
          upper bounds of flavor violating LFV decays. The off-diagonal Yukawa couplings generated from holomorphic and 
          non-holomorphic trilinear couplings are displayed in cyan and blue respectively. The black lines in each plot correspond to the 
          most stringent limits on $Y_{ij}$'s.}
      \label{Ymutau_plot}
  \end{center}
\end{figure}

\color{black}
One may further combine the results of figures \ref{fig:Aij_Yij} and \ref{Ymutau_plot} to conclusively draw upper limits on $A'_{ij}$'s obeying the observations related to relevant  LFV processes and charge breaking minima bounds. This in turn would determine maximum allowed branching ratios for our concerned cLFV process
which could be tested in the near future. For completeness
we summarize our results in the table \ref{benchmark-table}
which shows the allowed values of $A^{l}_{ij}$ and $A^{\prime l}_{ij}$ in general
and the resulting maximum values of different LFV decay branching ratios. Our results can be summarized as follows:
(i) For first two generations of lepton, rise off-diagonal holomorphic and NH
trilinear couplings i.e., $A^{(\prime)}_{e\mu}$ are visibly 
restricted by
the upper bound of $\mu \rightarrow e\gamma$.
(ii) The other two combinations of trilinear coupling parameters, namely, $e\tau$ and $\mu\tau$ are
regulated by the CMS 13 TeV results.
(iii) Finally the derived allowed ranges for 
$A_{e\mu}$ and $A^{\prime}_{e\mu}$ are much more restricted compared to $A^{(\prime)}_{e\tau}$ or  $A^{(\prime)}_{\mu\tau}$.

\scriptsize{
\begin{table}[!htb]
\label{benchmark-table}
	\centering
	\begin{tabular}{|c|c||c|c|c|c|}
		\hline\hline 
		Processes  &  Maximum BR  & Slepton Mass & Parameter & Maximum Ranges & Most Sensitive  \\ [0.5ex]
		           &              &  [GeV]       &           &  of  $A'_{ij}$ \& $A_{ij}$ [GeV]        &       to        \\
		\hline
		$\mu \rightarrow e\gamma$ &  $4.20 \times 10^{-13}$ & $1000$  &  & $--$, $[-100:100]$  & \\
		$\mu \rightarrow 3e$ & $7.52 \times 10^{-15}$ & $2000$ & $A'_{e\mu}$, $A_{e\mu}$  & $[-8:8],[-350:350]$ & Bounds from  \\
		$h \rightarrow e\mu$ & $5.07 \times 10^{-13}$ &  $3000$ &  &  $[-20,20], [-800:800]$ &        $Br(\mu \rightarrow e \gamma)$\\
		$H / A \rightarrow e \mu$ & $2.5 \times 10^{-10}$ & $5000$ & & $[-50,50], [-1500:1500]$ &\\
		\hline
		$\tau \rightarrow e \gamma$ & $1.66 \times 10^{-9}$ & $1000$ & & $--$, $[-3000:3000]$  &\\
		$\tau \rightarrow 3e$ & $1.53 \times 10^{-11}$  & $2000$ & $A'_{e\tau}$, $A_{e\tau}$ & $[-650:650], [-6500:6500]$ & LHC null results of \\
		$h \rightarrow e\tau$ & $1.21 \times 10^{-10}$  & $3000$ &  &$[-1400:1400]$, $[-8000:8000]$   &          $Br(h \rightarrow e\tau)$\\
		$H / A \rightarrow e \tau$ & $3.20 \times 10^{-5}$ & $5000$ &  & $[-3800:3800]$, $[-8000:8000]$ &  \\
		\hline
		$\tau \rightarrow \mu \gamma$ & $2.90 \times 10^{-9}$ & $1000$ & & $--$, $[-2700:2700]$ & \\
		$\tau \rightarrow 3\mu$ & $6.97 \times 10^{-12}$ & $2000$ &  $A'_{\mu\tau}$, $A_{\mu\tau}$ & $[-350:350], [-6000:6000]$ & LHC null results of   \\
		$h \rightarrow \mu\tau$ & $3.24 \times 10^{-10}$ & $3000$ &  & $[-800:800]$, $[-8000:8000]$  &           $Br(h \rightarrow \mu\tau)$\\
		$H / A \rightarrow \mu\tau$ & $5.54 \times 10^{-5}$ & $5000$ & & $[-2400:2400]$, $[-8000:8000]$ &\\
		\hline
		\hline
	\end{tabular}
	\label{benchmark-table}
	\caption{Allowed ranges of holomorphic and non-holomorphic trilinear couplings and the corresponding maximum decay branching ratios after avoiding 
	charge breaking minima, respecting the upper limits of various other LFV decays and LHC constraints on the off-diagonal Yukawa couplings from the non-observation of
	a scalar boson decaying into $e\mu/e\tau/\mu\tau$ channels.}
\end{table}
}

\normalsize{
\color{black}
\section{Conclusion}
The Minimal Supersymmetric Standard Model when extended with the most general soft SUSY breaking trilinear terms   
may lead to interesting phenomenologies. These additional terms that are non-holomorphic
in nature were analyzed in several studies 
in the past as well as in the recent years. We focus on introducing flavor violating lepton decays and Higgs
decaying to charged leptons involving flavor violation due to non-vanishing non-diagonal entries of the 
trilinear coupling matrices both of standard and non-standard types. In this analysis, we first upgrade the existing
analytical result involving trilinear couplings for avoiding the appearance of charge breaking minima of vacuum 
in MSSM. 
In other words, we extend the traditional analytical result for charge breaking in MSSM by including 
non-diagonal entries of the soft-breaking trilinear coupling matrices ($A^l_{ij}$). We extend the analysis further by
involving non-holomorphic trilinear coupling matrices ($A^{\prime l}_{ij}$). By considering {\it vevs} for
appropriate sleptons and non-vanishing values of $A^{\prime l}_{ij}$ we are able to delineate regions of 
parameter space that are associated with appearance of charge breaking minima of the vacuum.
On the contrary, we also find plenty of possibilities of evading the charge breaking conditions
even with reasonably large values of $A^{\prime l}_{ij}$ due to cancellation of terms in the analytical
result. We studied the effects of considering non-vanishing off-diagonal trilinear terms of both types,
one by one, on cLFV processes like $l_j \to l_i \gamma$ or $l_j \to 3l_i$ and all the variants of Higgs ($h,H,A$)
decays into $l_i {\bar l}_j$ involving flavor violation.
For simplicity, we do not consider any flavor violation effect from
the slepton mass matrices. In this phenomenological work, we include 
(i) the present and future experimental sensitivities of cLFV observables, and (ii) the 8~TeV and 13~TeV CMS
results that search SM Higgs boson decays into flavor violating modes, namely $e\tau$ or $\mu \tau$. 
We find that
NH couplings namely $A^{\prime l}_{ij}$ are 
better suited in achieving
larger rates for all flavor violating decay observables that can potentially be tested in the near future. In particular, $\mu \to e \gamma$ would be more favourable 
to test $A^{\prime l}_{ij}$ involving first two generation sleptons. On the other hand, 
MSSM Higgs decays (specially that of the heavier Higgs bosons) into LFV modes may strongly be influenced by
$A^\prime_{e\tau}$ or $A^\prime_{\mu\tau}$. For most of these
observables the standard trilinear couplings $A_{ij}$ turn out to be inadequate to
produce any significant contribution in relation to the present or future experimental measurements. This indeed emphasizes the usefulness 
of including the non-holomorphic trilinear terms for such analyses.

\section*{Acknowledgements}

SM would like to thank Abhishek Dey for many helpful discussions. The computations was supported in part by the SAMKHYA: The High Performance Computing
Facility provided by Institute of Physics, Bhubaneswar.

}


\begin{thebibliography}{999}
\bibitem{Khachatryan:2016vau} 
  G.~Aad {\it et al.} [ATLAS and CMS Collaborations],
  JHEP {\bf 1608}, 045 (2016)
  doi:10.1007/JHEP08(2016)045
  [arXiv:1606.02266 [hep-ex]].


\bibitem{Aad:2012tfa} 
  G.~Aad {\it et al.} [ATLAS Collaboration],
  Phys.\ Lett.\ B {\bf 716}, 1 (2012)
  doi:10.1016/j.physletb.2012.08.020
  [arXiv:1207.7214 [hep-ex]].


\bibitem{Chatrchyan:2012xdj} 
  S.~Chatrchyan {\it et al.} [CMS Collaboration],
  Phys.\ Lett.\ B {\bf 716}, 30 (2012)
  doi:10.1016/j.physletb.2012.08.021
  [arXiv:1207.7235 [hep-ex]].


\bibitem{petkov} 
  S. Petcov, "The processes  $\mu \rightarrow e \gamma$,
  $V^\prime \rightarrow V\gamma$ in the
Weinberg Salam model with neutrino mixing," Soviet Journal of
Nuclear Physics, vol. 25, p. 340, 1977.


\bibitem{Cheng:1976uq} 
  T.~P.~Cheng and L.~F.~Li,
  Phys.\ Rev.\ Lett.\  {\bf 38}, 381 (1977).
  doi:10.1103/PhysRevLett.38.381


\bibitem{Bilenky:1977du} 
  S.~M.~Bilenky, S.~T.~Petcov and B.~Pontecorvo,
  Phys.\ Lett.\  {\bf 67B}, 309 (1977).
  doi:10.1016/0370-2693(77)90379-3


\bibitem{Raidal:2008jk} 
  M.~Raidal {\it et al.},
  Eur.\ Phys.\ J.\ C {\bf 57}, 13 (2008)
  doi:10.1140/epjc/s10052-008-0715-2
  [arXiv:0801.1826 [hep-ph]].

\bibitem{SUSYreviews1}
For reviews on supersymmetry, see, { e.g.},
H. P. Nilles, Phys. Report (110, 1, 1984);
J.~D.~Lykken,
  hep-th/9612114;
J. Wess and J. Bagger, {\it Supersymmetry and Supergravity}, 2nd ed.,
(Princeton, 1991).


\bibitem{SUSYbook1}
M. Drees, P. Roy and R. M. Godbole,
{\it Theory and Phenomenology of Sparticles},
(World Scientific, Singapore, 2005).  

\bibitem{SUSYbook2}
  H.~Baer and X.~Tata,
  {\it{Weak scale supersymmetry: From superfields to scattering events}},
  Cambridge, UK: Univ. Pr. (2006) 537 p.
  
\bibitem{SUSYreviews2}
  D.~J.~H.~Chung, L.~L.~Everett, G.~L.~Kane, S.~F.~King,
J.~D.~Lykken and L.~T.~Wang,
  Phys.\ Rept.\  {\bf 407}, 1 (2005);
H. E. Haber and G. Kane, Phys. Report (117, 75, 1985) ;
S.~P.~Martin,
arXiv:hep-ph/9709356.

\bibitem{Gabbiani:1996hi} 
  F.~Gabbiani, E.~Gabrielli, A.~Masiero and L.~Silvestrini,
  Nucl.\ Phys.\ B {\bf 477}, 321 (1996)
  doi:10.1016/0550-3213(96)00390-2
  [hep-ph/9604387].


\bibitem{deSalas:2017kay} 
  P.~F.~de Salas, D.~V.~Forero, C.~A.~Ternes, M.~Tortola and J.~W.~F.~Valle,
  Phys.\ Lett.\ B {\bf 782}, 633 (2018)
  doi:10.1016/j.physletb.2018.06.019
  [arXiv:1708.01186 [hep-ph]].


\bibitem{Esteban:2016qun} 
  I.~Esteban, M.~C.~Gonzalez-Garcia, M.~Maltoni, I.~Martinez-Soler and T.~Schwetz,
  JHEP {\bf 1701}, 087 (2017)
  doi:10.1007/JHEP01(2017)087
  [arXiv:1611.01514 [hep-ph]].

\bibitem{seesaw:I}
P.~Minkowski,
Phys.\ Lett.\ B {\bf 67} (1977) 421;
%
M.~Gell-Mann, P.~Ramond and R.~Slansky, in {\it Complex Spinors and
  Unified Theories} eds. P.~Van.~Nieuwenhuizen and D.~Z.~Freedman,
  {\it Supergravity} (North-Holland, Amsterdam, 1979), 
  p.315 [Print-80-0576 (CERN)];
%
T.~Yanagida, in {\it Proceedings of the Workshop on the Unified Theory
and the Baryon Number in the Universe}, eds. O.~Sawada and
A.~Sugamoto (KEK, Tsukuba, 1979), p.95;
%
S.~L.~Glashow, in {\it Quarks and Leptons}, eds. M.~L\'evy {\it et
al.} (Plenum Press, New York, 1980), p.687;
%
R.~N.~Mohapatra and G.~Senjanovi\'c,
Phys.\ Rev.\ Lett.\  {\bf 44} (1980) 912.

\bibitem{seesaw:II}
%
R.~Barbieri, D.~V.~Nanopolous, G.~Morchio and F.~Strocchi, 
Phys.\ Lett.\ B {\bf 90} (1980) 91;
%
R.~E.~Marshak and R.~N.~Mohapatra, {\it Invited talk given at Orbis
  Scientiae, Coral Gables, Fla., Jan. 14-17, 1980}, VPI-HEP-80/02;
%
T.~P.~Cheng and L.~F.~Li,
Phys.\ Rev.\ D {\bf 22} (1980) 2860;
%
M.~Magg and C.~Wetterich,
Phys.\ Lett.\ B {\bf 94} (1980) 61;
%
G. Lazarides, Q. Shafi and C. Wetterich,
Nucl. Phys. B {\bf 181} (1981) 287; 
J.~Schechter and J.~W.~F.~Valle,
Phys.\ Rev.\ D {\bf 22} (1980) 2227;
%
R.~N.~Mohapatra and G.~Senjanovi\'c,
Phys.\ Rev.\ D {\bf 23} (1981) 165.

\bibitem{seesaw:III}
E. Ma,
Phys. Rev. Lett. {\bf 81} (1998) 1171 
[arXiv:hep-ph/9805219];
R. Foot, H. Lew, X. G. He and G. C. Joshi,
Z. Phys. C {\bf 44} (1989) 441.

\bibitem{susy-seesaw}
S.~F.~King,
  Phys.\ Lett.\  B {\bf 439} (1998) 350 
  [arXiv:hep-ph/9806440];
%
  S.~Davidson and S.~F.~King,
  Phys.\ Lett.\  B {\bf 445} (1998) 191 
  [arXiv:hep-ph/9808296].

\bibitem{Borzumati:1986qx} 
  F.~Borzumati and A.~Masiero,
  Phys.\ Rev.\ Lett.\  {\bf 57}, 961 (1986).
  doi:10.1103/PhysRevLett.57.961

\bibitem{hisano}
J.~Hisano, T.~Moroi, K.~Tobe and M.~Yamaguchi,
Phys.\ Rev.\ D {\bf 53} (1996) 2442 
[arXiv:hep-ph/9510309];
  J.~Hisano, T.~Moroi, K.~Tobe, M.~Yamaguchi and T.~Yanagida,
  Phys.\ Lett.\ B {\bf 357}, 579 (1995)
  doi:10.1016/0370-2693(95)00954-J
  [hep-ph/9501407];
J.~Hisano and D.~Nomura,
Phys.\ Rev.\ D {\bf 59} (1999) 116005 
[arXiv:hep-ph/9810479].

\bibitem{Hamzaoui:1998nu} 
  C.~Hamzaoui, M.~Pospelov and M.~Toharia,
  Phys.\ Rev.\ D {\bf 59}, 095005 (1999)
  doi:10.1103/PhysRevD.59.095005
  [hep-ph/9807350].


\bibitem{DiazCruz:1999xe} 
  J.~L.~Diaz-Cruz and J.~J.~Toscano,
  Phys.\ Rev.\ D {\bf 62}, 116005 (2000)
  doi:10.1103/PhysRevD.62.116005
  [hep-ph/9910233].


\bibitem{Isidori:2001fv} 
  G.~Isidori and A.~Retico,
  JHEP {\bf 0111}, 001 (2001)
  doi:10.1088/1126-6708/2001/11/001
  [hep-ph/0110121].


\bibitem{Babu:2002et} 
  K.~S.~Babu and C.~Kolda,
  Phys.\ Rev.\ Lett.\  {\bf 89}, 241802 (2002)
  doi:10.1103/PhysRevLett.89.241802
  [hep-ph/0206310].


 \bibitem{masieroall}
  A.~Masiero, S.~K.~Vempati and O.~Vives,
  Nucl.\ Phys.\ B {\bf 649} (2003) 189
  doi:10.1016/S0550-3213(02)01031-3
  [hep-ph/0209303];
  M.~Ciuchini, A.~Masiero, L.~Silvestrini, S.~K.~Vempati and O.~Vives,
  Phys.\ Rev.\ Lett.\  {\bf 92} (2004) 071801
  doi:10.1103/PhysRevLett.92.071801
  [hep-ph/0307191];
  A.~Masiero, S.~K.~Vempati and O.~Vives,
  Nucl.\ Phys.\ Proc.\ Suppl.\  {\bf 137} (2004) 156
  doi:10.1016/j.nuclphysbps.2004.10.058
  [hep-ph/0405017];
  A.~Masiero, S.~K.~Vempati and O.~Vives,
  New J.\ Phys.\  {\bf 6} (2004) 202
  doi:10.1088/1367-2630/6/1/202
  [hep-ph/0407325];
 L.~Calibbi, A.~Faccia, A.~Masiero and S.~K.~Vempati,
  Phys.\ Rev.\ D {\bf 74} (2006) 116002
  doi:10.1103/PhysRevD.74.116002
  [hep-ph/0605139].
\bibitem{Paradisi:2005tk} 
  P.~Paradisi,
  JHEP {\bf 0602}, 050 (2006)
  doi:10.1088/1126-6708/2006/02/050
  [hep-ph/0508054].


\bibitem{Brignole:2003iv} 
  A.~Brignole and A.~Rossi,
  Phys.\ Lett.\ B {\bf 566}, 217 (2003)
  doi:10.1016/S0370-2693(03)00837-2
  [hep-ph/0304081].


\bibitem{Brignole:2004ah} 
  A.~Brignole and A.~Rossi,
  Nucl.\ Phys.\ B {\bf 701}, 3 (2004)
  doi:10.1016/j.nuclphysb.2004.08.037
  [hep-ph/0404211].


\bibitem{Han:2000jz} 
  T.~Han and D.~Marfatia,
  Phys.\ Rev.\ Lett.\  {\bf 86}, 1442 (2001)
  doi:10.1103/PhysRevLett.86.1442
  [hep-ph/0008141].


\bibitem{Arganda:2004bz} 
  E.~Arganda, A.~M.~Curiel, M.~J.~Herrero and D.~Temes,
  Phys.\ Rev.\ D {\bf 71}, 035011 (2005)
  doi:10.1103/PhysRevD.71.035011
  [hep-ph/0407302].


\bibitem{Arganda:2005ji} 
  E.~Arganda and M.~J.~Herrero,
  Phys.\ Rev.\ D {\bf 73}, 055003 (2006)
  doi:10.1103/PhysRevD.73.055003
  [hep-ph/0510405].


\bibitem{Gomez:2015ila} 
  M.~E.~Gomez, S.~Heinemeyer and M.~Rehman,
  Eur.\ Phys.\ J.\ C {\bf 75}, no. 9, 434 (2015)
  doi:10.1140/epjc/s10052-015-3654-8
  [arXiv:1501.02258 [hep-ph]].


\bibitem{Gomez:2017dhl} 
  M.~E.~Gomez, S.~Heinemeyer and M.~Rehman,
  arXiv:1703.02229 [hep-ph].


\bibitem{Aloni:2015wvn} 
  D.~Aloni, Y.~Nir and E.~Stamou,
  JHEP {\bf 1604}, 162 (2016)
  doi:10.1007/JHEP04(2016)162
  [arXiv:1511.00979 [hep-ph]].


\bibitem{Arhrib:2012ax} 
  A.~Arhrib, Y.~Cheng and O.~C.~W.~Kong,
  Phys.\ Rev.\ D {\bf 87}, no. 1, 015025 (2013)
  doi:10.1103/PhysRevD.87.015025
  [arXiv:1210.8241 [hep-ph]].


\bibitem{Arana-Catania:2013xma} 
  M.~Arana-Catania, E.~Arganda and M.~J.~Herrero,
  JHEP {\bf 1309}, 160 (2013)
  Erratum: [JHEP {\bf 1510}, 192 (2015)]
  doi:10.1007/JHEP10(2015)192, 10.1007/JHEP09(2013)160
  [arXiv:1304.3371 [hep-ph]].


\bibitem{Abada:2014kba} 
  A.~Abada, M.~E.~Krauss, W.~Porod, F.~Staub, A.~Vicente and C.~Weiland,
  JHEP {\bf 1411}, 048 (2014)
  doi:10.1007/JHEP11(2014)048
  [arXiv:1408.0138 [hep-ph]].


\bibitem{Hammad:2016bng} 
  A.~Hammad, S.~Khalil and C.~S.~Un,
  Phys.\ Rev.\ D {\bf 95}, no. 5, 055028 (2017)
  doi:10.1103/PhysRevD.95.055028
  [arXiv:1605.07567 [hep-ph]].

  
\bibitem{Evans:2018ewb} 
  J.~L.~Evans, K.~Kadota and T.~Kuwahara,
  Phys.\ Rev.\ D {\bf 98}, no. 7, 075030 (2018)
  doi:10.1103/PhysRevD.98.075030
  [arXiv:1807.08234 [hep-ph]].

\bibitem{Deppisch}
  F.~Deppisch and J.~W.~F.~Valle,
  Phys.\ Rev.\ D {\bf 72} (2005) 036001
  doi:10.1103/PhysRevD.72.036001
  [hep-ph/0406040].
  
\bibitem{mypaper1}
  A.~Abada, D.~Das and C.~Weiland,
  JHEP {\bf 1203} (2012) 100
  doi:10.1007/JHEP03(2012)100
  [arXiv:1111.5836 [hep-ph]].
\bibitem{mypaper2}
  A.~Abada, D.~Das, A.~Vicente and C.~Weiland,
  JHEP {\bf 1209} (2012) 015
  doi:10.1007/JHEP09(2012)015
  [arXiv:1206.6497 [hep-ph]].

\bibitem{Arganda:2014dta} 
  E.~Arganda, M.~J.~Herrero, X.~Marcano and C.~Weiland,
  Phys.\ Rev.\ D {\bf 91}, no. 1, 015001 (2015)
  doi:10.1103/PhysRevD.91.015001
  [arXiv:1405.4300 [hep-ph]].

\bibitem{Arganda:2015naa} 
  E.~Arganda, M.~J.~Herrero, X.~Marcano and C.~Weiland,
  Phys.\ Rev.\ D {\bf 93}, no. 5, 055010 (2016)
  doi:10.1103/PhysRevD.93.055010
  [arXiv:1508.04623 [hep-ph]].


\bibitem{Arganda:2015uca} 
  E.~Arganda, M.~J.~Herrero, R.~Morales and A.~Szynkman,
  JHEP {\bf 1603}, 055 (2016)
  doi:10.1007/JHEP03(2016)055
  [arXiv:1510.04685 [hep-ph]].


\bibitem{Barbieri:1994pv} 
  R.~Barbieri and L.~J.~Hall,
  Phys.\ Lett.\ B {\bf 338}, 212 (1994)
  doi:10.1016/0370-2693(94)91368-4
  [hep-ph/9408406].


\bibitem{Barbieri:1995tw} 
  R.~Barbieri, L.~J.~Hall and A.~Strumia,
  Nucl.\ Phys.\ B {\bf 445}, 219 (1995)
  doi:10.1016/0550-3213(95)00208-A
  [hep-ph/9501334].


\bibitem{Hall:1985dx} 
  L.~J.~Hall, V.~A.~Kostelecky and S.~Raby,
  Nucl.\ Phys.\ B {\bf 267}, 415 (1986).
  doi:10.1016/0550-3213(86)90397-4


\bibitem{Dev:2009aw} 
  P.~S.~B.~Dev and R.~N.~Mohapatra,
  Phys.\ Rev.\ D {\bf 81}, 013001 (2010)
  doi:10.1103/PhysRevD.81.013001
  [arXiv:0910.3924 [hep-ph]].


\bibitem{Brignole:1997dp} 
  A.~Brignole, L.~E.~Ibanez and C.~Munoz,
  Adv.\ Ser.\ Direct.\ High Energy Phys.\  {\bf 18}, 125 (1998)
  [hep-ph/9707209].


  \bibitem{Jack:nh} 
  I.~Jack and D.~R.~T.~Jones,
  Phys.\ Lett.\ B {\bf 457}, 101 (1999)
  doi:10.1016/S0370-2693(99)00530-4
  [hep-ph/9903365].
\bibitem{Martin:nh} 
  S.~P.~Martin,
  Phys.\ Rev.\ D {\bf 61}, 035004 (2000)
  doi:10.1103/PhysRevD.61.035004
  [hep-ph/9907550].
\bibitem{Jack:nh1} 
  I.~Jack and D.~R.~T.~Jones,
  Phys.\ Rev.\ D {\bf 61}, 095002 (2000)
  doi:10.1103/PhysRevD.61.095002
  [hep-ph/9909570].
\bibitem{Haber:wh} 
  H.~E.~Haber and J.~D.~Mason,
  Phys.\ Rev.\ D {\bf 77}, 115011 (2008)
  doi:10.1103/PhysRevD.77.115011
  [arXiv:0711.2890 [hep-ph]].

\bibitem{Hetherington:2001bk} 
  J.~P.~J.~Hetherington,
  JHEP {\bf 0110}, 024 (2001)
  doi:10.1088/1126-6708/2001/10/024
  [hep-ph/0108206].


\bibitem{Chattopadhyay:2016ivr} 
  U.~Chattopadhyay and A.~Dey,
  JHEP {\bf 1610}, 027 (2016)
  doi:10.1007/JHEP10(2016)027
  [arXiv:1604.06367 [hep-ph]].


\bibitem{Un:2014afa} 
  C.~S.~Ün, Ş.~H.~Tanyıldızı, S.~Kerman and L.~Solmaz,
  Phys.\ Rev.\ D {\bf 91}, no. 10, 105033 (2015)
  doi:10.1103/PhysRevD.91.105033
  [arXiv:1412.1440 [hep-ph]].


\bibitem{Ross:2016pml} 
  G.~G.~Ross, K.~Schmidt-Hoberg and F.~Staub,
  Phys.\ Lett.\ B {\bf 759}, 110 (2016)
  doi:10.1016/j.physletb.2016.05.053
  [arXiv:1603.09347 [hep-ph]].


\bibitem{Ross:2017kjc} 
  G.~G.~Ross, K.~Schmidt-Hoberg and F.~Staub,
  JHEP {\bf 1703}, 021 (2017)
  doi:10.1007/JHEP03(2017)021
  [arXiv:1701.03480 [hep-ph]].


\bibitem{Chattopadhyay:2017qvh} 
  U.~Chattopadhyay, D.~Das and S.~Mukherjee,
  JHEP {\bf 1801}, 158 (2018)
  doi:10.1007/JHEP01(2018)158
  [arXiv:1710.10120 [hep-ph]].


  
\bibitem{Crivellin:2018mqz} 
  A.~Crivellin, Z.~Fabisiewicz, W.~Materkowska, U.~Nierste, S.~Pokorski and J.~Rosiek,
  JHEP {\bf 1806}, 003 (2018)
  doi:10.1007/JHEP06(2018)003
  [arXiv:1802.06803 [hep-ph]].
  
  
\bibitem{Crivellin:2011jt} 
  A.~Crivellin, L.~Hofer and J.~Rosiek,
  JHEP {\bf 1107}, 017 (2011)
  doi:10.1007/JHEP07(2011)017
  [arXiv:1103.4272 [hep-ph]].
  
  
\bibitem{Rosiek:1995kg} 
  J.~Rosiek,
  hep-ph/9511250.


\bibitem{Pham:1998fq} 
  X.~Y.~Pham,
  Eur.\ Phys.\ J.\ C {\bf 8}, 513 (1999)
  doi:10.1007/s100529901088
  [hep-ph/9810484].


\bibitem{Lees:2010ez} 
  J.~P.~Lees {\it et al.} [BaBar Collaboration],
  Phys.\ Rev.\ D {\bf 81}, 111101 (2010)
  doi:10.1103/PhysRevD.81.111101
  [arXiv:1002.4550 [hep-ex]].


\bibitem{Hayasaka:2010np} 
  K.~Hayasaka {\it et al.},
  Phys.\ Lett.\ B {\bf 687}, 139 (2010)
  doi:10.1016/j.physletb.2010.03.037
  [arXiv:1001.3221 [hep-ex]].


\bibitem{Aaij:2014azz} 
  R.~Aaij {\it et al.} [LHCb Collaboration],
  JHEP {\bf 1502}, 121 (2015)
  doi:10.1007/JHEP02(2015)121
  [arXiv:1409.8548 [hep-ex]].


\bibitem{Amhis:2014hma} 
  Y.~Amhis {\it et al.} [Heavy Flavor Averaging Group (HFAG)],
  arXiv:1412.7515 [hep-ex].


\bibitem{Dedes:2002rh} 
  A.~Dedes, J.~R.~Ellis and M.~Raidal,
  Phys.\ Lett.\ B {\bf 549}, 159 (2002)
  doi:10.1016/S0370-2693(02)02900-3
  [hep-ph/0209207].


  \bibitem{chowdhury95}
  D. Choudhury, F. Eberlein, A. Konig, J. Louis and S. Pokorski,Phys. Lett.B342(1995) 180

\bibitem{Casas:1996de} 
  J.~A.~Casas and S.~Dimopoulos,
  Phys.\ Lett.\ B {\bf 387}, 107 (1996)
  doi:10.1016/0370-2693(96)01000-3
  [hep-ph/9606237].


 \bibitem{Frere1983}
  J. Frere, D. Jones and S. Raby,
Fermion Masses and Induction of the Weak Scale by Supergravity,
Nucl.Phys. B222(1983) 11

\bibitem{Casas:1995pd} 
  J.~A.~Casas, A.~Lleyda and C.~Munoz,
  Nucl.\ Phys.\ B {\bf 471}, 3 (1996)
  doi:10.1016/0550-3213(96)00194-0
  [hep-ph/9507294].

\bibitem{Gunion1988}  
  J. Gunion, H. Haber and M. Sher,
Charge/Color Breaking Minima and a-Parameter Bounds in
Supersymmetric Models,
Nucl.Phys. B306(1988) 1.

\bibitem{Dress1985}
M. Drees, M. Gluck and K. Grassie,
A New Class of False Vacua in Low-energy N=1
Supergravity Theories,
Phys.Lett. B157(1985) 164


\bibitem{Komatsu1988}
H. Komatsu,
New Constraints on Parameters in the Minimal Supersymmetric Model,
Phys.Lett. B215(1988) 323

\bibitem{Langacker:1994bc} 
  P.~Langacker and N.~Polonsky,
  Phys.\ Rev.\ D {\bf 50}, 2199 (1994)
  doi:10.1103/PhysRevD.50.2199
  [hep-ph/9403306].


\bibitem{Strumia:1996pr} 
  A.~Strumia,
  Nucl.\ Phys.\ B {\bf 482}, 24 (1996)
  doi:10.1016/S0550-3213(96)00554-8
  [hep-ph/9604417].


\bibitem{Hollik:2016dcm} 
  W.~G.~Hollik,
  JHEP {\bf 1608}, 126 (2016)
  doi:10.1007/JHEP08(2016)126
  [arXiv:1606.08356 [hep-ph]].


\bibitem{Chattopadhyay:2014gfa} 
  U.~Chattopadhyay and A.~Dey,
  JHEP {\bf 1411}, 161 (2014)
  doi:10.1007/JHEP11(2014)161
  [arXiv:1409.0611 [hep-ph]].


\bibitem{Chowdhury:2013dka} 
  D.~Chowdhury, R.~M.~Godbole, K.~A.~Mohan and S.~K.~Vempati,
  JHEP {\bf 1402}, 110 (2014)
  Erratum: [JHEP {\bf 1803}, 149 (2018)]
  doi:10.1007/JHEP03(2018)149, 10.1007/JHEP02(2014)110
  [arXiv:1310.1932 [hep-ph]].


\bibitem{Beuria:2017gtf} 
  J.~Beuria and A.~Dey,
  JHEP {\bf 1710}, 154 (2017)
  doi:10.1007/JHEP10(2017)154
  [arXiv:1708.08361 [hep-ph]].


\bibitem{Kusenko:1996jn} 
  A.~Kusenko, P.~Langacker and G.~Segre,
  Phys.\ Rev.\ D {\bf 54}, 5824 (1996)
  doi:10.1103/PhysRevD.54.5824
  [hep-ph/9602414].


\bibitem{Kusenko:1996xt} 
  A.~Kusenko and P.~Langacker,
  Phys.\ Lett.\ B {\bf 391}, 29 (1997)
  doi:10.1016/S0370-2693(96)01470-0
  [hep-ph/9608340].


\bibitem{Kusenko:1996vp} 
  A.~Kusenko,
  Nucl.\ Phys.\ Proc.\ Suppl.\  {\bf 52A}, 67 (1997)
  doi:10.1016/S0920-5632(96)00535-X
  [hep-ph/9607287].


\bibitem{Kusenko:1995jv} 
  A.~Kusenko,
  Phys.\ Lett.\ B {\bf 358}, 51 (1995)
  doi:10.1016/0370-2693(95)00994-V
  [hep-ph/9504418].


\bibitem{Brandenberger:1984cz} 
  R.~H.~Brandenberger,
  Rev.\ Mod.\ Phys.\  {\bf 57}, 1 (1985).
  doi:10.1103/RevModPhys.57.1


\bibitem{LeMouel:2001ym} 
  C.~Le Mouel,
  Phys.\ Rev.\ D {\bf 64}, 075009 (2001)
  doi:10.1103/PhysRevD.64.075009
  [hep-ph/0103341].


\bibitem{Chattopadhyay:2018tqv} 
  U.~Chattopadhyay, A.~Datta, S.~Mukherjee and A.~K.~Swain,
  JHEP {\bf 1810}, 202 (2018)
  doi:10.1007/JHEP10(2018)202
  [arXiv:1809.05438 [hep-ph]].


\bibitem{TheMEG:2016wtm} 
  A.~M.~Baldini {\it et al.} [MEG Collaboration],
  Eur.\ Phys.\ J.\ C {\bf 76}, no. 8, 434 (2016)
  doi:10.1140/epjc/s10052-016-4271-x
  [arXiv:1605.05081 [hep-ex]].


\bibitem{Baldini:2013ke} 
  A.~M.~Baldini {\it et al.},
  arXiv:1301.7225 [physics.ins-det].


\bibitem{Blondel:2013ia} 
  A.~Blondel {\it et al.},
  arXiv:1301.6113 [physics.ins-det].


\bibitem{Perrevoort:2016nuv} 
  A.~K.~Perrevoort [Mu3e Collaboration],
  EPJ Web Conf.\  {\bf 118}, 01028 (2016)
  doi:10.1051/epjconf/201611801028
  [arXiv:1605.02906 [physics.ins-det]].


\bibitem{Bellgardt:1987du} 
  U.~Bellgardt {\it et al.} [SINDRUM Collaboration],
  Nucl.\ Phys.\ B {\bf 299}, 1 (1988).
  doi:10.1016/0550-3213(88)90462-2


\bibitem{Perrevoort:2018ttp} 
  A.~K.~Perrevoort [Mu3e Collaboration],
  SciPost Phys.\ Proc.\  {\bf 1}, 052 (2019)
  doi:10.21468/SciPostPhysProc.1.052
  [arXiv:1812.00741 [hep-ex]].


\bibitem{Hayasaka:2007vc} 
  K.~Hayasaka {\it et al.} [Belle Collaboration],
  Phys.\ Lett.\ B {\bf 666}, 16 (2008)
  doi:10.1016/j.physletb.2008.06.056
  [arXiv:0705.0650 [hep-ex]].


\bibitem{Aubert:2009ag} 
  B.~Aubert {\it et al.} [BaBar Collaboration],
  Phys.\ Rev.\ Lett.\  {\bf 104}, 021802 (2010)
  doi:10.1103/PhysRevLett.104.021802
  [arXiv:0908.2381 [hep-ex]].


\bibitem{Aaij:2013fia} 
  R.~Aaij {\it et al.} [LHCb Collaboration],
  Phys.\ Lett.\ B {\bf 724}, 36 (2013)
  doi:10.1016/j.physletb.2013.05.063
  [arXiv:1304.4518 [hep-ex]].


\bibitem{Aushev:2010bq} 
  T.~Aushev {\it et al.},
  arXiv:1002.5012 [hep-ex].


\bibitem{collaboration:2010ipa} 
  K.~Hayasaka [Belle Collaboration],
  PoS ICHEP {\bf 2010}, 241 (2010)
  doi:10.22323/1.120.0241
  [arXiv:1011.6474 [hep-ex]].


\bibitem{OLeary:2010hau} 
  B.~O'Leary {\it et al.} [SuperB Collaboration],
  arXiv:1008.1541 [hep-ex].


\bibitem{Khachatryan:2015kon} 
  V.~Khachatryan {\it et al.} [CMS Collaboration],
  Phys.\ Lett.\ B {\bf 749}, 337 (2015)
  doi:10.1016/j.physletb.2015.07.053
  [arXiv:1502.07400 [hep-ex]].


\bibitem{Aad:2015gha} 
  G.~Aad {\it et al.} [ATLAS Collaboration],
  JHEP {\bf 1511}, 211 (2015)
  doi:10.1007/JHEP11(2015)211
  [arXiv:1508.03372 [hep-ex]].


  \bibitem{LFVCMS13TeV}
  Search for Lepton Flavour Violating Decays of the Higgs Boson in the mu-tau final state at 13 TeV\\
  https://cds.cern.ch/record/2159682

\bibitem{Sirunyan:2017xzt} 
  A.~M.~Sirunyan {\it et al.} [CMS Collaboration],
  JHEP {\bf 1806}, 001 (2018)
  doi:10.1007/JHEP06(2018)001
  [arXiv:1712.07173 [hep-ex]].


\bibitem{Aad:2019ugc} 
  G.~Aad {\it et al.} [ATLAS Collaboration],
  doi:10.1016/j.physletb.2019.135069
  arXiv:1907.06131 [hep-ex].

  \bibitem{pdg}
  PDG 2018\\
  M. Tanabashi et al. (Particle Data Group), Phys. Rev. D 98, 030001 (2018).

\bibitem{Harnik:2012pb} 
  R.~Harnik, J.~Kopp and J.~Zupan,
  JHEP {\bf 1303}, 026 (2013)
  doi:10.1007/JHEP03(2013)026
  [arXiv:1209.1397 [hep-ph]].


\bibitem{Khachatryan:2016rke} 
  V.~Khachatryan {\it et al.} [CMS Collaboration],
  Phys.\ Lett.\ B {\bf 763}, 472 (2016)
  doi:10.1016/j.physletb.2016.09.062
  [arXiv:1607.03561 [hep-ex]].
  
\bibitem{Aad:2019ojw} 
  G.~Aad {\it et al.} [ATLAS Collaboration],
  Phys.\ Lett.\ B {\bf 801}, 135148 (2020)
  doi:10.1016/j.physletb.2019.135148
  [arXiv:1909.10235 [hep-ex]].


\bibitem{Aad:2012an} 
  G.~Aad {\it et al.} [ATLAS Collaboration],
  Phys.\ Rev.\ D {\bf 86}, 032003 (2012)
  doi:10.1103/PhysRevD.86.032003
  [arXiv:1207.0319 [hep-ex]].


\bibitem{Aad:2012cfr} 
  G.~Aad {\it et al.} [ATLAS Collaboration],
  JHEP {\bf 1302}, 095 (2013)
  doi:10.1007/JHEP02(2013)095
  [arXiv:1211.6956 [hep-ex]].


\bibitem{Chatrchyan:2014nva} 
  S.~Chatrchyan {\it et al.} [CMS Collaboration],
  JHEP {\bf 1405}, 104 (2014)
  doi:10.1007/JHEP05(2014)104
  [arXiv:1401.5041 [hep-ex]].


\bibitem{Aad:2015vsa} 
  G.~Aad {\it et al.} [ATLAS Collaboration],
  JHEP {\bf 1504}, 117 (2015)
  doi:10.1007/JHEP04(2015)117
  [arXiv:1501.04943 [hep-ex]].


\bibitem{Sirunyan:2017khh} 
  A.~M.~Sirunyan {\it et al.} [CMS Collaboration],
  Phys.\ Lett.\ B {\bf 779}, 283 (2018)
  doi:10.1016/j.physletb.2018.02.004
  [arXiv:1708.00373 [hep-ex]].


\bibitem{Aaboud:2017sjh} 
  M.~Aaboud {\it et al.} [ATLAS Collaboration],
  JHEP {\bf 1801}, 055 (2018)
  doi:10.1007/JHEP01(2018)055
  [arXiv:1709.07242 [hep-ex]].


\bibitem{Sirunyan:2018zut} 
  A.~M.~Sirunyan {\it et al.} [CMS Collaboration],
  JHEP {\bf 1809}, 007 (2018)
  doi:10.1007/JHEP09(2018)007
  [arXiv:1803.06553 [hep-ex]].


\bibitem{Khachatryan:2014wca} 
  V.~Khachatryan {\it et al.} [CMS Collaboration],
  JHEP {\bf 1410}, 160 (2014)
  doi:10.1007/JHEP10(2014)160
  [arXiv:1408.3316 [hep-ex]].


\bibitem{Aad:2014vgg} 
  G.~Aad {\it et al.} [ATLAS Collaboration],
  JHEP {\bf 1411}, 056 (2014)
  doi:10.1007/JHEP11(2014)056
  [arXiv:1409.6064 [hep-ex]].


\bibitem{Arganda:2019gnv} 
  E.~Arganda, X.~Marcano, N.~I.~Mileo, R.~A.~Morales and A.~Szynkman,
  Eur.\ Phys.\ J.\ C {\bf 79}, no. 9, 738 (2019)
  doi:10.1140/epjc/s10052-019-7249-7
  [arXiv:1906.08282 [hep-ph]].


\bibitem{Aad:2016blu} 
  G.~Aad {\it et al.} [ATLAS Collaboration],
  Eur.\ Phys.\ J.\ C {\bf 77}, no. 2, 70 (2017)
  doi:10.1140/epjc/s10052-017-4624-0
  [arXiv:1604.07730 [hep-ex]].


\bibitem{Aad:2015pfa} 
  G.~Aad {\it et al.} [ATLAS Collaboration],
  Phys.\ Rev.\ Lett.\  {\bf 115}, no. 3, 031801 (2015)
  doi:10.1103/PhysRevLett.115.031801
  [arXiv:1503.04430 [hep-ex]].


\bibitem{Aaboud:2018jff} 
  M.~Aaboud {\it et al.} [ATLAS Collaboration],
  Phys.\ Rev.\ D {\bf 98}, no. 9, 092008 (2018)
  doi:10.1103/PhysRevD.98.092008
  [arXiv:1807.06573 [hep-ex]].


\bibitem{Aaij:2018mea} 
  R.~Aaij {\it et al.} [LHCb Collaboration],
  Eur.\ Phys.\ J.\ C {\bf 78}, no. 12, 1008 (2018)
  doi:10.1140/epjc/s10052-018-6386-8
  [arXiv:1808.07135 [hep-ex]].


\bibitem{Staub:2013tta} 
  F.~Staub,
  Comput.\ Phys.\ Commun.\  {\bf 185}, 1773 (2014)
  doi:10.1016/j.cpc.2014.02.018
  [arXiv:1309.7223 [hep-ph]].


\bibitem{Staub:2015kfa} 
  F.~Staub,
  Adv.\ High Energy Phys.\  {\bf 2015}, 840780 (2015)
  doi:10.1155/2015/840780
  [arXiv:1503.04200 [hep-ph]].


\bibitem{Porod:2011nf} 
  W.~Porod and F.~Staub,
  Comput.\ Phys.\ Commun.\  {\bf 183}, 2458 (2012)
  doi:10.1016/j.cpc.2012.05.021
  [arXiv:1104.1573 [hep-ph]].


\bibitem{Olive:2016xmw} 
  C.~Patrignani {\it et al.} [Particle Data Group],
  Chin.\ Phys.\ C {\bf 40}, no. 10, 100001 (2016).
  doi:10.1088/1674-1137/40/10/100001

  \bibitem{loopcorrection}
 G.~Degrassi, S.~Heinemeyer, W.~Hollik, P.~Slavich and G.~Weiglein,
  Eur.\ Phys.\ J.\ C {\bf 28}, 133 (2003),
  [hep-ph/0212020];
 B.~C.~Allanach, A.~Djouadi, J.~L.~Kneur, W.~Porod and P.~Slavich,
  JHEP {\bf 0409}, 044 (2004), 
  [hep-ph/0406166];
S.~P.~Martin,
  Phys.\ Rev.\ D {\bf 75}, 055005 (2007), 
  [hep-ph/0701051];
 R.~V.~Harlander, P.~Kant, L.~Mihaila and M.~Steinhauser,
  Phys.\ Rev.\ Lett.\  {\bf 100}, 191602 (2008), 
  [Phys.\ Rev.\ Lett.\  {\bf 101}, 039901 (2008)], 
  [arXiv:0803.0672 [hep-ph]];
 S.~Heinemeyer, O.~Stal and G.~Weiglein,
  Phys.\ Lett.\ B {\bf 710}, 201 (2012), 
  [arXiv:1112.3026 [hep-ph]];
 A.~Arbey, M.~Battaglia, A.~Djouadi and F.~Mahmoudi,
  JHEP {\bf 1209}, 107 (2012), 
  [arXiv:1207.1348 [hep-ph]];
  M.~Chakraborti, U.~Chattopadhyay and R.~M.~Godbole,
  Phys.\ Rev.\ D {\bf 87}, no. 3, 035022 (2013), 
  [arXiv:1211.1549 [hep-ph]].


\bibitem{Aaboud:2017ayj} 
  M.~Aaboud {\it et al.} [ATLAS Collaboration],
  JHEP {\bf 1712}, 085 (2017)
  doi:10.1007/JHEP12(2017)085
  [arXiv:1709.04183 [hep-ex]].


\bibitem{Sher:2002ew} 
  M.~Sher,
  Phys.\ Rev.\ D {\bf 66}, 057301 (2002)
  doi:10.1103/PhysRevD.66.057301
  [hep-ph/0207136].


\bibitem{Aaboud:2016cre} 
  M.~Aaboud {\it et al.} [ATLAS Collaboration],
  Eur.\ Phys.\ J.\ C {\bf 76}, no. 11, 585 (2016)
  doi:10.1140/epjc/s10052-016-4400-6
  [arXiv:1608.00890 [hep-ex]].


\bibitem{Calibbi:2017uvl} 
  L.~Calibbi and G.~Signorelli,
  Riv.\ Nuovo Cim.\  {\bf 41}, no. 2, 71 (2018)
  doi:10.1393/ncr/i2018-10144-0
  [arXiv:1709.00294 [hep-ph]].
  

  
\bibitem{Sirunyan:2019shc} 
  A.~M.~Sirunyan {\it et al.} [CMS Collaboration],
  arXiv:1911.10267 [hep-ex].


  
\end{thebibliography}
\end{document}